\documentclass[aps,prc,showpacs,amsmath,amssymb,twocolumn,superscriptaddress,letterpaper,nofootinbib, showkeys]{revtex4-2}
\usepackage{hyperref} 
\usepackage{graphicx} 
\usepackage{dcolumn} 
\usepackage{bm} 
\usepackage{xcolor}
\usepackage[mathlines]{lineno} 
\newcommand{\lambdadp}{{\lambda'}_{\!\! d}}
\newcommand{\lambdanp}{{\lambda'}_{\!\! n}}
\newcommand{\lambdapp}{{\lambda'}_{\!\! p}}
\begin{document}
\preprint{}
\title{Deep-inelastic electron-deuteron scattering with spectator nucleon tagging \\
at the electron-ion collider. Extracting free nucleon structure}
\author{Alexander Jentsch}
\affiliation{Department of Physics, Brookhaven National Laboratory, Upton, New York 11973, USA}
\email{ajentsch@bnl.gov}
\author{Zhoudunming Tu}
\affiliation{Department of Physics, Brookhaven National Laboratory, Upton, New York 11973, USA}
\affiliation{Center for Frontiers in Nuclear Science, Stony Brook, New York 11794, USA}
\email{zhoudunming@bnl.gov}
\author{Christian Weiss}
\affiliation{Theory Center, Jefferson Lab, Newport News, Virginia 23606, USA}
\email{weiss@jlab.org}
\date{\today}
%
%
\begin{abstract}
\begin{description}
\item[Background] Deep-inelastic scattering (DIS) on the deuteron with spectator nucleon tagging represents a unique method for extracting the free neutron structure functions and exploring the nuclear modifications of bound protons and neutrons. The detection of the spectator (with typical momenta $\lesssim$ 100 MeV/$c$ in the deuteron rest frame) controls the nuclear configuration during the DIS process and enables a differential analysis of nuclear effects. At the future electron-ion collider (EIC) such measurements will be performed using far-forward detectors. 
\item[Purpose] Simulate deuteron DIS with proton or neutron tagging with the baseline EIC far-forward detector design. Quantify detector acceptance and resolution effects. Study feasibility of free nucleon structure extraction using pole extrapolation in the spectator momentum.
\item[Methods] DIS events with proton and neutron spectators are generated using the BeAGLE
Monte Carlo generator. The spectator nucleon momentum is reconstructed including effects of detector acceptance and resolution. Pole extrapolation is performed under realistic conditions. The free nucleon structure extraction is validated by comparing with the input model.
\item[Results] Proton and neutron spectator detection is possible over the full transverse momentum range $0 < p_T < 100$ MeV/$c$ needed for pole extrapolation. Resolution effects on the distributions before corrections are 
$\sim$ 10\% for proton and $\sim$ 30\% for neutron spectators. The overall accuracy of nucleon structure extraction is expected to be at the few-percent level.
\item[Conclusions] Free neutron structure extraction through proton tagging and pole extrapolation
is feasible with the baseline EIC far-forward detector design. The corresponding extraction of free 
proton structure through neutron tagging provides a reference point for future 
studies of nuclear modifications.
\end{description}
\end{abstract}
%
\maketitle
\tableofcontents
\newpage
\section{Introduction}
Deep-inelastic lepton scattering (DIS) represents a principal tool for exploring the short-range structure of hadrons and nuclei and studying the expressions of quantum chromodynamics in the perturbative and non-perturbative regimes. DIS measurements are performed on the proton, light ions ($2 \leq A \lesssim 12$), and heavy ions, with complementary physics purposes. DIS measurements on light ions pursue several specific objectives. One objective is to extract the
DIS observables of the neutron, to enable the flavor separation of the nucleon's partonic
structure expressed in the parton distribution functions (PDFs) \cite{Ethier:2020way,Gao:2017yyd,Aidala:2012mv}, 
generalized parton distributions (GPDs) \cite{Goeke:2001tz,Diehl:2003ny,Belitsky:2005qn,Guidal:2013rya},
and transverse momentum dependent structures (TMDs) \cite{Bacchetta:2006tn,Signori:2013mda,Barone:2010zz}. 
A second objective is to study the nuclear modifications of partonic structure (EMC effect at $x > 0.3$, antishadowing at $x \sim 0.1$), to explain their dynamical origin and connection
with conventional nuclear interactions \cite{Frankfurt:1988nt,Geesaman:1995yd,Malace:2014uea,Hen:2016kwk}. A third objective is to measure coherent and diffractive scattering on light nuclei, to characterize
the quark/gluon	structure of the nucleus in novel ways \cite{Fucini:2020vpr},
and to observe the onset of the coherent phenomena expected in heavy nuclei at small $x$ (shadowing, diffraction) \cite{Frankfurt:2011cs,Kopeliovich:2012kw}.

The future electron-ion collider (EIC) will enable a comprehensive program of DIS measurements
on light ions. The accelerator design provides light ion beams of several species, including the deuteron $d \equiv$ $^2$H, $^{3}\rm{He}$, $^{4}\rm{He}$. It supports electron-ion collisions in a broad range of center-of-mass energies, $\sim$ 20--100 GeV/nucleon for electron-deuteron, at luminosities $\sim 10^{33}$--$10^{34}$ cm$^{-2}$ s$^{-1}$; for more details on the capabilities for protons, light ion, and heavy ion beams see Ref.~\cite{ref:EICCDR}.
Ion polarization will be available for the $^{3}\rm{He}$ beams, and possibly also for the $d$.
DIS-type measurements on light ions will be performed with inclusive, semi-inclusive, and exclusive final states. The EIC central detector will provide excellent coverage for the scattered electron and the current fragmentation region of the DIS final states. In addition, a suite of optimized far-forward detectors (pseudorapidity $\eta > 4.5$) will enable detection of the nuclear breakup state and/or the identification of coherent nuclear events. A description of the proposed reference detector and detailed discussion of the requirements can be found in the recently completed EIC Yellow Report~\cite{AbdulKhalek:2021gbh}.

The main challenge in the interpretation of DIS measurements on light nuclei lies in the treatment
of nuclear binding effects. The nucleus participates in the DIS process in a variety of nuclear
configurations characterized by the nucleon momenta, spins, interactions, and non-nucleonic degrees of freedom; in a quantum-mechanical superposition described by the nuclear wave function. The nuclear binding effects one needs to account for generally depend on the nuclear configuration.
In neutron structure extraction one needs to correct for dilution from scattering on the protons
and eliminate effects of nucleon motion and interactions. In studies of the EMC effect
one wants to connect the observed modifications of partonic structure with a particular range of
nucleon momenta or distances and the interactions between them (e.g. a possible connection with short-range nucleon-nucleon correlations~\cite{Hen:2011}). With inclusive nuclear DIS measurements, where no detection of the nuclear breakup state is performed, one has no information on the nuclear configurations during the DIS process and must model the nuclear binding effects in all possible configurations and sum over them, resulting in large theoretical uncertainties. This problem can be overcome with tagged measurements, where one detects part or all of the nuclear breakup state, so that one can use the breakup observables to infer the nuclear configuration during the DIS process. In this way one can effectively control the nuclear configuration during the DIS process and treat the nuclear effects
in defined configurations. In neutron structure extraction, one can select configurations where
the neutron is effectively free. In studies of the EMC effect, one can select configurations with definite nucleon momenta/distances and control the strength of nucleon interactions. The method has great potential but presents new challenges: for theory, the description of the nuclear breakup and final-state interactions; for experiment, the detection
of spectator protons, neutrons, and/or other nuclear fragments at very high pseudorapidity.

The tagging method is particularly effective in DIS measurements on the deuteron. The deuteron wave
function in nucleonic degrees of freedom ($pn$) is simple and well-known up to nucleon momenta
$\sim$ 300 MeV/$c$; non-nucleonic degrees of freedom such as $\Delta$ isobars are suppressed \cite{Frankfurt:1981mk}.
The detection of the spectator nucleon (proton or neutron) identifies the active nucleon and completely
fixes the nuclear configuration in the DIS process. Deuteron DIS with proton spectator tagging at low
momenta $p_p \lesssim$ 100 MeV/$c$ selects DIS events on the neutron in average $pn$ configurations
in the deuteron, where some nuclear modifications are present. By performing an extrapolation
in the proton spectator momentum one can reach configurations where the nucleons are at asymptotically 
large separations and effectively free, and in this way extract the free 
neutron structure function (so-called pole extrapolation) \cite{Sargsian:2005rm,Strikman:2017koc,Cosyn:2020kwu}.
Because of the symmetry between the proton and neutron in the deuteron, one can use the same technique to
extract the free proton structure functions with neutron spectator tagging, which allows one to
validate the method by comparing with measurements on the proton target. In addition, deuteron DIS with
proton or neutron tagging at higher momenta $p_{p, n} \sim$ few 100 MeV/$c$ selects small-size $pn$
configurations with significant interactions and allows one to study the EMC effect as a function
of the configuration size. Other applications include tagged DIS on the polarized deuteron
(vector and tensor polarization) \cite{Frankfurt:1983qs,Cosyn:2019hem,Cosyn:2020kwu}
and tagged diffractive scattering at small $x$ \cite{Frankfurt:2003jf,Frankfurt:2006am}.

Deuteron DIS with proton tagging was measured in fixed-target experiments at JLab with a 6 GeV electron beam energy
using the CLAS spectrometer and the BoNuS proton detector \cite{Baillie:2011za,Tkachenko:2014byy}.
The results are used to constrain the $F_{2n}/F_{2d}$ structure function ratio at large $x$. Measurements at
12 GeV electron beam energy are planned with the BoNuS and ALERT detectors \cite{Bonus12,Armstrong:2017zqr}.
The BoNuS setup detects only protons with momenta $p_p \gtrsim$ 70 MeV/$c$ (slower protons cannot
escape the target), which makes pole extrapolation difficult and requires a model-dependent
extraction of free neutron structure. Other DIS experiments with proton and neutron tagging at
larger momenta $p_{p, n} \sim$ few 100 MeV/$c$ explore the EMC effect and its possible connection with nucleon short-range correlations \cite{Klimenko:2005zz,Hen:2011,Hen:2014vua}.

In tagged DIS at the EIC, the spectator nucleon (proton or neutron) from the deuteron breakup emerges in the outgoing ion beam direction, with a momentum given by the boost of its momentum in the
deuteron rest frame,
\begin{align}
p_p(\textrm{longit}) & \; \approx \; \frac{p_d}{2} \left[ 1 + \frac{p_p(\textrm{longit, rest frame})}{m_N} \right],
\label{longitudinal_intro}
\\[1ex]
p_p(\textrm{transv}) & \; = \; p_p(\textrm{transv, rest frame}),
\end{align}
and similarly for $p \rightarrow n$. 
Here ``longitudinal'' and ``transverse'' refer to the outgoing ion beam direction.
The spectator longitudinal momentum is given by half the deuteron beam momentum, $p_d/2$, times a
factor of order unity determined by the ratio of the longitudinal rest-frame momentum and the
nucleon mass $m_N$; the spectator transverse momentum is given by the transverse rest-frame momentum. 
In this kinematics the spectator nucleon can be detected with the far-forward detectors integrated into the
outgoing ion beamline \cite{ref:EICCDR,AbdulKhalek:2021gbh}. The detection of spectator nucleons from
nuclear breakup has been a priority of the EIC far-forward detector design since its inception.
Protons are detected with a magnetic dipole spectrometer integrated in the first dipole after
the interaction point, as well as with Roman Pots and Off-Momentum Detectors along the beam path.
The setup provides excellent coverage for spectator protons over a broad range of 
$p_p(\textrm{longit})/p_d \approx \frac{1}{2}$ and $0 < p_p(\textrm{transv}) \lesssim\rm{1 \ GeV}/c$.
Neutrons are detected with a Zero-Degree Calorimeter with comparable coverage.

The unique far-forward detection capabilities of the EIC, combined with the kinematic coverage for DIS,
will enable new types of tagged DIS measurements on the deuteron that have not been possible at
existing facilities. Free neutron structure can be extracted through proton tagging
with $p_p(\textrm{transv}) \lesssim$ 100 MeV/$c$ and pole extrapolation in the spectator momentum. Free proton
structure can be determined through neutron tagging and pole extrapolation, validating the
extraction method, since free proton structure can be measured at the EIC in $ep$ collisions in similar kinematics and with the same detector configuration. Nuclear modifications can be studied in detail using both proton and neutron tagging at larger transverse momenta. The physics potential of these measurements calls for a dedicated study. A preliminary assessment of the feasibility and physics impact of tagged DIS at the EIC was made in an earlier Research and Development project \cite{JLabLDRD,Cosyn:2016oiq}; this assessment can now be taken to the next
level through full detector simulations with the actual EIC far-forward detector design.

In this series of articles we report a comprehensive study of DIS on the deuteron with spectator proton and neutron tagging at the EIC with the baseline far-forward detector design. The objectives are to explore the physics potential of tagged measurements, quantify the detector effects, and provide guidance for optimization of the far-forward detector design. We generate deuteron DIS events using the BeAGLE
Monte Carlo (MC) generator~\cite{Beagle}, reconstruct the spectator nucleon momentum including detector acceptance
and resolution and beam-related effects, and perform the physics analysis under realistic conditions. In the present article
we study the extraction of free nucleon structure from tagged DIS with pole extrapolation: both free neutron structure from proton tagging and proton structure from neutron tagging.
These applications involve far-forward proton and neutron detection at low transverse momenta 
$p_{p, n}(\textrm{transv}) \lesssim$ 100 MeV/$c$, where the acceptance is generally high and uniform, but good momentum
resolution is critical. The focus is on studying the performance of the pole extrapolation technique, quantifying the detector resolution effects, and validating the
extraction via comparisons with the input model. In a subsequent article we turn to the exploration of
nuclear modifications and the tagged EMC effect through tagging at higher transverse momenta 
$p_{p, n}(\textrm{transv}) \sim$ few 100 MeV/$c$, where the detector acceptance becomes critical \cite{Jentsch:inprep}.

The outline of the article is as follows. In Sec.~\ref{sec:theory} we summarize the kinematic
variables and experimental observables in tagged DIS, the theoretical description of deuteron structure,
and the procedure for free nucleon structure extraction through pole extrapolation.
In Sec.~\ref{sec:simulations} we describe the BeAGLE MC event generator, the EIC far-forward detectors,
and the procedure used to quantify the impact of detector acceptance and resolution effects.
In Sec.~\ref{sec:analysis} we present the steps of the simulated analysis, including the deuteron
reduced cross section measurement, the removal of deuteron structure, 
and the extraction of free nucleon structure through pole extrapolation;
we also validate the result of the free nucleon structure extraction by comparing with the model input.
In Sec.~\ref{sec:discussion} we discuss the experimental and theoretical uncertainties of the proposed
measurement and explain which of those can be quantified with the present simulations and which
require future detailed studies. In Sec.~\ref{sec:conclusions} we summarize our conclusions.
In Sec.~\ref{sec:extensions} we discuss possible extensions of the method to other processes of interest.

Appendix~\ref{app:deuteron} summarizes the deuteron structure model used in the event generation
and physics analysis. Appendix~\ref{app:resolution} describes the far-forward detector acceptances and resolutions obtained from full simulations, which are used to model the detector response in the present study. These materials can be used in simulations of other nuclear breakup processes at EIC. 
\section{Process and theory} 
\label{sec:theory}
\subsection{Kinematic variables}
\label{subsec:kinematic}
We begin by summarizing the variables and observables of tagged DIS measurements,
the theoretical description in terms of nuclear and nucleonic structure, and the procedure
for extracting free nucleon structure through pole extrapolation. The theoretical framework
is described in Refs.~\cite{Strikman:2017koc,Cosyn:2020kwu}; here we adapt the formalism 
to the experimental analysis.

\begin{figure}[t]
\includegraphics[width=.24\textwidth]{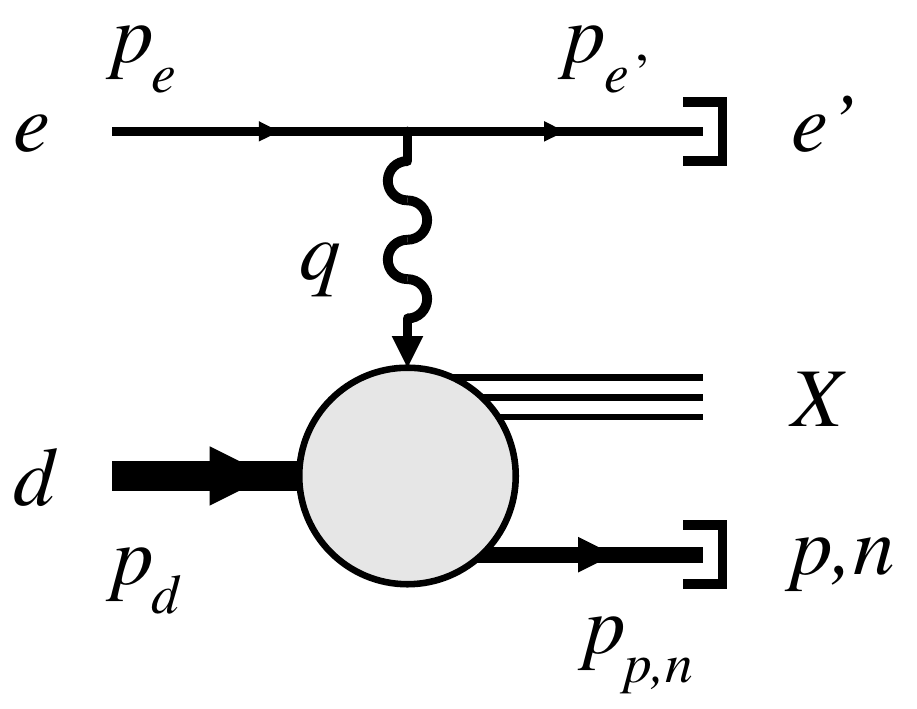}
\caption[]{DIS on the deuteron with detection of a proton (or neutron) in the nuclear
fragmentation region, $e + d \rightarrow e' + X + p (n)$ (``tagged DIS'').}
\label{fig:deut_tagged}
\end{figure}

We consider unpolarized inclusive electron scattering on the deuteron, with detection of the scattered electron
and an identified proton or neutron in the nuclear fragmentation region (see Fig.~\ref{fig:deut_tagged}),
\begin{align}
e(p_e) + d(p_d) &\rightarrow e'(p_{e'}) + X + p(p_p) \;\; [\textrm{or} \; n(p_n)].
\label{tagged_dis}
\end{align}
The 4-momenta of the particles are denoted as indicated in Eq.~(\ref{tagged_dis}) and Fig.~\ref{fig:deut_tagged}.
The 4-momentum transfer is defined as the difference of the initial and final electron 4-momenta,
\begin{align}
q \; & \equiv \; p_e - p_{e'} .
\end{align}
The DIS process is characterized by the invariant momentum transfer $Q^2 \equiv -q^2$ and the scaling
variables
\begin{align}
x \; &\equiv \;  \displaystyle \frac{Q^2}{(p_d q)}, 
& 0 < x < 2,
\label{x_def}
\\[1ex]
y \; &\equiv \; \displaystyle \frac{(p_d q)}{(p_d p_e)},
& 0 < y < 1,
\label{y_def}
\end{align}
which satisfy the relation
\begin{align}
Q^2 \; &= \;  {\textstyle\frac{1}{2}} xy (s_{ed} - M_d^2) ,
\label{s_Q2_relation}
\end{align}
where $s_{ed} \equiv (p_e + p_d)^2$ is the invariant squared center-of-mass energy of the $ed$ collision
and $M_d$ is the deuteron mass.
The variable $x$ in Eq.~(\ref{x_def}) is the Bjorken variable computed with $1/2$ times the deuteron 4-momentum
and can be interpreted as the Bjorken variable for scattering on a nucleon in an ``unbound'' deuteron in which
each nucleon carries half the deuteron 4-momentum ($x$ is a kinematic variable and does not depend
on this interpretation; the effects of nuclear binding on the scattering of the nucleon are discussed below).
The variable $y$ in Eq.~(\ref{y_def}) can be interpreted as the fractional energy loss of the electron
in the scattering on the deuteron with 4-momentum $p_d$, or, equivalently, in the scattering from a
nucleon with $p_d/2$ in an unbound deuteron.
%
%

We study the process Eq.~(\ref{tagged_dis}) with proton or neutron detection in the final state. 
For simplicity we write the following formulas for the case of 
proton detection; the formulas for neutron detection can be obtained by simple exchange $p \leftrightarrow n$. 
Situations where additional considerations are needed in obtaining the neutron formulas are indicated in the text.

The momentum of the detected proton (or neutron) in Eq.~(\ref{tagged_dis}) depends on the reference
frame and can be characterized in various ways. For theoretical analysis it is convenient to use a frame
in which the momentum transfer $\bm{q}$ and the deuteron momentum $\bm{p}_d$ are collinear and define the $z$-axis (so-called collinear frame). In this frame one describes the nucleon momentum in terms of its light-front components,
\begin{align}
p_p^+ \; &\equiv \; p_p^0 + p_p^z,
\hspace{2em} \bm{p}_{pT} \; \equiv \; (p_p^x, p_p^y), 
\label{p_p_plus}
\end{align}
and expresses the proton plus momentum component as a fraction of 1/2 the deuteron plus momentum
\begin{align}
p_p^+ \; \equiv \; \alpha_p p_d^+/2,
\hspace{2em}
0 < \alpha_p < 2;
\label{alpha_p_def}
\end{align}
the value of $p_d^+$ is arbitrary and can be changed by a boost along the $z$-axis. The light-front
variables $\alpha_p$ and $\bm{p}_{pT}$ then characterize the proton momentum in any frame that can be
connected to the collinear frame by a Lorentz transformation. Lorentz-invariant expressions of
$\alpha_p$ and $\bm{p}_{pT}$, which allow one to compute the variables directly from the 4-vector
components of $p_d, q$ and $p_p$ in any frame without going through a Lorentz transformation,
are given in Sec.III H of Ref.\cite{Cosyn:2020kwu}.

The kinematic limit of $\alpha_p$ in the tagged DIS process is dictated by the conservation
of light-front plus momentum in the collinear frame,
\begin{align}
\alpha_p \; &< \; 2(1 - \xi) \; \approx \; \; 2(1 - x),
\label{alpha_limit}
\end{align}
where
\begin{align}
\xi \; &\equiv \; \frac{2 x}{1 + \sqrt{1 + x^2 M_d^2/Q^2}}
\; = \; x \; + \; \mathcal{O}\left(\frac{x^2 M_d^2}{Q^2}\right).
\end{align}
Equation~(\ref{alpha_limit}) expresses the fact that the spectator nucleon can have plus momentum
at most as large as the total plus momentum of the DIS final state produced on the deuteron
(the initial deuteron plus momentum, less the plus momentum removed by the virtual photon).
For $x \ll 1$ the upper limit of $\alpha_p$ is close to 2; for $x \sim 1$ it is significantly below 2.
The invariant phase space element in the spectator momentum is expressed in terms of the variables
$\alpha_p$ and $\bm{p}_{pT}$ as
\begin{align}
d\Gamma_p \; &\equiv \; [2 (2\pi)^3]^{-1}
\frac{d^3 p_p}{E_p} \; = \; [2 (2\pi)^3]^{-1} \frac{d\alpha_p}{\alpha_p} d^2p_{pT}
\nonumber \\
& = \; [2 (2\pi)^3]^{-1} \frac{d\alpha_p}{\alpha_p} \frac{dp_{pT}^2}{2} d\phi_p ,
\label{phase_space_alpha_pt}
\end{align}
where $p_{pT}^2 \equiv |\bm{p}_{pT}|^2$ is the squared modulus and
$\phi_p$ the azimuthal angle of $\bm{p}_{pT}$. The case of neutron detection is described by the same
formulas with $p \rightarrow n$.
\subsection{Differential cross section}
\label{subsec:cross_section}
The basic observable in tagged DIS Eq.~(\ref{tagged_dis}) is the cross section 
\begin{align}
d\sigma [ed \rightarrow e'Xp]\hspace{2em} (\textrm{or}\; p \rightarrow n),
\label{cross_section_orig}
\end{align}
differential in the momentum of the scattered electron and the observed proton (or neutron);
the energy and momentum of the unobserved hadronic final state $X$ follow from 4-momentum conservation
and do not count as independent variables.
The general structure of the tagged cross section and its parametrization in terms of invariant structure functions
are described in Refs.~\cite{Strikman:2017koc,Cosyn:2020kwu}. Here we represent the
electroproduction cross section Eq.~(\ref{cross_section_orig}) in terms of a reduced
photoproduction cross section, as is customary in proton electroproduction at HERA; see e.g.\
Ref.~\cite{Aaron:2009bp}. In this representation
\begin{align}
d\sigma [ed \rightarrow e'Xp] \; =& \; 
\textrm{Flux}(x, Q^2) \, dx \, dQ^2 \, \frac{d\phi_{e^{'}}}{2\pi}
\nonumber \\
& \times \, \sigma_{{\rm red}, d}(x, Q^2; \alpha_p, p_{pT}, \phi_p) \, d\Gamma_p .
\label{eq:disXSec}
\end{align}
The first factor represents the virtual photon flux produced by the electron scattering process;
it depends only on the electron variables and is differential in $x, Q^2$, and the
azimuthal angle of the scattered electron around the electron beam direction, $\phi_{e'}$.
The function is given by
\begin{align}
&\textrm{Flux}(x, Q^2) \; \equiv \; \frac{2\pi\alpha_{\rm em}^{2}y^{2}}{Q^{4}(1-\epsilon)x}
\nonumber \\
&= \; \frac{2\pi\alpha_{\rm em}^{2}[1-(1-y)^{2}]}{Q^{4}x} \left[ 1 +
\mathcal{O}\left(\frac{x^2 m_N^2}{Q^2}\right) \right] ,
\label{eq:fluxFact}
\end{align}
where $\alpha_{\rm em}$ is the fine structure constant and $\epsilon$ is the virtual photon
polarization parameter,
\begin{align}
1 - \epsilon \; = \; \frac{y^2}{1 + (1 - y)^2}
+ \mathcal{O}\left(\frac{x^2 m_N^2}{Q^2}\right);
\end{align}
the exact expression including power corrections can be found in
Refs.~\cite{Strikman:2017koc,Cosyn:2020kwu}. The flux factor defined in
Eq.~(\ref{eq:fluxFact}) is identical to the one in electron-nucleon scattering with
a nucleon beam of 1/2 the deuteron beam momentum and with $x$ as the standard
nucleon Bjorken variable (up to completely negligible kinematic corrections proportional to
the deuteron binding energy); this definition allows for an easy comparison
with the formulas and results in electron-proton scattering at HERA.

The second factor in Eq.~(\ref{eq:disXSec}) represents the reduced cross section
for tagged deuteron DIS; it depends on both the electron variables $x, Q^2$ and the
tagged nucleon variables $\alpha_p, p_{pT}$, and $\phi_p$, and is proportional to the
differential phase space of the tagged nucleon momentum, $d\Gamma_p$ in Eq.~(\ref{phase_space_alpha_pt}).
The function $\sigma_{{\rm red}, d}$ contains the hadronic information in the tagged DIS cross section.
Its dependence on $y$ (or $\epsilon$) and on $\phi_p$ is dictated by relativistic covariance
and can be made explicit by expanding it in structure functions,\footnote{
The $F_L$ structure function in Eq.~(\ref{reduced_cross_section}) is defined as in
Ref.~\cite{Aaron:2009bp} and differs from the one in Ref.~\cite{Cosyn:2020kwu} by a factor $x$:
$F_{Ld}(\textrm{here}) = x F_{Ld}(\textrm{Ref.\cite{Cosyn:2020kwu}})$.}
\begin{align}
& \sigma_{{\rm red}, d}(x, Q^2; \alpha_p, p_{pT}, \phi_p)
\nonumber \\[1ex]
& = \; F_{2d}(x, Q^2; \alpha_p, p_{pT})
- (1 - \epsilon) F_{Ld}(x, Q^2; \alpha_p, p_{pT})
\nonumber \\[1ex]
& + \; \textrm{$\phi_p$-dependent structures}.
\label{reduced_cross_section}
\end{align}
The tagged structure functions $F_{2d}$ and $F_{Ld}$ depend on the tagged proton momentum
only through the light-front fraction $\alpha_p$ and the transverse momentum modulus $p_{pT}$.
Equation~(\ref{reduced_cross_section}) presents only the terms in the reduced cross section
that do not explicitly depend on $\phi_p$; the full structure including the $\phi_p$
dependence is given in Ref.~\cite{Cosyn:2021inprep}. In the analysis performed here we consider
only the cross section averaged over $\phi_p$, in which the $\phi_p$ dependent structures
average to zero,
\begin{align}
&\overline{\sigma}_{{\rm red}, d}(x, Q^2; \alpha_p, p_{pT})
\nonumber
\\[1ex]
& \equiv \; \int\frac{d\phi_p}{2\pi} \; \sigma_{{\rm red}, d}(x, Q^2; \alpha_p, p_{pT}, \phi_p)
\nonumber
\\[1ex]
& = \; F_{2d}(x, Q^2; \alpha_p, p_{pT}) - (1 - \epsilon) F_{Ld}(x, Q^2; \alpha_p, p_{pT})
\label{reduced_cross_section_phi_average}
\end{align}
The case of neutron tagging is described by the same formulas with $p \rightarrow n$.

The tagged deuteron structure functions $F_{2d}, F_{Ld}$ etc.\ contain the basic information that can
be extracted from tagged DIS measurements. They represent a special case of semi-inclusive DIS
structure functions \cite{Bacchetta:2006tn}, with the target being the deuteron nucleus and the observed
hadron being a nucleon in the nuclear fragmentation region. Note that no assumptions regarding a composite
nuclear structure in terms of nucleons or a particular reaction mechanism are made in the
general decomposition of Eqs.~(\ref{reduced_cross_section}) and (\ref{reduced_cross_section_phi_average}).
A relation between the tagged deuteron structure functions and the neutron structure functions
$F_{2n}$ and $F_{Ln}$ (or the proton structure functions in the case of neutron tagging)
can be established only in the context of a theoretical description
combining nuclear and nucleonic structure.
\subsection{Deuteron structure description}
\label{subsec:deuteron}
The theoretical treatment of tagged DIS starts from the picture of the nucleus as a composite system
of nucleons and describes the cross section by combining nuclear and nucleonic structure.
The objectives are to predict the tagged structure functions in terms of the nucleon structure functions
and calculable nuclear structure elements, and to enable the extraction of the nucleon structure functions
from the tagged DIS data.

Nuclear binding modifies the deep-inelastic structure of the nucleus relative to the sum of free nucleons in several ways: (i) The motion of the nucleons in the nucleus shifts the effective kinematics in the electron-nucleon scattering process, (ii) The interactions between the nucleons affect the partonic structure seen by the high-energy probe. These effects can be interpreted alternatively as a modification of bound nucleon structure or the presence of non-nucleonic degrees of freedom and have been the object of
extensive studies; see Refs.~\cite{Frankfurt:1988nt,Arneodo:1992wf,Geesaman:1995yd,Malace:2014uea,Hen:2016kwk} for a review. In the present study of tagged DIS, the interaction effects are eliminated by pole extrapolation, which selects large-size $pn$ configurations in the deuteron where the nucleons are effectively free. This avoids the need for an explicit description of these effects and greatly simplifies the theoretical treatment.

The nucleonic structure of the deuteron is described at fixed light-front time $x^+ \equiv t + z$
(light-front quantization). This quantization scheme is unique in the sense
that the energy off-shellness of the electron-nucleon scattering subprocess (the energy difference
between the initial and final state) remains finite in the limit of large incident energy,
so that one can describe the electron-nucleon
subprocess in terms of the on-shell scattering amplitude and construct a composite description \cite{Frankfurt:1981mk}.
The nucleon 4-momenta in the deuteron are characterized by their ``plus'' and
transverse components in the collinear frame [see Sec.~\ref{subsec:kinematic} and
Eqs.~(\ref{p_p_plus}) and (\ref{alpha_p_def})]
\begin{align}
&p_p^+ = \alpha_p p_d^+/2, & \bm{p}_{pT},
\label{lf_momentum_def}
\\[1ex]
&p_n^+ = (2 - \alpha_p) p_d^+/2, & \bm{p}_{nT} = - \bm{p}_{pT} ;
\label{neutron_momentum}
\end{align}
the proton and neutron momenta are related by light-front momentum conservation (the transverse momentum
of the deuteron bound state in the collinear frame is zero, $\bm{p}_{dT} = 0$). The ``minus'' components
of the 4-momenta play the role of energies and are fixed by the mass-shell conditions $p_{p, n}^2 = m_N^2$,
\begin{align}
p_p^- = (|\bm{p}_{pT}|^2 + m_N^2)/p_p^+ ,
\\[1ex]
p_n^- = (|\bm{p}_{nT}|^2 + m_N^2)/p_n^+ .
\end{align}
Here $m_N \equiv (m_p + m_n)/2$ denotes the average nucleon mass; we assume isospin
symmetry and neglect the difference of proton and neutron masses, see Appendix~\ref{app:proton_neutron}.
The superposition of $pn$ configurations in the deuteron is described by the light-front wave function
\begin{align}
\Psi_d (\alpha_p, \bm{p}_{pT}),
\label{wf_general}
\end{align}
which is normalized such that
\begin{align}
\int \frac{d\alpha_p \; d^2 p_{pT}}{\alpha_p (2 - \alpha_p)} \;
|\Psi_d (\alpha_p, \bm{p}_{pT})|^2 
\; = \; 1.
\label{wf_normalization}
\end{align}
Here we suppress the nucleon and deuteron spin variables for brevity; the full expressions
including spins are given in Appendix~\ref{app:spin} and Ref.~\cite{Cosyn:2020kwu}.
The deuteron light-front wave function Eq.~(\ref{wf_general}) can be obtained by solving the
light-front bound state equation with realistic $pn$ interactions \cite{Frankfurt:1981mk}.
In the present study we use an approximation where the light-front wave function is
constructed from the well-known non-relativistic wave function of the deuteron bound state,
see Appendix~\ref{app:nonrelativistic}. This approximation is accurate at nucleon rest-frame momenta
$|\bm{p}_{p, n}| \lesssim$ 100 MeV/$c$ used in low-momentum tagging. In particular, the approximation
correctly implements the analytic properties of the deuteron wave function and the ``nucleon pole''
used in the extraction of free nucleon structure with pole extrapolation,
see Appendix~\ref{app:pole}. Altogether, the deuteron light-front structure is theoretically well understood and can reliably be constructed in the momentum range probed in the present study.

%
%
\begin{figure}[t]
\includegraphics[width=.98\columnwidth]{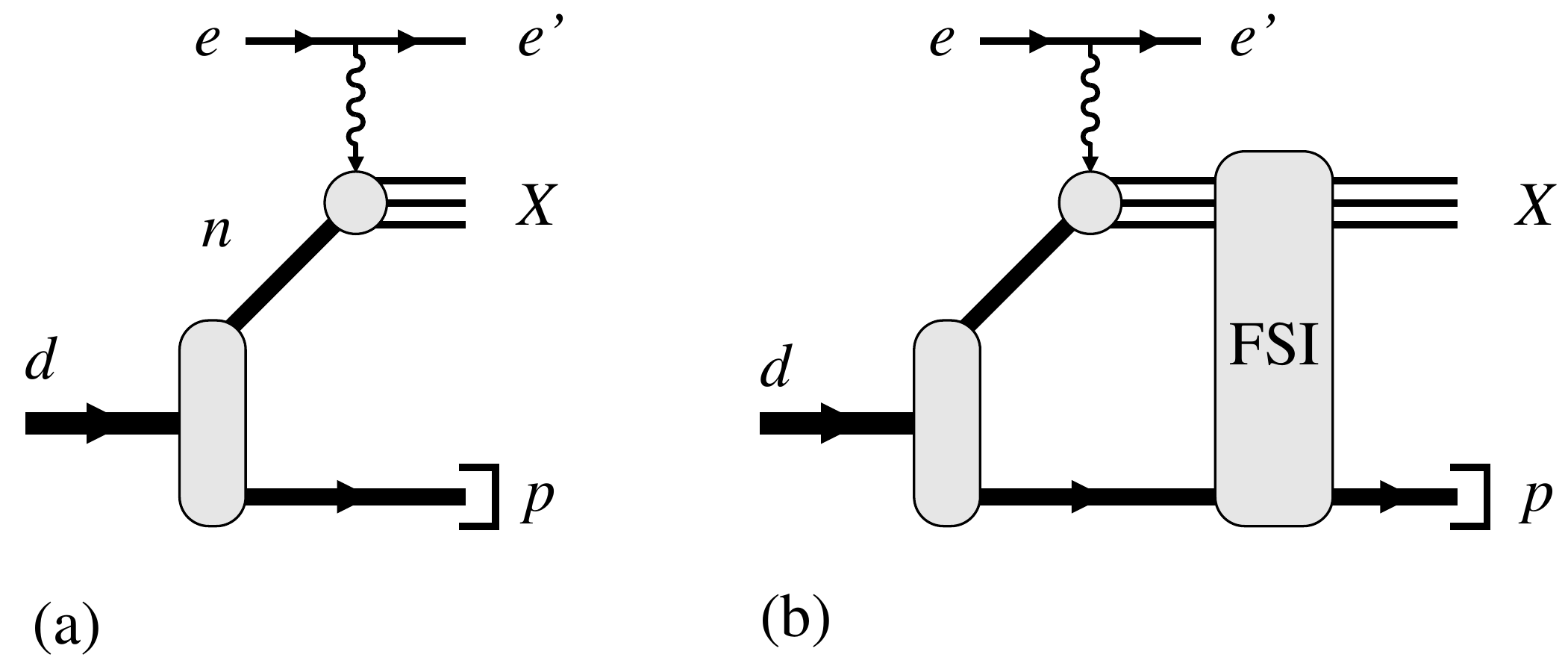}
\caption[]{Theoretical description of tagged DIS. (a)~Impulse approximation.
(b) Final-state interactions.}
\label{fig:deut_impulse_fsi}
\end{figure}
The tagged DIS cross section is calculated in the impulse approximation (see Fig.~\ref{fig:deut_impulse_fsi}a).
It takes into account the motion of the active nucleon in the deuteron and its correlation with the
spectator kinematics as governed by the deuteron wave function, but does not include dynamical initial-state modifications or final-state interactions (Fig.~\ref{fig:deut_impulse_fsi}b).
In the impulse approximation the tagged deuteron structure functions
for the case of proton tagging are obtained as \cite{Strikman:2017koc,Cosyn:2020kwu}
\begin{align}
& F_{2d}(x, Q^2; \alpha_p, p_{pT})
\nonumber
\\[1ex]
&= \;  [2(2\pi)^3] \; \mathcal{S}_{d} (\alpha_p, p_{pT}) \; F_{2n} (x_n, Q^2)
\nonumber
\\[1ex]
&+ \; \textrm{initial-state modifications}
\nonumber
\\[1ex]
&+ \; \textrm{final-state interactions},
\label{F2_ia}
\\[2ex]
& F_{Ld}(x, Q^2; \alpha_p, p_{pT}) 
\nonumber
\\[1ex]
&= \;  [2(2\pi)^3] \; \mathcal{S}_{d} (\alpha_p, p_{pT}) \; F_{Ln} (x_n, Q^2)
\nonumber
\\[1ex]
&+ \; \textrm{initial-state modifications}
\nonumber
\\[1ex]
&+ \; \textrm{final-state interactions}.
\label{FL_ia}
\end{align}
Here $\mathcal{S}_{d}$ is the deuteron light-front spectral function. It depends on the
tagged proton momentum variables $\alpha_p$ and $p_{pT} \equiv |\bm{p}_{pT}|$
(unpolarized deuteron) and represents the density of the deuteron light-front wave function
times a flux factor depending on $\alpha_p$,
\begin{align}
\mathcal{S}_d (\alpha_p, p_{pT})
\; \equiv \; \frac{|\Psi_d(\alpha_p, \bm{p}_{pT})|^2}{2 - \alpha_p} .
\label{spectral_impulse}
\end{align}
$F_{2n}$ and $F_{Ln}$ are the DIS structure functions of the neutron. They are evaluated
at the effective scaling variable
\begin{align}
x_n \; & \equiv \; \frac{x}{2 - \alpha_p} ,
\label{x_n}
\end{align}
which results from the fact that the neutron plus momentum in the deuteron is determined by
that of the spectator proton, see Eq.~(\ref{neutron_momentum}). This shows how detection of
the spectator fixes the nuclear configuration in the DIS process. The momentum transfer
$Q^2$ in the neutron structure functions is equal to the electron variable $Q^2$.
The quoted expressions for the neutron structure function arguments $x_n$ and $Q^2$
are valid up to power corrections $\sim m_N^2/Q^2$, which are negligible in our kinematics.

From the impulse approximation results for the tagged structure functions
we obtain the tagged reduced cross section as
\begin{align}
& \bar\sigma_{{\rm red}, d}(x, Q^2; \alpha_p, p_{pT})
\nonumber
\\[1ex]
&= \;  [2(2\pi)^3] \; \mathcal{S}_{d} (\alpha_p, p_{pT}) \; \sigma_{{\rm red}, n} (x_n, Q^2)
\nonumber
\\[1ex]
&+ \; \textrm{initial-state modifications}
\nonumber
\\[1ex]
&+ \; \textrm{final-state interactions},
\label{sigma_red_ia}
\end{align}
where $\sigma_{{\rm red}, n}$ is the reduced cross section for DIS on the neutron,
\begin{align}
\sigma_{{\rm red}, n}(x_n, Q^2)
\; = \; F_{2n}(x_n, Q^2) - (1 - \epsilon) F_{Ln}(x_n, Q^2).
\label{reduced_cross_section_neutron}
\end{align}
Here we have used the fact that the $\epsilon$ parameter for scattering on the deuteron (with
scaling variable $x$) is equal to that for scattering on the neutron (with $x_n$) up to power
corrections $\propto y^2 m_N^2/Q^2$, which can be neglected in DIS kinematics. Equation~(\ref{sigma_red_ia})
concisely summarizes the impulse approximation result as relevant to unpolarized tagged DIS without L/T separation.

The case of neutron tagging is described by analogous formulas.
The reduced deuteron cross section for neutron tagging in the impulse approximation is
\begin{align}
& \bar\sigma_{{\rm red}, d}(x, Q^2; \alpha_n, p_{nT})
\nonumber
\\[1ex]
&= \;  [2(2\pi)^3] \; \mathcal{S}_{d} (\alpha_n, p_{nT}) \; \sigma_{{\rm red}, p} (x_p, Q^2)
\; + \; \textrm{mod.}
\label{sigma_red_ia_neutron}
\end{align}
where the spectral function is now
\begin{align}
& \mathcal{S}_d (\alpha_n, p_{nT})
\; \equiv \; \frac{|\Psi_d(\alpha_n, \bm{p}_{nT})|^2}{2 - \alpha_n} ,
\label{spectral_neutron}
\end{align}
and the reduced proton cross section is
\begin{align}
& \sigma_{{\rm red}, p}(x_p, Q^2)
\; = \; F_{2p}(x_p, Q^2) - (1 - \epsilon) F_{Lp}(x_p, Q^2),
\end{align}
evaluated at the effective scaling variable
\begin{align}
& x_p \; \equiv \; \frac{x}{2 - \alpha_n}.
\label{x_p}
\end{align}
The spectral function for neutron tagging, Eq.~(\ref{spectral_neutron}), is given by the same
mathematical function as for proton tagging, Eq.~(\ref{spectral_impulse}), only evaluated at the neutron
momentum variables $\alpha_n$ and $\bm{p}_{nT}$. The symmetry properties of the deuteron light-front wave
function and spectral function under proton-neutron interchange are summarized in Appendix~\ref{app:proton_neutron};
see in particular Eqs.~(\ref{proton_neutron_wf}) and (\ref{proton_neutron_spectral}).
\subsection{Nucleon structure extraction}
\label{subsec:extraction}
In the present work we study the extraction of free nucleon structure from tagged DIS measurements.
This requires separating deuteron and nucleon structure in the measured cross section,
and -- if possible -- suppressing the effects of initial-state modifications and final-state interactions.
This can be accomplished using the dependence of the tagged cross section on the spectator nucleon momentum.
In the following we discuss two methods:

\subparagraph{Method I: Integration
over spectator momentum.} This method uses proton tagging only to identify events with an active neutron,
but does not measure the spectator momentum, so it integrates over the spectator or active nucleon kinematics. 
Initial-state modifications of neutron structure are not suppressed; their strength is comparable to
that in inclusive nuclear DIS. This is the traditional method for tagged DIS analysis. The overall
uncertainty is dominated by the unknown nuclear modifications. We simulate such measurements with the EIC
only as a reference point, to enable comparisons with other methods and internal validation.

\subparagraph{Method II: Pole extrapolation in spectator momentum.} This method uses the analytic properties
of the deuteron wave function to select large-size $pn$ configurations
in the deuteron, in which both initial-state modifications and final-state interactions are suppressed.
It enables a model-independent extraction of free neutron structure. The resulting uncertainty is
determined by the quality of the measurement and the extrapolation procedure. This novel method demands
good detector coverage and resolution at small proton momenta $|\bm{p}_{pT}| \ll 100$ MeV/$c$ and
may become possible with the EIC. We simulate such measurements with the EIC as the potential method of choice for free neutron structure extraction with proton tagging (and proton structure with neutron tagging) and quantify its uncertainties.
\subsection{Integration over spectator momentum}
\label{subsec:integration}
Taking the impulse approximation expressions of the tagged deuteron structure functions, Eqs.~(\ref{F2_ia}) and
(\ref{FL_ia}), and computing the integral over the spectator momentum, we obtain
\begin{subequations}
\label{int}
\begin{align}
&\int d\Gamma_p \; F_{2d}(x, Q^2; \alpha_p, p_{pT}) 
\nonumber 
\\[1ex]
&= \;
\int\frac{d\alpha_p}{\alpha_p} \, d^2 p_{pT} \; 
\mathcal{S}_{d}(\alpha_p, p_{pT}) \; F_{2n} (x_n, Q^2)
\label{F2d_int_1}
\\[1ex]
&\approx \; F_{2n} (x, Q^2) \int\frac{d\alpha_p}{\alpha_p} \, d^2 p_{pT} \; 
\mathcal{S}_{d}(\alpha_p, p_{pT})
\label{F2d_int_2}
\\[1ex]
&= \;  F_{2n} (x, Q^2) ,
\label{F2d_int}
\end{align}
\end{subequations}
and similarly for $F_{Ld}$ and $F_{Ln}$. The expressions of Eq.~(\ref{int})
are valid up to initial-state modifications of nucleon structure, which are not included in the IA.
In Eq.~(\ref{F2d_int_1}) we have used the explicit form of the proton phase space element $d\Gamma_p$
in terms of the light-front momentum variables $\alpha_p$ and $\bm{p}_{pT}$,
Eq.~(\ref{phase_space_alpha_pt}).
In Eq.~(\ref{F2d_int_1}) the variable $x_n$ in the argument of $F_{2n}$ 
depends on the integration variable $\alpha_p$, so that the integral represents a convolution
of the spectral function and the neutron structure function. The integrand is concentrated around
$\alpha_p = 1$ because of the shape of the spectral function. Equation~(\ref{F2d_int_2}) is
obtained in the ``peaking approximation,'' where one neglects the $\alpha_p$ dependence of $F_{2n}$
under the integral and approximates
\begin{align}
F_{2n}(x_n, Q^2) \; \approx \; F_{2n}(x, Q^2) .
\end{align}
Equation~(\ref{F2d_int}) is then obtained by using the integral relation (``sum rule'')
of deuteron spectral function Eq.~(\ref{spectral_impulse}),
\begin{align}
\int\frac{d\alpha_p}{\alpha_p} \, d^2 p_{pT} \; 
\mathcal{S}_{d}(\alpha_p, p_{pT})
\; = \; 1,
\end{align}
which follows from the normalization condition of the deuteron light-front wave
function Eq.~(\ref{wf_normalization}). 

Combining the relations of Eq.~(\ref{int}) for $F_{2d}$ and $F_{Ld}$, we find that the integral of
the reduced tagged cross section from Eq.~(\ref{reduced_cross_section}) over the spectator momentum is
equal to the reduced cross section for scattering on the neutron
\begin{align}
&\int d\Gamma_p \; \sigma_{{\rm red}, d}(x, Q^2; \alpha_p, p_{pT}, \phi_p)
\nonumber
\\[1ex]
&\approx \; F_{2n} (x, Q^2) - (1 - \epsilon) F_{2L} (x, Q^2)
\nonumber
\\[1ex]
&\equiv \; \sigma_{{\rm red}, n}(x, Q^2).
\end{align}
Including the flux factor Eq.~(\ref{eq:fluxFact}), we see that the tagged electron-deuteron cross section in the impulse approximation, integrated over the spectator momentum, is equal to the electron-neutron cross section
in nominal kinematics (neutron with 1/2 the deuteron beam momentum; see Sec.~\ref{subsec:cross_section})
\begin{align}
&d\sigma [ed \rightarrow e'Xp({\rm integrated})]
\nonumber
\\[1ex]
&= \; \textrm{Flux}(x, Q^2) \; dx \, dQ^2 \,
\frac{d\phi_{e'}}{2\pi} \;
\nonumber
\\[1ex]
& \hspace{1em} \times \;
\int d\Gamma_p \; \sigma_{{\rm red}, d}(x, Q^2; \alpha_p, p_{pT}, \phi_p)
\nonumber
\\[1ex]
&= \; \textrm{Flux}(x, Q^2) \; dx \, dQ^2 \,
\frac{d\phi_{e'}}{2\pi} \, \times \, \sigma_{{\rm red}, n}(x, Q^2).
\label{integrated_cross_section}
\end{align}
Equation~(\ref{integrated_cross_section}) provides a simple connection between the tagged deuteron and
neutron cross sections in the IA. Its value for neutron structure extraction is limited by the fact that
it is specific to the impulse approximation and does not include initial-state modifications, which are generally as large as
in untagged inclusive scattering. However, in simulations with an IA-based physics model,
Eq.~(\ref{integrated_cross_section}) serves as a simple way of recovering the neutron structure input,
and we use it in this sense in our validation in Sec.~\ref{subsec:validation}
\subsection{Pole extrapolation in spectator momentum\label{sub:poleremoval}}
\label{subsec:pole}
The deuteron light-front wave function is an analytic function of the nucleon momentum variables.
It can be considered both at physical (real) and unphysical (imaginary) values of the momentum,
and its behavior is governed by singularities in the unphysical region. The dominant feature
at low momenta is the ``nucleon pole'' singularity, which results from the free motion of the
nucleons outside the range of the nucleon-nucleon interactions. It is of the form \cite{Cosyn:2020kwu}
\begin{align}
\Psi_{d} (\alpha_p, \bm{p}_{pT})
= \; \frac{R}{p_{pT}^2 + a_T^2} \; + \; \textrm{(less singular)} .
\label{pole_wf}
\end{align}
The pole in $p_{pT}^2 \equiv |\bm{p}_{pT}|^2$ occurs at $p_{pT}^2 = - a_T^2 < 0$ in the unphysical region.
The position is given by the squared transverse mass of the $pn$ configuration,
\begin{align}
a_T^2 \; \equiv \; a_T^2(\alpha_p) \; = \; (\alpha_p - 1)^2 m_N^2 + \alpha_p (2 - \alpha_p) a^2 ,
\label{a_T_def}
\end{align}
where 
\begin{align}
a^2 \equiv m_N \epsilon_d
\label{a2_def}
\end{align}
in which $\epsilon_d$ is the deuteron binding energy. $a^2$ defines the position
of the nucleon pole in the nonrelativistic wave function and provides a measure
of the natural size of the deuteron (see Appendix~\ref{app:pole}).
The transverse mass Eq.~(\ref{a_T_def}) depends on $\alpha_p$ and attains its minimal value
$a_T^2 = a^2$ at $\alpha_p = 1$. The residue of the pole in Eq.~(\ref{pole_wf}) is given by
\begin{align}
& R \; \equiv \; R(\alpha_p) \; \equiv \; \alpha_p (2 - \alpha_p) \, \sqrt{m_N} \, \Gamma ,
\label{R_def}
\end{align}
where $\Gamma$ is the residue of the nucleon pole of the nonrelativistic deuteron wave function
(see Appendix~\ref{app:pole}).

The nucleon pole singularity in the deuteron light-front wave function Eq.~(\ref{pole_wf}) has a simple physical
interpretation. In the transverse coordinate representation of the wave function, it describes $pn$ configurations
with asymptotically large transverse size $r_T \rightarrow \infty$ in the deuteron \cite{Cosyn:2020kwu}.
At such distances the nucleons are outside of the range of the nucleon-nucleon interactions, and their
motion is essentially free. The nucleon pole thus represents a universal feature of the deuteron as a
weakly bound system. It can be derived from the structure of the bound state equation and is found in all
models that describe the deuteron as a bound state with a finite-range nucleon-nucleon interaction.
The pole position Eq.~(\ref{a_T_def}) follows from kinematic considerations and is known exactly.
The residue Eq.~(\ref{R_def}) can be inferred from non-relativistic deuteron structure calculations
and low-energy measurements and is known with an accuracy $\lesssim$ 1\% (see Appendix~\ref{app:pole}
and Table~\ref{tab:pole}).

Tagged DIS at physical transverse momenta $p_{pT}^2 > 0$ always samples finite-size $pn$ configurations
in the deuteron, where nucleon interactions are generally present. However, analytic continuation
to unphysical momenta $p_{pT}^2 \rightarrow - a_T^2$ can effectively access infinite-size configurations
$r_T \rightarrow \infty$, where nucleon interactions are absent. Final-state interactions of the DIS
products with the spectator are also suppressed in such configurations. This allows one to practically
realize DIS on an unbound nucleon in the deuteron and to extract free neutron structure.

%
%
\begin{figure}[t]
\includegraphics[width=.98\columnwidth]{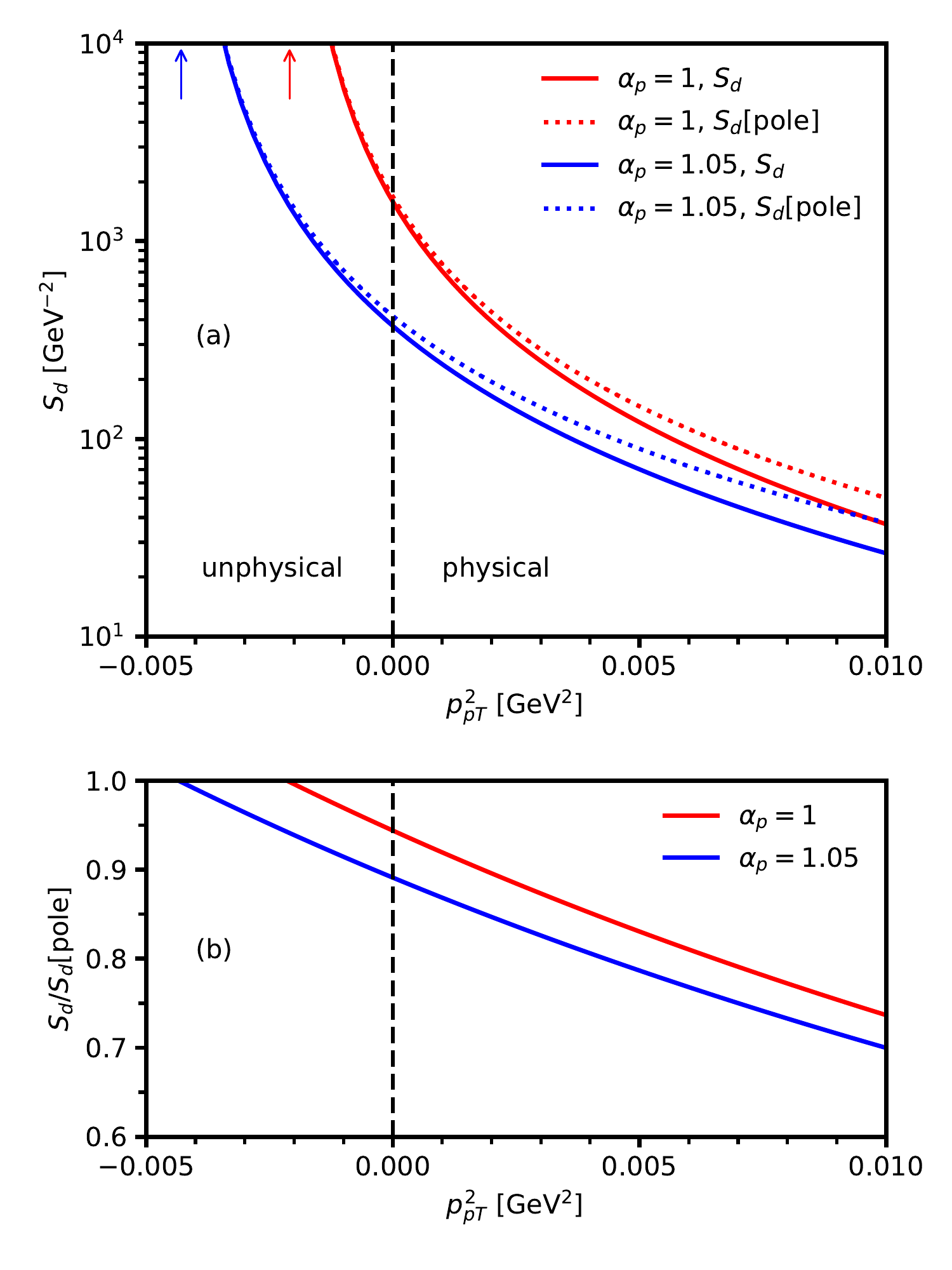}
\caption[]{Deuteron spectral function and its nucleon pole contribution.
(a) Spectral function $\mathcal{S}_d(\alpha_p, p_{pT})$, Eq.~(\ref{spectral_impulse}) (solid lines)
and its pole term $\mathcal{S}_d(\alpha_p, p_{pT})$[pole], Eq.~(\ref{pole_spectral}) (dashed lines)
as a function of $p_{pT}^2$, for two fixed values of $\alpha_p$. The plot shows the
functions in the physical ($p_{pT}^2 > 0$) and unphysical regions ($p_{pT}^2 < 0$).
The positions of the poles at $p_{pT}^2 = -a_T^2(\alpha_p)$ are marked by arrows 
for the two values of $\alpha_p$.
(b) Ratio of the full spectral function and the pole term, Eq.~(\ref{spectral_pole_ratio}),
as a function of $p_{pT}^2$, for the same fixed values of $\alpha_p$.}
\label{fig:spectral_pole}
\end{figure}
In the light-front spectral function Eq.~(\ref{spectral_impulse}), the nucleon pole Eq.~(\ref{pole_wf})
gives rise to a singularity of the form,
\begin{align}
\mathcal{S}_{d} (\alpha_p, p_{pT})
\; &= \; \frac{C}{(p_{pT}^2 + a_T^2)^2} \; + \; \textrm{(less singular)}
\nonumber
\\[1ex]
& \equiv \; \mathcal{S}_{d} (\alpha_p, p_{pT})[\textrm{pole}],
\label{pole_spectral}
\end{align}
where the residue is 
\begin{align}
& C \; \equiv \; C(\alpha_p) \; \equiv \; \alpha_p^2 (2 - \alpha_p) \, m_N \Gamma^2 .
\label{C_def}
\end{align}
The nucleon pole Eq.~(\ref{pole_spectral}) dominates the behavior of the spectral function at low transverse
momenta in the physical region. Figure~\ref{fig:spectral_pole}a shows the spectral function and its pole term
as functions of $p_{pT}^2$ in the physical ($p_{pT}^2 > 0$) and unphysical regions ($p_{pT}^2 < 0$),
for fixed values of $\alpha_p$. (This numerical example uses the two-pole
parametrization of the wave function of Appendix~\ref{app:twopole}.)
One observes that the pole term accounts for most of the value
and the variation of the spectral function in the physical region $0 < p_{pT}^2 \lesssim$ 0.01 GeV$^2$;
the spectral function varies by an order-of-magnitude over this interval.
Figure~\ref{fig:spectral_pole}b shows the ratio of the spectral function and the pole term,
\begin{align}
\frac{\mathcal{S}_{d} (\alpha_p, p_{pT})}{\mathcal{S}_{d} (\alpha_p, p_{pT})[\textrm{pole}]} ,
\label{spectral_pole_ratio}
\end{align}
as a function of $p_{pT}^2$ in the same interval. One notes that the deviations of the full spectral
function from the pole term are $\lesssim 30\%$ for $0 < p_{pT}^2 \lesssim$ 0.01 GeV$^2$, and that
dividing the full spectral function by the pole term removes most of the $p_{pT}^2$ dependence.
In particular, the plots also illustrate that, when following the dependence into the unphysical
region $p_{pT}^2 < 0$ and approaching the pole at $p_{pT}^2 \rightarrow -a_T^2$, the pole term represents
the entire spectral function, as implied by Eq.~(\ref{pole_spectral}), and the ratio becomes unity,
\begin{align}
\frac{\mathcal{S}_{d} (\alpha_p, p_{pT})}{\mathcal{S}_{d} (\alpha_p, p_{pT})[\textrm{pole}]}
\; \rightarrow \; 1
\hspace{2em} (p_{pT}^2 \rightarrow -a_T^2) .
\label{spectral_pole_limit}
\end{align}

The existence of the nucleon pole and its properties enable a unique method for neutron structure extraction
from DIS on the deuteron with proton tagging (``pole extrapolation''):
\begin{itemize}
\item[(i)] Measure the tagged DIS cross section Eq.~(\ref{eq:disXSec})
at fixed $\alpha_p$ and small physical transverse momenta, remove the flux factor,
and extract the $\phi_p$-integrated reduced cross section:
\begin{align}
& \sigma_{{\rm red}, d}(x, Q^2; \alpha_p, p_{pT}, \phi_p)
\nonumber \\
& = \; [\textrm{Flux}(x, Q^2)]^{-1} \; \frac{d\sigma [ed \rightarrow e'Xp]}
{dx \, dQ^2 \, (d\phi_{e'}/2\pi) d\Gamma_p} ,
\label{pole_analysis_reduced}
\\[1ex]
&\overline{\sigma}_{{\rm red}, d}(x, Q^2; \alpha_p, p_{pT})
\nonumber \\
& = \; \int\frac{d\phi_p}{2\pi} \; \sigma_{{\rm red}, d}(x, Q^2; \alpha_p, p_{pT}, \phi_p).
\label{pole_analysis_reduced_average}
\end{align}
\item[(ii)] Divide the reduced cross section by the theoretically known pole factor of
the spectral function, Eq.(\ref{pole_spectral}): 
\begin{align}
& \frac{\overline{\sigma}_{{\rm red}, d}(x, Q^2; \alpha_p, p_{pT})}
{\mathcal{S}_{d} (\alpha_p, p_{pT})[\textrm{pole}]}.
\label{pole_analysis_ratio}
\end{align}
In the impulse approximation, Eqs.~(\ref{F2_ia}) and (\ref{FL_ia}), this ratio is theoretically equal to
\begin{align}
[...] \; &= \; \frac{\mathcal{S}_{d} (\alpha_p, p_{pT}) \; \sigma_{{\rm red}, n}(x_n, Q^2)}
{\mathcal{S}_{d} (\alpha_p, p_{pT})[\textrm{pole}]}
\nonumber
\\[1ex]
&+ \; \textrm{initial-state modifications}
\nonumber
\\[2ex]
&+ \; \textrm{final-state interactions}.
\label{IA_ratio}
\end{align}
\item[(iii)] Extrapolate the ratio Eq.~(\ref{pole_analysis_ratio}) in $p_{pT}^2$ to the point $p_{pT}^2 = -a_T^2$
(pole position) in the unphysical region by a low-order polynomial fit:
\begin{align}
& \sigma_{{\rm red}, n}(x_n, Q^2)
\nonumber \\
& = \; \lim [p_{pT}^2 \rightarrow -a_T^2] \; \;
\frac{\overline{\sigma}_{{\rm red}, d}(x, Q^2; \alpha_p, p_{pT})}{\mathcal{S}_{d} (\alpha_p, p_{pT})[\textrm{pole}]}.
\label{pole_analysis_extrapolation}
\end{align}
At the pole both initial-state modifications and final-state interactions in Eq.~(\ref{IA_ratio}) vanish,
and the impulse approximation becomes exact \cite{Sargsian:2005rm,Strikman:2017koc}. 
Furthermore, at the pole the ratio of the full spectral function with respect to its pole
becomes unity, Eq.~(\ref{spectral_pole_limit}). The procedure Eq.~(\ref{pole_analysis_extrapolation}) therefore returns
the {\it free neutron reduced cross section} without nuclear modifications.
\end{itemize}
\section{Simulation tools and detectors}
\label{sec:simulations}
\subsection{BeAGLE Monte Carlo generator} 
\label{subsec:beagle}
We now describe the simulation tools used in the present study, the EIC far-forward detector design,
and the specific considerations in the reconstruction of the far-forward spectator momentum.

BeAGLE is a general purpose lepton-nucleus ($eA$) event generator, which combines PYTHIA~6.4
\cite{Sjostrand:2006za}, DPMJET~3.0 \cite{Roesler:2000he}, and the
FLUKA model \cite{Bohlen:2014buj,Ferrari:2005zk}. A detailed description of the entire program can
be found in Ref.~\cite{Beagle}. In the modeling of scattering on the deuteron and other light ions,
DPMJET and FLUKA are not used, and the high-energy scattering process is treated in the IA
(no final-state interactions).
The parts of BeAGLE used in the present analysis are the electron-nucleon DIS process modeled
by PYTHIA 6.4 and the deuteron light-front spectral function describing the kinematic distribution of
the spectator nucleon; see Sec.~\ref{subsec:deuteron}, Appendix~\ref{app:deuteron}, and Ref.~\cite{Tu:2020ymk}.
The generator thus implements the theoretical framework for tagged deuteron DIS as described in
Sec.~\ref{sec:theory} and can be used for simulations of nucleon structure extraction with pole extrapolation.
The present version of BeAGLE uses the parametrization of Ref.~\cite{CiofidegliAtti:1995qe} to
generate the deuteron spectral function; the nucleon pole parameters for this parametrization
are given in Appendix~\ref{app:pole} and Table~\ref{tab:pole}. The code version (git tag)
used in the present study is BeAGLE 1.01.03; the deuteron structure implementation in this version is the
same as in BeAGLE 1.0 used in Ref.~\cite{Tu:2020ymk}.

BeAGLE describes electron-neutron scattering in the same way as electron-proton scattering in PYTHIA 6,
adjusting for the different isospin in the initial state (different proton and neutron PDFs).
For technical reasons the proton and neutron PDFs in the generator include an empirical
nuclear modification modeled on that of the alpha particle $A=4$ \cite{Beagle}. This feature is irrelevant
for the present study and does not affect the results, as we look at the extracted neutron and proton
structure functions only relative to the model input, not in absolute terms.
Because	BeAGLE describes electron-deuteron scattering in the impulse approximation, it gives the same result when
extracting nucleon structure using integration over the spectator momentum or pole extrapolation
(Methods I and II of Secs.~\ref{subsec:integration}). We use this feature to validate the results
of the pole extrapolation simulations in Sec.~\ref{subsec:validation}.

The treatment of the kinematics of the electron-nucleon scattering process in BeAGLE requires some explanation.
For technical reasons BeAGLE evaluates the electron-nucleon DIS cross section at the Bjorken variable of the
unbound nucleon, $x$, Eq.~(\ref{x_def}), not at the effective variable of the bound nucleon, $x_n$, Eq.~(\ref{x_n}).
The tagged DIS cross section used in BeAGLE therefore differs from the true impulse approximation value by
the factor (for events with tagged proton and active neutron)
\begin{align}
\frac{\sigma [en \rightarrow e'X](x, Q^2)}{\sigma [en \rightarrow e'X](x_n, Q^2)} ,
\label{beagle_ratio}
\end{align}
where $\sigma [en \rightarrow e'X]$ is the electron-neutron DIS cross section. This difference needs to be taken
into account in simulations of cross section measurements with BeAGLE.
It can easily be corrected by multiplying the cross sections extracted from BeAGLE
with a correction factor given by the inverse of Eq.~(\ref{beagle_ratio}):
\begin{subequations}
\label{beagle_corr}
\begin{align}
\textrm{Corr}(x, Q^2; \alpha_p)
\; &\equiv \; \frac{\sigma [en \rightarrow e'X](x_n, Q^2)}{\sigma [en \rightarrow e'X](x, Q^2)}
\label{beagle_corr_1}
\\
&= \; \frac{\textrm{Flux}(x_n, Q^2) \; \sigma_{{\rm red}, n}(x_n, Q^2)}
{\textrm{Flux}(x, Q^2) \; \sigma_{{\rm red}, n}(x, Q^2)}
\label{beagle_corr_2}
\\
&= \; \frac{x \; \sigma_{{\rm red}, n}(x_n, Q^2)}
{x_n \; \sigma_{{\rm red}, n}(x, Q^2)}
\label{beagle_corr_3}
\\
&= \; (2 - \alpha_p) \; \frac{\sigma_{{\rm red}, n}(x_n, Q^2)}
{\sigma_{{\rm red}, n}(x, Q^2)} .
\label{beagle_corr_4}
\end{align}
\end{subequations}
In Eq.~(\ref{beagle_corr_2}) we have expressed the electron-neutron DIS cross sections in terms of the flux factors
Eq.~(\ref{eq:fluxFact}) and the reduced cross sections Eq.~(\ref{reduced_cross_section_neutron});
in Eq.~(\ref{beagle_corr_3}) we have used that the flux factor at fixed $Q^2$ is proportional to $1/x$;
in Eq.~(\ref{beagle_corr_4}) we have replaced the ratio $x/x_n$ by $2 - \alpha_p$ using Eq.~(\ref{x_n}).
Thus the correction factor is given by a simple expression in terms of the tagged proton $\alpha_p$
and the neutron reduced cross section ratio.

The correction factor Eq.~(\ref{beagle_corr}) satisfies
\begin{align}
\textrm{Corr}(x, Q^2; \alpha_p = 1) \; \equiv \; 1,
\end{align}
because $x_n = x$ at $\alpha_p = 1$; see Eq.~(\ref{x_n}).
For $|1 - \alpha_p| \ll 1$ we can expand the factor around $\alpha_p = 1$ and obtain
\begin{subequations}
\label{beagle_corr_expanded}
\begin{align}
&\textrm{Corr}(x, Q^2; \alpha_p)
\nonumber \\[1ex]
&= \; 1 \; + \; \left[ 1 - \frac{x \frac{d}{dx} \sigma_{{\rm red}, n}(x, Q^2)}{
\sigma_{{\rm red}, n}(x, Q^2)} \right] (1 - \alpha_p)
\label{beagle_corr_expanded_1}
\\[1ex]
&\approx \; 1 \; + \; (1 + \lambda) (1 - \alpha_p) .
\label{beagle_corr_expanded_2}
\end{align}
\end{subequations}
In Eq.~(\ref{beagle_corr_expanded_1}) the first term in the bracket, 1, is the ``kinematic''
correction resulting from the flux factors; the second term is the ``dynamical'' correction
resulting from the reduced cross sections. The form Eq.~(\ref{beagle_corr_expanded_2}) applies
at $x \ll 0.1$, where the reduced cross section depends on $x$ approximately as
$\sigma_{{\rm red}, n}(x, Q^2) \propto x^{-\lambda}$, with $\lambda \equiv \lambda(Q^2)$ \cite{Aaron:2009bp}.
The HERA measurements find values $\lambda \approx$ 0.15--0.2 \cite{Aaron:2009bp}, showing that the dynamical
correction is small and the kinematic correction dominates at $x\ll 0.1$. Note that the
dynamical correction is generally large at $x \gtrsim 0.1$, where the nucleon structure
functions and the reduced cross section strongly depend on $x$.
\subsection{Kinematics and event sample}
\label{subsec:event_sample}
In the present study we use the EIC configuration with 18 GeV electrons colliding with 110 GeV/nucleon deuterons,
corresponding to an electron-nucleon squared center-of-mass energy of $s_{eN} \equiv s_{ed}/2 = (89\, \textrm{GeV})^2$. The simulations can easily be adapted to other beam energy configurations \cite{ref:EICCDR}. 

The kinematic phase space used in the
analysis is $Q^{2}>10~\rm{GeV^{2}}$ (DIS region, lower values can be considered as well)
and $0.01<y<0.95$ (standard limits for event reconstruction using the electron method).
Tagged DIS and nucleon structure extraction are simulated in the range
$10^{-2} \lesssim x \lesssim 10^{-1}$. The main physical interest is in the measurements
at $x \gtrsim 0.1$, where the neutron and proton structure functions are significantly different, 
and where the free nucleon structure extraction with tagging provides a baseline for studies of 
nuclear modifications (EMC effect, antishadowing). 
Because the nucleon DIS process and the deuteron breakup are described
independently in BeAGLE, the simulations of forward spectator detection do not depend significantly 
on the choice of $x$ and $Q^2$. We therefore include in the simulations also events at $x \ll 0.1$, 
where the statistical sample is large, but use them only for studying the detector performance.

In the analysis, $10^8$ (100 million) electron-deuteron DIS events were generated above
$Q^{2} > 10~ \rm{GeV^{2}}$, corresponding to an integrated luminosity of 1~$\rm fb^{-1}$
for electron-nucleon (proton or neutron) collisions. Of these, approximately half are
events with an active proton, and half with an active neutron. The integrated luminosity represents a baseline number for DIS studies at EIC \cite{AbdulKhalek:2021gbh}. In the kinematic region considered here the uncertainties of the measured DIS cross section and extracted nucleon structure functions
are dominated by systematic effects; the large event sample was chosen only to enable accurate
phase space integration in the study of systematic effects.
\subsection{EIC far-forward detectors}
\label{subsec:detectors}
%
%
\begin{figure*}[t]
\includegraphics[width=.7\textwidth]{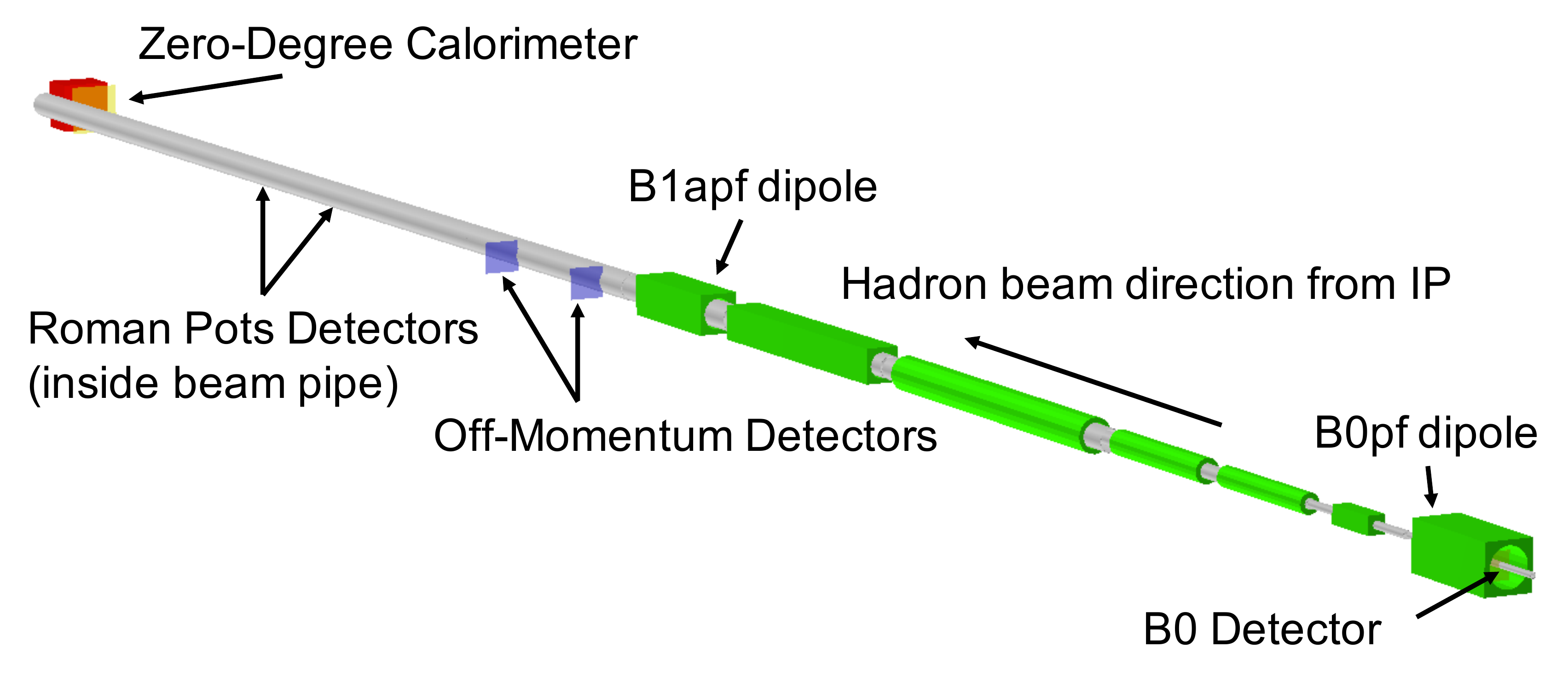}
\caption{The layout of the EIC far-forward area showing the four detector subsystems. A few of
the relevant beam-line magnets are also labeled for reference. The rectangular boxes are dipole magnets, and the cylinders are focusing quadrupoles. A schematic beam pipe is included in the drawing. The interaction point (IP)
is at the bottom right, and the hadron beam direction is noted in the figure.}
\label{fig:figure_IR}
\end{figure*}
In the present study we perform full detector simulations using the subsystems specified in 
the EIC reference detector design, with the far-forward detector configuration presented in the 
EIC Yellow Report~\cite{AbdulKhalek:2021gbh} and implemented in the EicRoot framework \cite{ref:EICROOT}.
EicRoot makes use of the ROOT Virtual Monte Carlo structure and GEANT4~\cite{GEANT4}
for detector simulations and contains classes for performing tracking and reconstruction tasks.
A three-dimensional rendering of the layout and the subsystems is shown in Fig.~\ref{fig:figure_IR}.

The far-forward detector subsystems are optimized to make best use of the available space for detectors
and maximize the geometric acceptance. The geometric acceptance for far-forward nucleons (protons or neutrons)
is a function of two variables: the polar angle of the outgoing nucleon at the interaction point relative
to the ion beam axis, and the fractional longitudinal momentum of the outgoing nucleon relative to the
deuteron beam momentum 
[see Eq.~(\ref{longitudinal_intro})],
\begin{align}
\theta_p \; &\equiv \; \frac{p_p(\textrm{transv})}{p_p(\textrm{longit})},
\label{theta_def}
\\[1ex]
\zeta_p \; &\equiv \; \frac{p_p(\textrm{longit})}{p_d} \hspace{2em} (\textrm{same for $n$}).
\label{zeta_def}
\end{align}
Here $p_d$ is the total deuteron momentum (not the momentum per nucleon),
so that a proton with the nominal longitudinal momentum $p_p(\textrm{longit}) = p_d/2$ has $\zeta_p = 1/2$.
Protons produced in the collision travel through the magnetic fields of the beam-line magnets and
experience bending in the dipoles inversely proportional to their longitudinal momentum. The proton acceptance therefore
depends on both $\zeta_p$ and $\theta_p$. Note that protons from deuteron breakup have a magnetic
rigidity $\sim$1/2 of that of the deuteron beam and experience different bending;
this effect is taken into account in the acceptance simulations; it is the main reason why 
the Off-Momentum Detectors have been added to the far-forward region (see below). 
Neutrons are not affected by the magnetic fields and propagate from the interaction point 
on straight trajectories, with acceptance only limited by the magnet apertures. 
Therefore, the neutron acceptance does not depend on $\zeta_n$ and is only a function of $\theta_n$.
Table~\ref{tab:FFDetectors_acceptance} summarizes the geometric acceptance for
far-forward protons and neutrons achieved with the present design \cite{AbdulKhalek:2021gbh}.
We note that, in the $\theta$ and $\zeta$ range considered in the present study, the acceptance does not
significantly depend on the azimuthal angle of the produced nucleon around the ion beam direction,
and we assume it to be uniform in the azimuthal angle (see Appendix~\ref{app:resolution}).
\begin{table}[t]
\begin{center}
\begin{tabular}{|l|c|c|c|c|}
\hline
Detector & Used for & $\theta$ accep. [mrad] & $\zeta$ accep. \\
\hline
 B0 tracker    & $p$ & 5.5--20.0  & N/A \\
 Off-Momentum  & $p$ & 0.0--5.0   & 0.45--0.65  \\
 Roman Pots    & $p$ & 0.0--5.0   & 0.6--0.95$^*$ \\
 Zero-Degree Calorim. & $n$ & 0.0--4.0   & N/A \\
\hline
\end{tabular}
\end{center}
\caption{\label{tab:FFDetectors_acceptance} Summary of the geometric acceptance for far-forward protons and neutrons
in polar angle $\theta$ and longitudinal momentum fraction $\zeta$, Eqs.~(\ref{theta_def}) and (\ref{zeta_def}),
provided by the baseline EIC far-forward detector design \cite{AbdulKhalek:2021gbh}. 
$^*$The Roman Pots acceptance at high values of $\zeta$ depends on the optics choice for the machine.} 
\end{table}

For most of the DIS kinematics considered in the present study, the virtual photon direction is
close to the ion beam direction, so that the nucleon longitudinal and transverse momenta relative to the ion beam axis
approximately coincide with those in the collinear frame, and one can infer the $\zeta$ and $\theta$ values
directly from the collinear frame variables (see Sec.~\ref{subsec:kinematic})
\begin{align}
\theta_p \; \approx \; 2 p_{pT}/p_d, \hspace{2em} 
\zeta_p \; \approx \; \alpha_p/2
\hspace{2em} (\textrm{same for $n$}).
\label{theta_zeta_from_collinear}
\end{align}
The tagged measurements for nucleon structure extraction use spectator detection at rest-frame momenta
$p_{p, n} \lesssim$ 100 MeV/$c$, corresponding to $0.9 \lesssim \alpha_{p, n} \lesssim 1.1$ and
$p_{pT, nT} \lesssim$ 100 MeV/$c$. With the beam momentum $p_d/2$ = 110 GeV/$c$ this implies
forward detection in the range
\begin{align}
\theta_{p, n} \; \lesssim \; 1 \, \textrm{mrad},
\hspace{2em}
0.45 \; \lesssim \; \zeta_{p, n} \; \lesssim \; 0.55 .
\end{align}
Note that the same measurements at a lower beam energy would cover a proportionally wider range in
$\theta$ and $\zeta - \frac{1}{2}$.

In the following we summarize the main features of the subsystems as relevant to the present study,
in the order in which they appear when moving away from the interaction point, see Fig.~\ref{fig:figure_IR}.
Details can be found in Refs.~\cite{ref:EICCDR,AbdulKhalek:2021gbh}. 

\subparagraph{B0 spectrometer.} The B0 spectrometer consists of four layers of silicon
tracking planes embedded in the first dipole magnet after the interaction point (B0pf). This subsystem
is designed for reconstructing charged particles with angles $5.5 < \theta < 20.0$ mrad,
such as large-angle protons from nuclear breakup. It is not used in the present study.

\subparagraph{Off-Momentum Detectors.} The Off-Momentum Detectors are designed to optimally tag
charged particles with a magnetic rigidity $\sim 1/2$ that of the beam. The present design
achieves an angular acceptance $0.0 < \theta < 5.0$ mrad, similar to that of the Roman Pot detectors
tagging particles with rigidity $\sim 1$ (see below). In the Yellow Report \cite{AbdulKhalek:2021gbh}
and the present study, the Off-Momentum Detectors were placed just after the B1apf dipole magnet; the final design of the beam pipe and vacuum system may require them to be placed elsewhere.
The Off-Momentum Detectors are the subsystem mainly used for proton tagging in the present study.
The $p_{pT}$ resolution is $\sim$20\% at $p_{pT} = 100$ MeV/$c$, with the smearing mainly resulting
from the transfer matrix used for reconstruction. The Off-Momentum Detectors have in general worse overall resolution than the Roman Pots because the transfer matrix for off-momentum particles requires a more sophisticated implementation than what was available at the time of this study.

\subparagraph{Roman Pots.} The Roman Pots detector is situated $\sim$27
meters downstream from the interaction point and consists of silicon sensors placed in Roman Pot
vessels or in shields without pots. They are injected into the beam line vacuum a few millimeters
from the hadron beam. The Roman Pots subsystem is used for capturing charged particles with small
scattering angles $0.0 < \theta < 5.0$ mrad and rigidities similar to that of the beam; the acceptance
for protons from deuteron breakup is $\zeta_p \gtrsim 0.6$.
In the present study of low-momentum tagging, the spectator protons rarely
impinge on the Roman Pots; the majority are captured by the Off-Momentum Detectors. However, the
Roman Pots become important in tagging experiments with $\alpha_p \gtrsim 1.2$, as are used in
studies of nuclear modifications or short-range correlations in the deuteron.
For the Roman Pots detector, the $p_{pT}$ resolution is $\sim$10\% at $p_{pT} = 100$ MeV/$c$,
with the smearing being driven primarily by the beam angular divergence.

\subparagraph{Zero-Degree Calorimeter.} Because neutrons are not affected by the magnetic field,
the geometric acceptance is determined by the apertures of the various beam elements that the
far-forward neutrons have to traverse before detection. The present EIC interaction region design
promises far-forward neutron acceptance at angles $\theta < 4.0~\rm{mrad}$, which allows for
tagging neutrons in a variety of final states of interest for physics studies, including the present one.
Detection of the outgoing neutrons requires hadronic calorimetry far enough down stream from
the interaction point to allow the neutrons to exit the beam pipe and impinge on the detector. This is achieved
with a Zero-Degree Calorimeter, which will also have an electromagnetic calorimeter for
tagging photons, and a layer of silicon for vetoing charged particles. In the present study
we assume a hadronic calorimeter with the same performance as in the Yellow Report~\cite{AbdulKhalek:2021gbh},
\begin{align}
\frac{\Delta E}{E} = \frac{50\%}{\sqrt{E}} \oplus 5\%, \hspace{2em} \frac{\Delta\theta}{\theta} =
\frac{3\rm{mrad}}{\sqrt{E}}
\label{ZDC_resolution}
\end{align}
\subsection{Momentum reconstruction}
\label{subsec:smearing}
The measurement of the momentum of the spectator nucleon (proton or neutron) is essential for the
analysis of tagged DIS and a main concern of the present study. The reconstruction of the 
far-forward nucleon momentum is impacted by several detector and beam effects \cite{AbdulKhalek:2021gbh}. 
First, there are effects intrinsic to the detectors themselves, e.g., finite pixel sizes or hadronic calorimeter energy resolution. Second, the momentum reconstruction algorithm for the Roman Pots and Off-Momentum Detectors depends on a transport matrix, which connects the spatial coordinates of the particles detected in the relevant far-forward detectors with their momenta at the interaction point. Currently this transport matrix is implemented as a linear transformation, which works very well for the Roman Pots system where trajectories are transported from the interaction point to the detector linearly, but does not accurately describe particles with momenta very different from the nominal beam momentum, as seen in the Off-Momentum Detectors. The strategy employed in this study was to calculate the matrix for the Off-Momentum Detectors for proton trajectories where $\zeta \sim 0.5$, allowing for minimal reconstruction smearing in the region of interest for the present study of pole extrapolation. Third, there are beam-related effects, such as
the beam angular divergence, the beam momentum spread, and the vertex smearing induced by the crab cavities used to prevent a luminosity drop in bunch collisions at the EIC crossing angle of $25$ mrad. 
All these effects are included in the simulations in the present study, 
to assess the impact on the physics measurements.

The detector and beam effects on the momentum reconstruction are quantified by processing the
BeAGLE events with the GEANT implementation of the far-forward detector systems. Because of the large
size of the event sample required for the present physics study (see Sec.~\ref{subsec:event_sample}),
running the entire generated BeAGLE sample through the GEANT full simulations proved impractical.
Instead, a representative sub-sample was processed through the full simulations to generate resolution functions for the reconstruction of momentum and energy for both protons and neutrons.
These distributions were then read as input into the analysis code and used to smear the energy
and individual momentum components as the various quantities were calculated to inject the effects
of detector reconstruction into the analysis. For reference, the distributions are presented
in Appendix~\ref{app:resolution}; they can be used in other physics studies 
requiring far-forward proton or neutron detection.
\section{Analysis and results}
\label{sec:analysis}
\subsection{Deuteron cross section measurement}
We now present the simulated analysis of tagged DIS and nucleon structure extraction
and the lessons learned and results obtained from the study.
The analysis follows the steps described in Sec.~\ref{subsec:pole} and uses the method
of pole extrapolation. The material is the BeAGLE event sample for electron-deuteron DIS of
Sec.~\ref{subsec:event_sample}, consisting of tagged proton and neutron events;
the simulated analysis applies the detector acceptance and the smearing distributions representing the
detector and beam effects on the spectator nucleon momentum reconstruction of Sec.~\ref{subsec:smearing}.
In each step we consider both proton and neutron tagging and compare the two channels. 

\begin{figure}[t]
\includegraphics[width=0.4\textwidth]{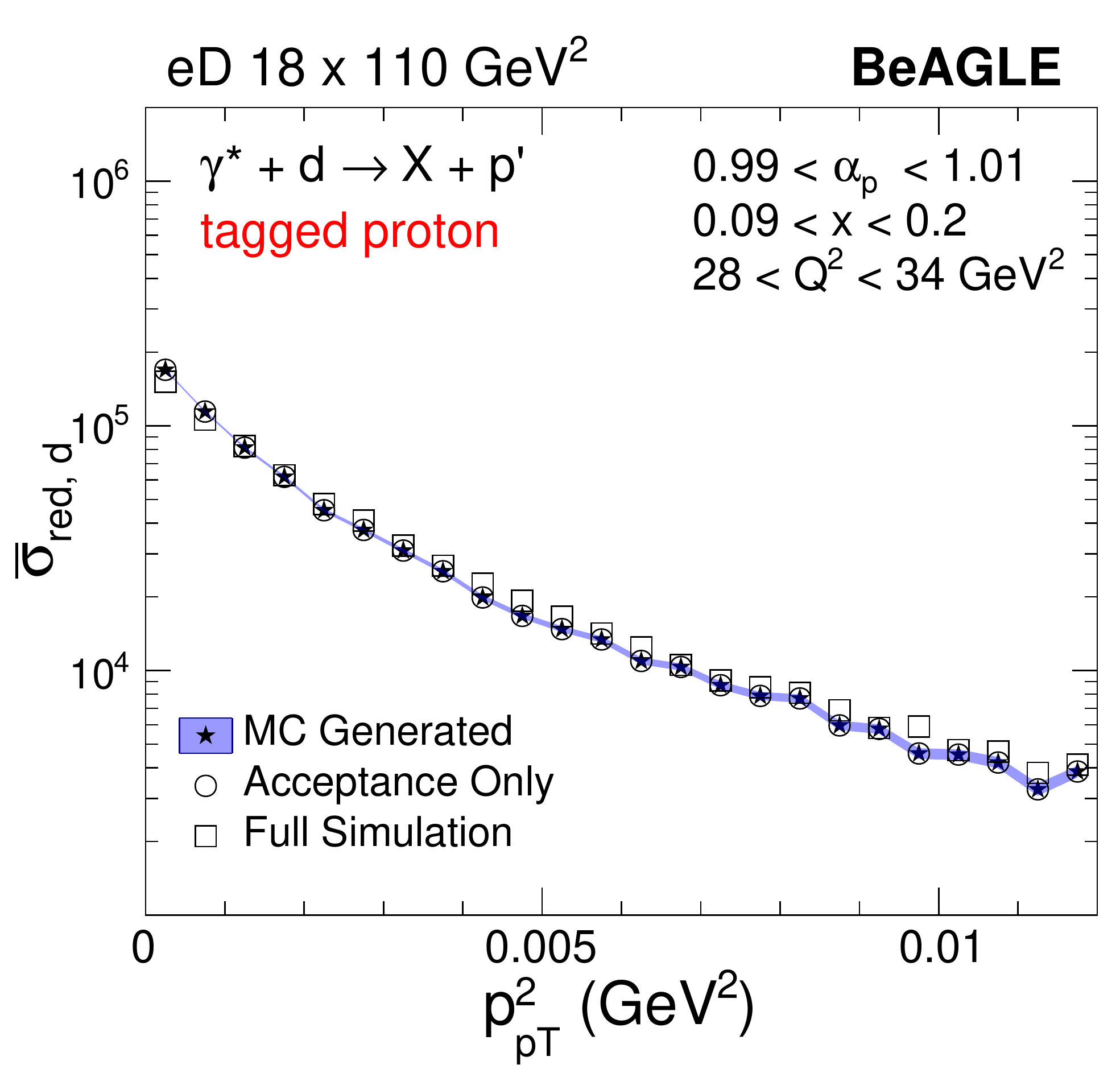}
\includegraphics[width=0.4\textwidth]{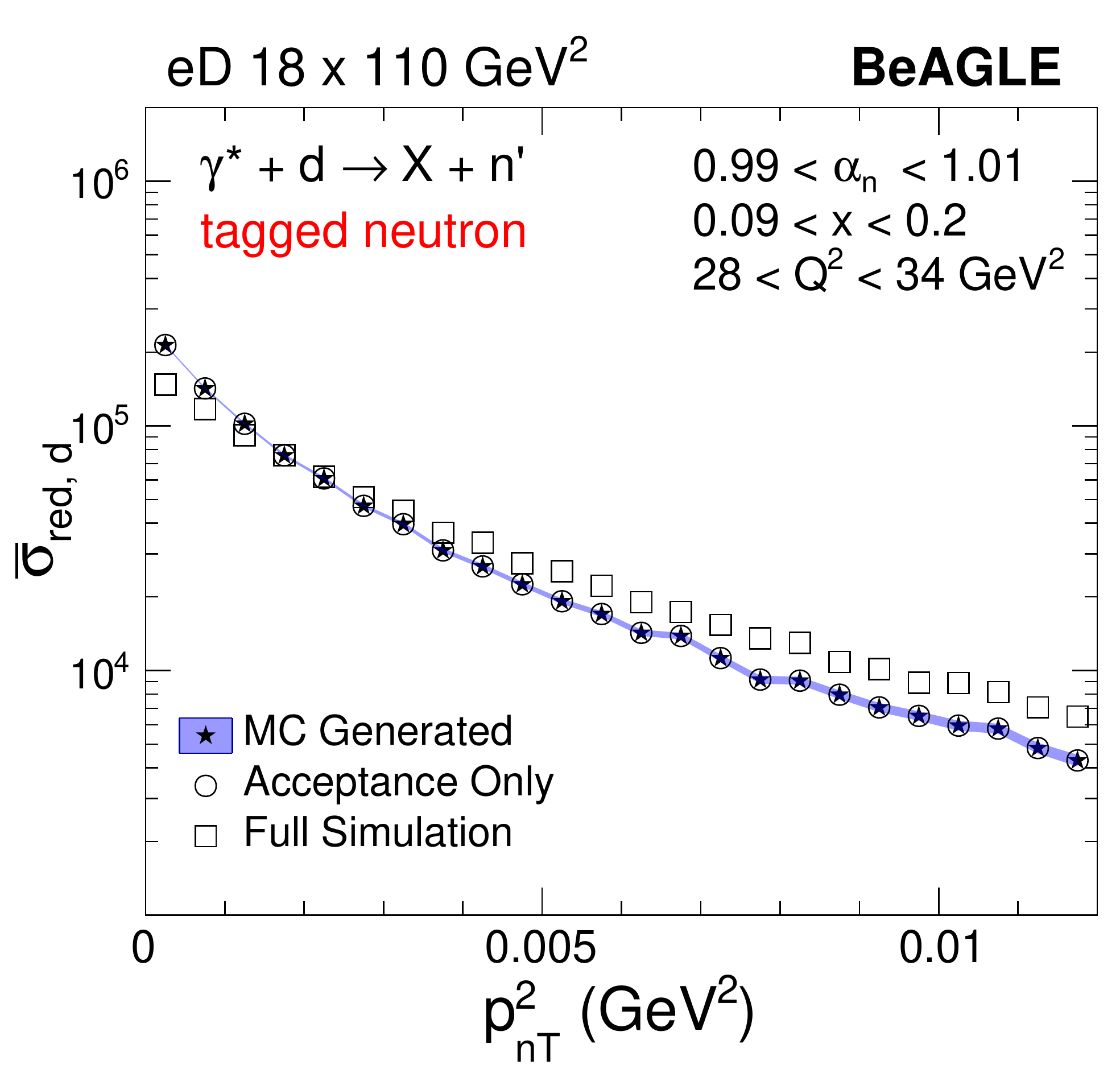}
\caption{The reduced cross section of deuteron DIS with proton and neutron tagging, 
Eq.~(\ref{pole_analysis_reduced_average}), as a function of $p_{pT}^2 \, (p_{nT}^2)$,
as extracted from simulated measurements at EIC. Stars and bands: Truth distributions from 
BeAGLE. Circles: Distributions reconstructed with detector acceptance only. Squares: 
Distributions reconstructed with full simulations.}
\label{fig:figure_red_deut}
\end{figure}

In the first step, we measure the tagged DIS cross section and extract the reduced cross section
by removing the flux factor, as specified in Eqs.~(\ref{pole_analysis_reduced}) and
(\ref{pole_analysis_reduced_average}) for proton tagging and the corresponding formulas
for neutron tagging. Figure~\ref{fig:figure_red_deut} shows the extracted $\phi_p$ ($\phi_n$) -averaged
reduced cross sections $\bar{\sigma}_{{\rm red}, d}$, as functions of the spectator transverse
momentum $p_{pT}^2$ ($p_{nT}^2$). The plots show the generator-level/MC distributions based on the BeAGLE 
events, the distributions reconstructed with acceptance effects only, and the distributions reconstructed
with the full simulations. The example covers the kinematic range is $28 < Q^{2} < 34~\rm{GeV^{2}}$, $0.09< x < 0.2$, 
and $0.99 < \alpha_p (\alpha_n) < 1.01$; similar results are obtained in other ranges. 
Comparing the truth and acceptance-only results in Fig.~\ref{fig:figure_red_deut}, 
one sees that the acceptances for both proton and neutron spectators are close to 100\% in the transverse momentum
range covered here. Comparing the acceptance-only and the full simulations, one sees the impact of the detector and beam
smearing effects on the reconstruction, typically $\sim$few percent for proton tagging and up to $\sim$30\%
for neutron tagging. In the case of neutron detection, 
the Zero-Degree Calorimeter energy resolution is the dominant source of momentum smearing.
%
%
\subsection{Implementation of pole removal}
\label{subsec:analysis_pole_removal}
In the second step of the analysis, we divide the deuteron reduced cross section by the pole
factor of the deuteron spectral function to extract the ratio Eq.~(\ref{pole_analysis_ratio}),
which gives access to the nucleon reduced cross section. This ``pole removal'' is the most
critical step of the experimental analysis and requires careful study.
The pole factor in Eq.~(\ref{pole_analysis_ratio}) is a theoretical function that
needs to be evaluated at the experimentally reconstructed spectator momentum. Because of the steep momentum dependence of the reduced cross section and the pole factor,
the uncertainties in the spectator momentum reconstruction
can have a large numerical effect on the result.

%
%
\begin{figure}[t]
\includegraphics[width=0.4\textwidth]{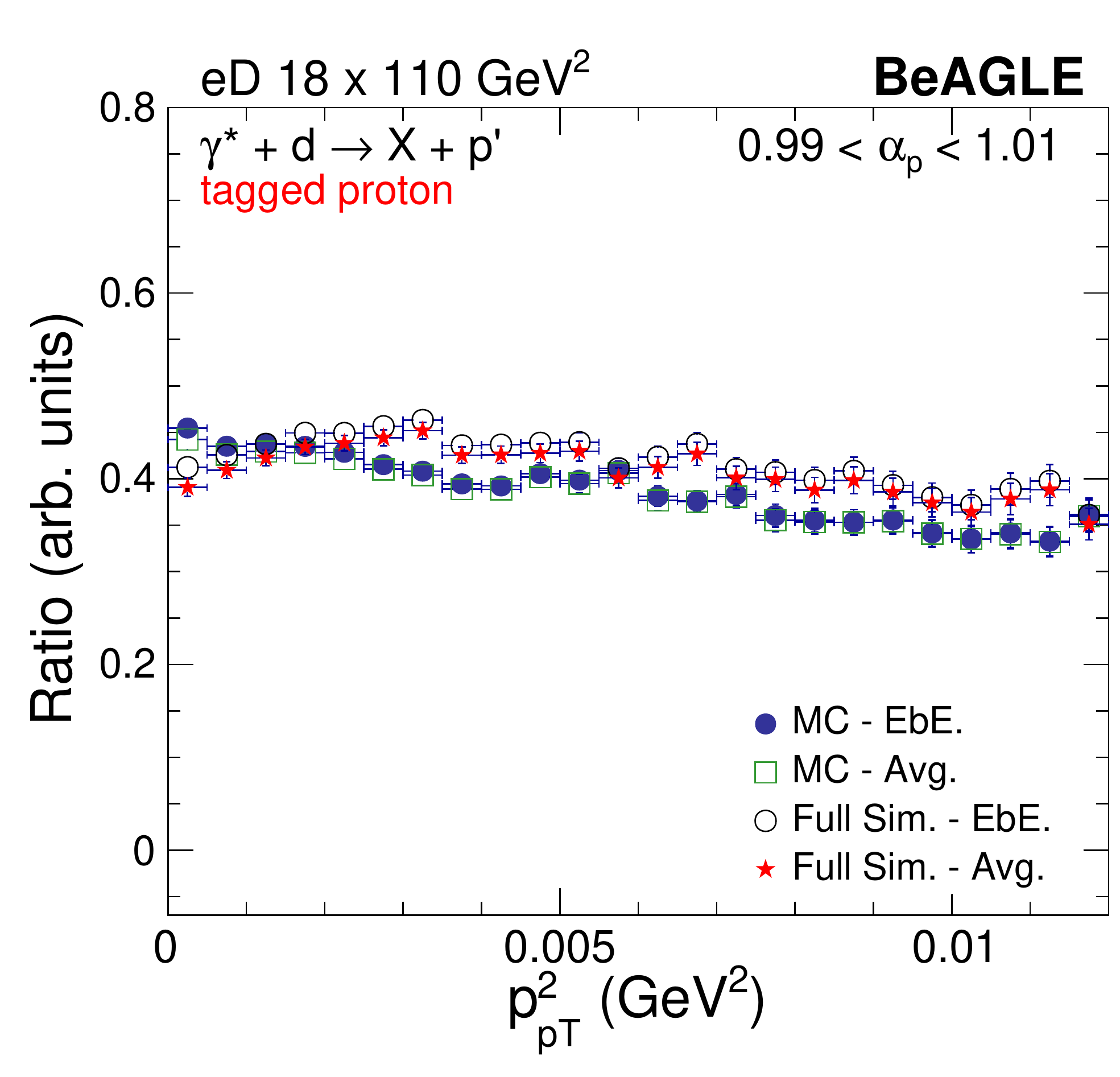}
\includegraphics[width=0.4\textwidth]{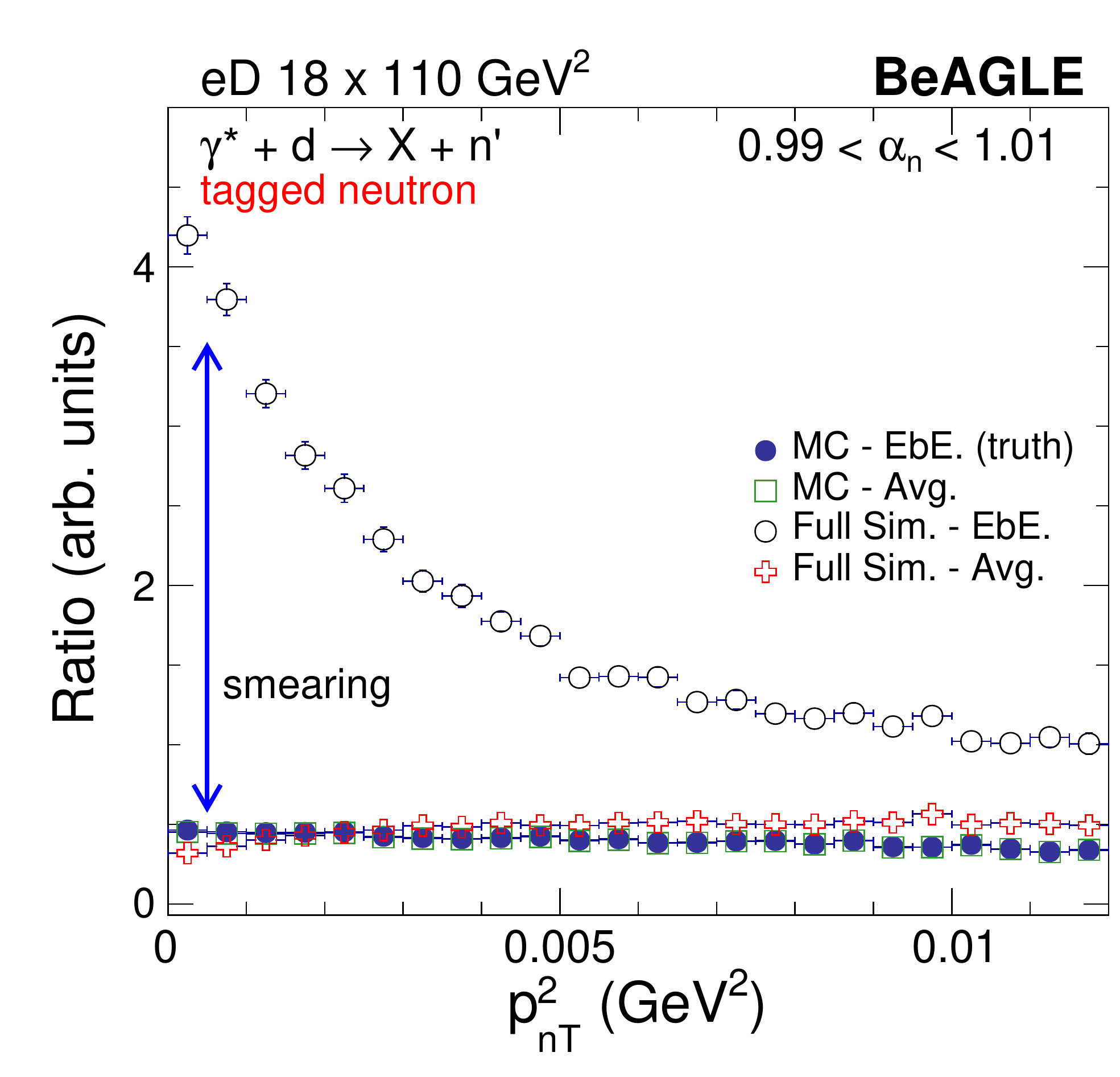}
\caption{Comparison of the event-by-event (EbE) and event-averaged (Avg) approaches to pole removal in
proton and neutron tagging. The plots show the ratio of the deuteron cross section and
the pole factor, Eq.~(\ref{pole_analysis_ratio}), extracted in different ways:
Solid circles: MC events (exact momenta), event-by-event approach.
Open squares: MC events, event-averaged approach.
Open circles: Full simulations (reconstructed momenta with smearing), event-by-event approach.
Crosses: Full simulations, event-averaged approach. The results shown in the plots
correspond to a typical ($x, Q^2$) bin.}
\label{fig:figure_poleremoval}
\end{figure}
There are two possible approaches to implementing the pole removal in the experimental analysis:
(i) compute the ratio Eq.~(\ref{pole_analysis_ratio}) on an event-by-event basis, i.e.,
evaluate the pole factor at the actual momentum of the event; (ii) compute the ratio on an
event-averaged basis, i.e., evaluate the pole factor at an average momentum in a finite bin.
Both have apparent advantages and disadvantages. The event-by-event approach is theoretically more accurate 
because of the steep momentum dependence of the functions; however, in the experimental analysis
the reconstructed momenta are subject to large uncertainties due to detector and beam effects.
The event-averaged approach can be corrected statistically for detector and beam effects;
however, it retains uncertainties from the finite bin size. The trade-offs between these effects
are generally different for proton and neutron tagging can be explored in our simulations.

We have performed a detailed study of the two approaches to pole removal for both proton and
neutron tagging. Figure~\ref{fig:figure_poleremoval} compares the results of the two approaches in
a typical $x, Q^2$ and $\alpha$ bin. The plots show the ratio Eq.~(\ref{pole_analysis_ratio})
extracted with the event-by-event and average approaches, first in an analysis using the original MC events
(exact momenta), and second in an analysis with full simulations (momentum
smearing from detector and beam effects). In the case of proton tagging (upper plot), one sees
that the event-by-event and average results are in good agreement when the analysis is performed with MC events,
showing the theoretical consistency of the two approaches. The event-by-event and average results are also
in reasonable agreement when the analysis is performed with the full simulations, showing that the
overall impact of the proton momentum smearing is moderate. In the case of neutron tagging (lower plot),
the situation is very different. One observes that that the event-by-event and average results are again in
good agreement when the analysis is performed with MC events, as expected. However, when the
the analysis if performed with full simulations, the results of the event-by-event approach differ qualitatively
from those of the event-averaged approach. The differences are caused by the substantial smearing of
the reconstructed neutron momentum variables $\alpha_{n}$ and $p_{nT}$, which has a large numerical
effect on the calculated pole factor. In particular, the smearing of $\alpha_{n}$
[caused mainly by the Zero-Degree Calorimeter energy resolution Eq.~(\ref{ZDC_resolution})] 
has a major numerical effect in the
evaluation of the pole position $a_T^2(\alpha_n)$ [given by Eq.~(\ref{a_T_def})
with $\alpha_p \rightarrow \alpha_n$] and causes $O(1)$ differences in the calculated pole factor.
In the event-by-event approach these smearing effects cannot be corrected, and the results are fully exposed
to the detector and beam uncertainties. In the event-averaged method the smearing effects can be
corrected, and the results are in reasonable agreement with those obtained from MC events without smearing.

%
%
\begin{figure}[t]
\includegraphics[width=0.4\textwidth]{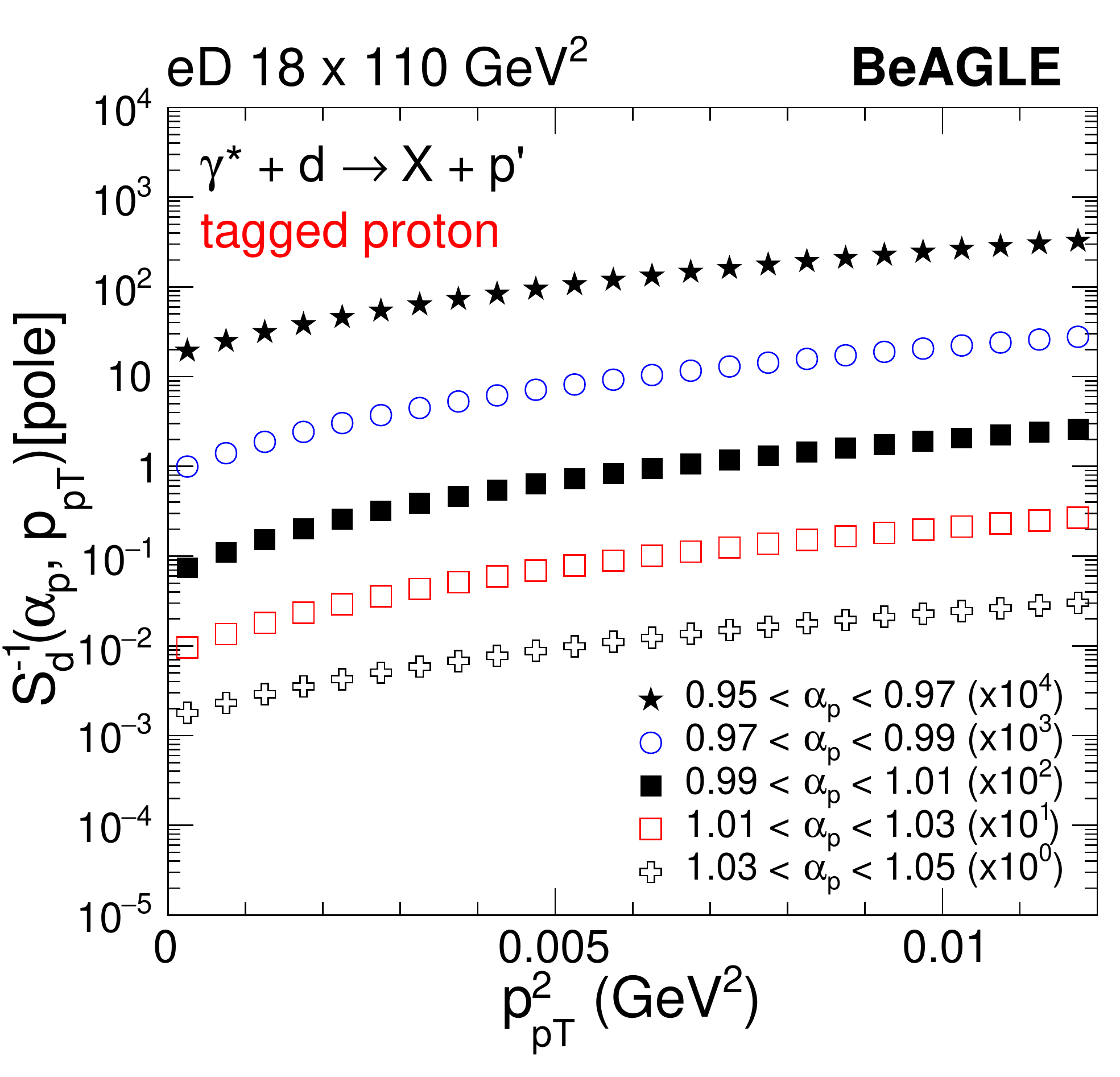}
\caption{The inverse pole factor in proton tagging, $1/\mathcal{S}_d(\alpha_p, p_{pT})$,
Eq.~(\ref{pole_spectral}), as a function of $p_{pT}^2$, for various values of $\alpha_p$.
The function was reconstructed from full simulations using the event-averaged approach.
The results for different $\alpha_p$ are offset by powers of 10 for visibility.}
\label{fig:figure_averagepole}
\end{figure}
Our study shows that for neutron tagging the event-averaged approach is the only realistic method for
performing the pole removal. For proton tagging an event-by-event approach might be considered; however, its performance
depends on the actual detector resolutions, and a final assessment is not possible at this stage.
Overall, the event-averaged approach to pole removal is more realistic and more robust against
detector effects, and we adopt it in the present analysis.

Figure~\ref{fig:figure_averagepole} shows the reconstructed pole factor in proton tagging using the
event-averaged approach. The plot shows $1/\mathcal{S}_{d}$ as a function of $p^{2}_{pT}$ in several bins 
of $\alpha_p$; this is the function that the deuteron cross section in Fig.~\ref{fig:figure_red_deut} 
is multiplied with in order to extract the neutron cross section. One sees that the experimentally 
reconstructed pole factor is a smooth function and follows the theoretical function shown in Fig.~\ref{fig:spectral_pole}.
\subsection{Nucleon structure from pole extrapolation}
\label{subsec:analysis_pole}
%
%
\begin{figure}[t]
\includegraphics[width=0.4\textwidth]{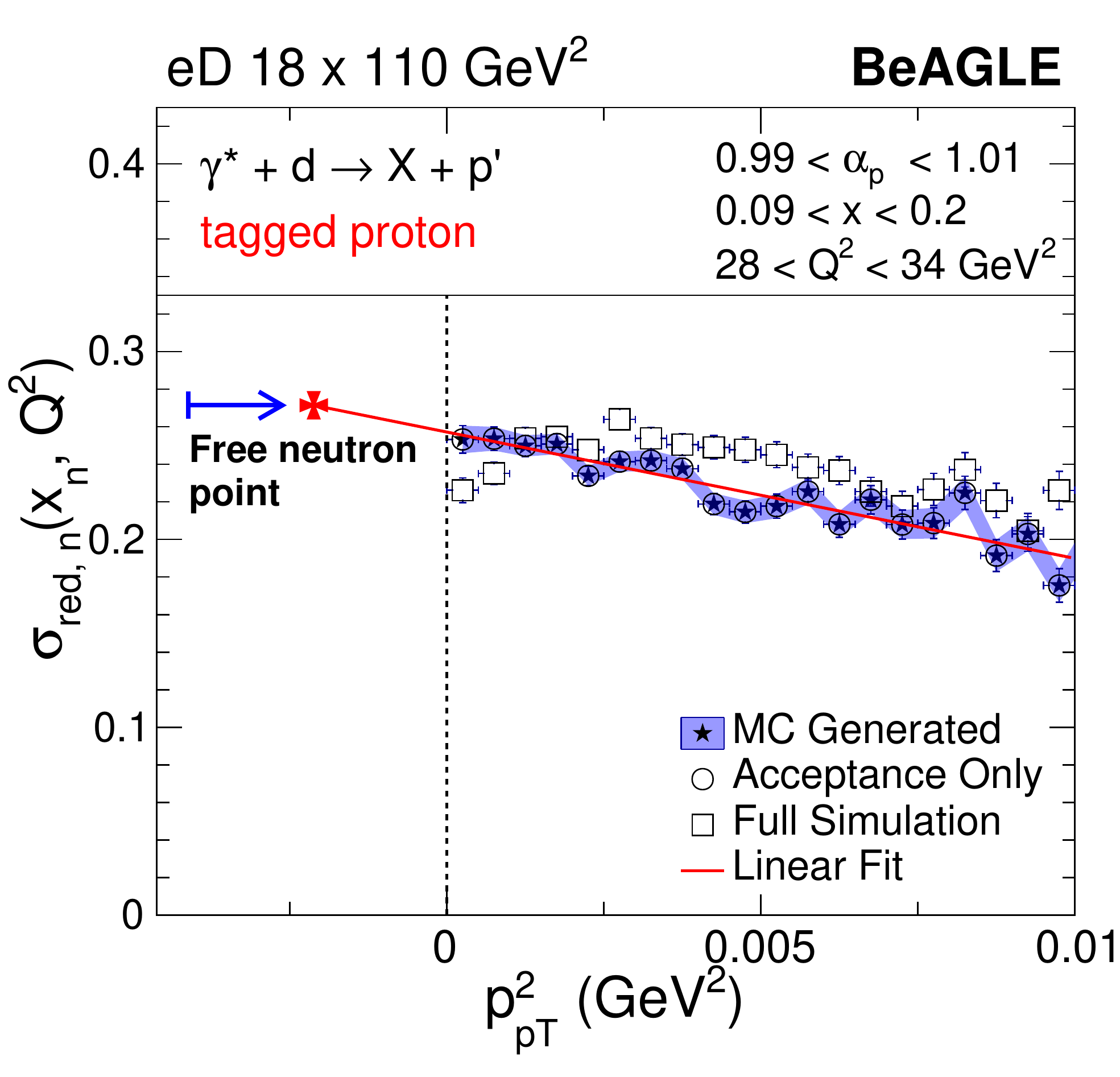}
\includegraphics[width=0.4\textwidth]{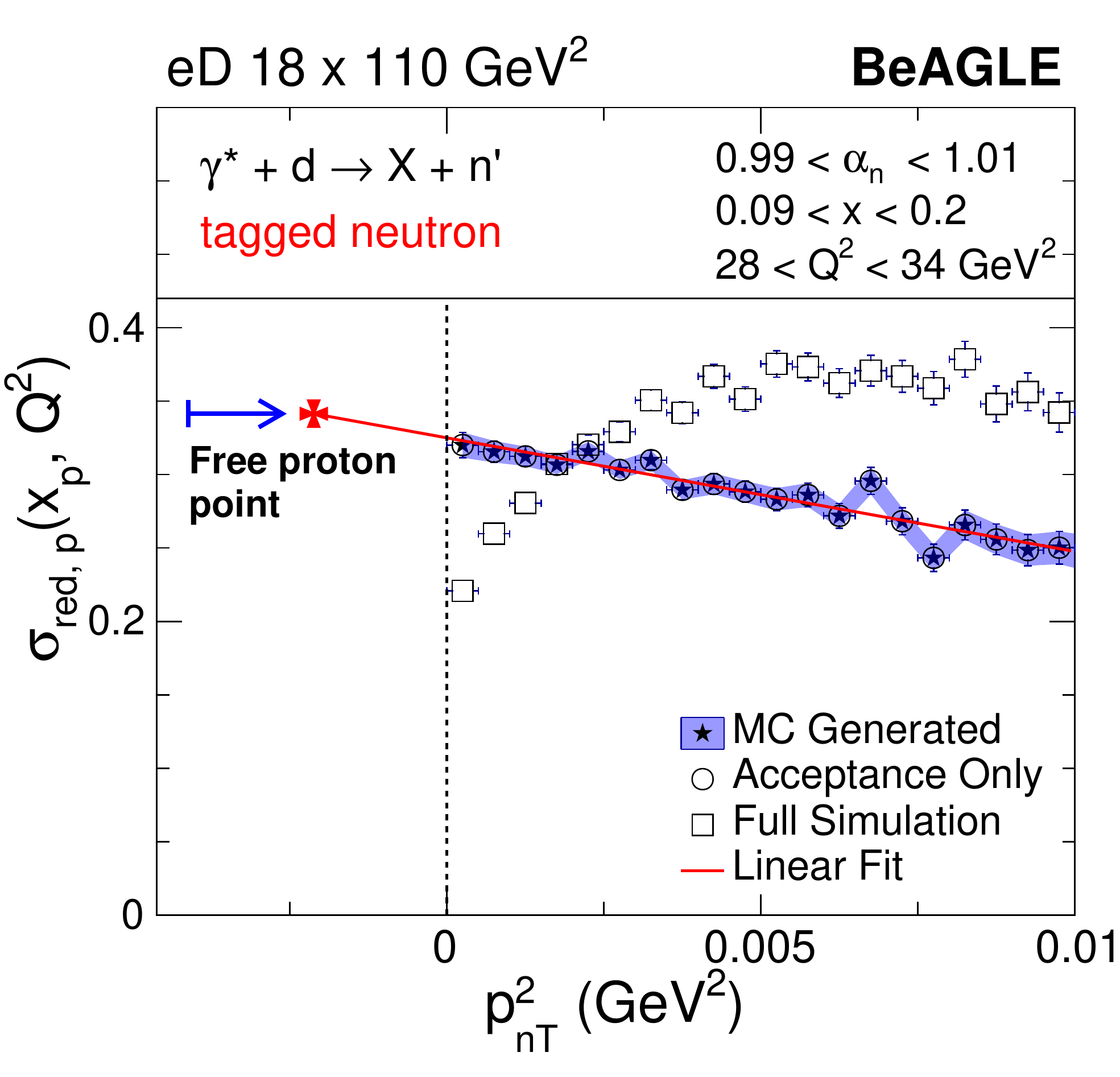}
\caption{Pole extrapolation and free nucleon cross section extraction in spectator tagging.
Top: Neutron cross section with proton tagging. Bottom: Proton cross section with neutron tagging. 
The data show the deuteron reduced cross sections divided by the pole factor, Eq.~(\ref{pole_analysis_extrapolation}),
as functions of $p^2_{pT} (p^2_{nT})$. Stars and bands: MC data (generator-level). Circles: Reconstructed with acceptance only.
Squares: Full simulations including acceptance and smearing effects (these data show the raw smearing
effects and have not been corrected). The lines shows the first-degree polynomial fits used for the pole extrapolation.
The fit functions are evaluated at the pole position Eq.~(\ref{a_T_def}), where they give the free
nucleon reduced cross sections (denoted by the arrows).}
\label{fig:figure_red_nucleon}
\end{figure}
In the third step of the analysis, we extrapolate the deuteron cross section after pole removal
to the nucleon pole $p_{pT}^2 (p_{nT}^2) \rightarrow -a_T^2$, where it gives the free 
nucleon cross section, see Eq.~(\ref{pole_analysis_extrapolation}).
Figure~\ref{fig:figure_red_nucleon} shows the simulated data and the extrapolation procedure 
for both proton and neutron tagging. The bands show the $p_{pT}^2 (p_{nT}^2)$ dependence of 
the cross section after pole removal, Eq.~(\ref{pole_analysis_ratio}), as obtained from the
MC data with acceptance effects only (no smearing). One sees that the dependence of this quantity on $p_T^2$ is very weak, because most of the $p_T^2$ dependence of the tagged cross section has been removed by the 
pole factor (see also Fig.~\ref{fig:spectral_pole}), and that the data indicate a regular distribution
around a smooth curve. The extrapolation to negative $p_T^2$ can therefore be performed with a 
low-order polynomial fit. The degree of the fitting polynomial and the choice of $p_T^2$ range 
for the fit are a matter of optimization and determine the fit uncertainty (see Sec.~\ref{sec:discussion}); 
the example in the figure is representative and shows a first-order fit over the range 
$0 < p_T^2 < (\textrm{100 MeV}/c)^2$. The free nucleon reduced cross section and its uncertainty
are obtained by evaluating the fit at the pole momentum $p_{pT}^2 (p_{nT}^2) = -a_T^2$.
Note that the extrapolation relies essentially on the EIC far-forward acceptance extending 
down to $p_T^2 = 0$ for both protons and neutrons; any acceptance limit $p_T^2 > 0$ 
would increase the extrapolation distance and uncertainty.

In Figure~\ref{fig:figure_red_nucleon} the extrapolation is performed with the MC data with acceptance
effects only. The plots also show the distributions obtained from the full simulations, which include
the effects of momentum smearing in the cross section and the pole factor. One sees that these distributions
differ from the generator-level distributions by $\sim$10\% in the case of proton tagging, and $\sim$30\% in neutron tagging.
In an actual experiment the smearing effects will be corrected by an unfolding procedure, which is 
expected to eliminate most of the differences. Performing the extrapolation with the original MC distributions 
therefore presents a realistic picture of nucleon structure extraction in the actual experiment.

%
%
\begin{figure}[t]
\includegraphics[width=0.4\textwidth]{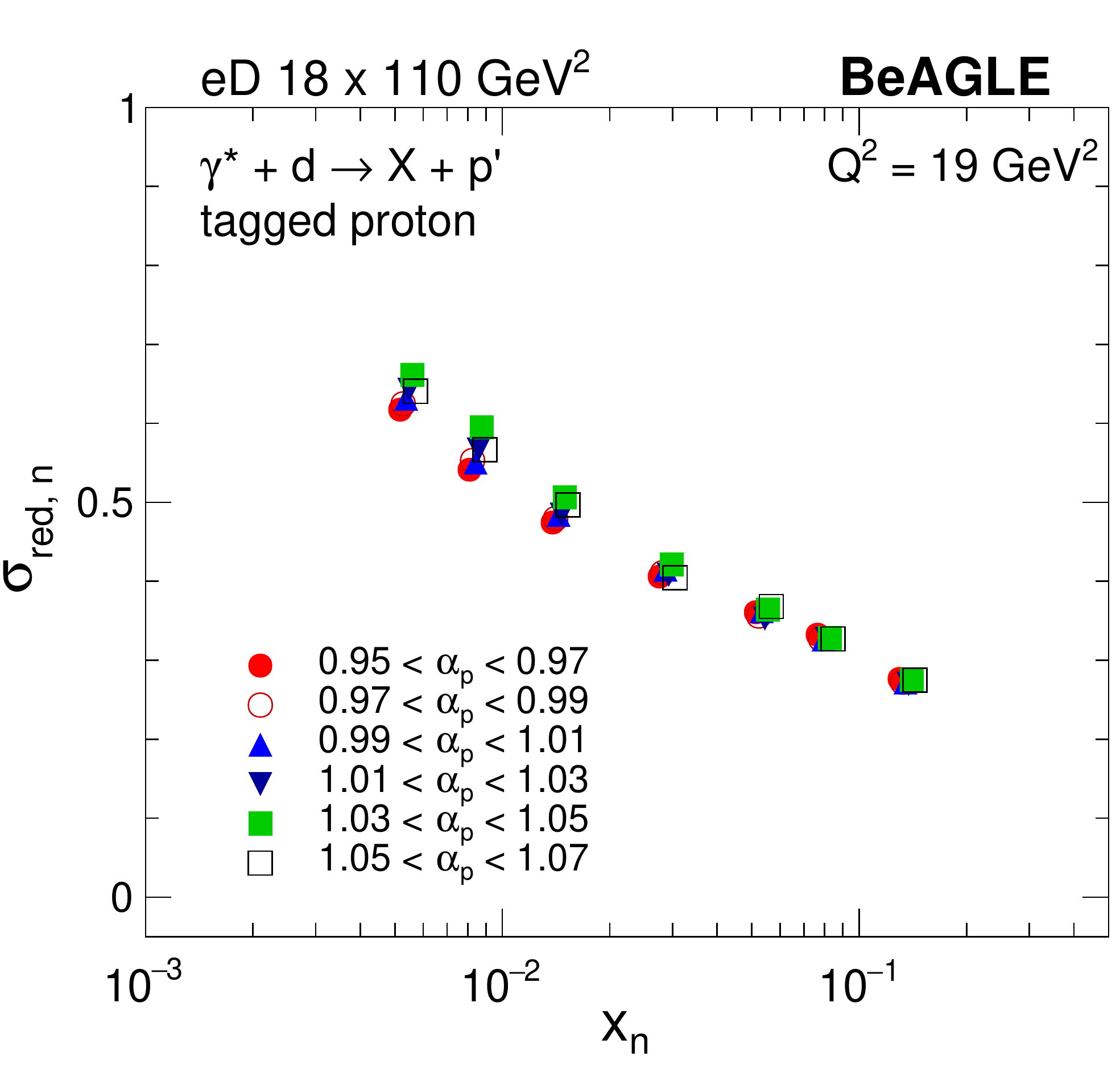}
\includegraphics[width=0.4\textwidth]{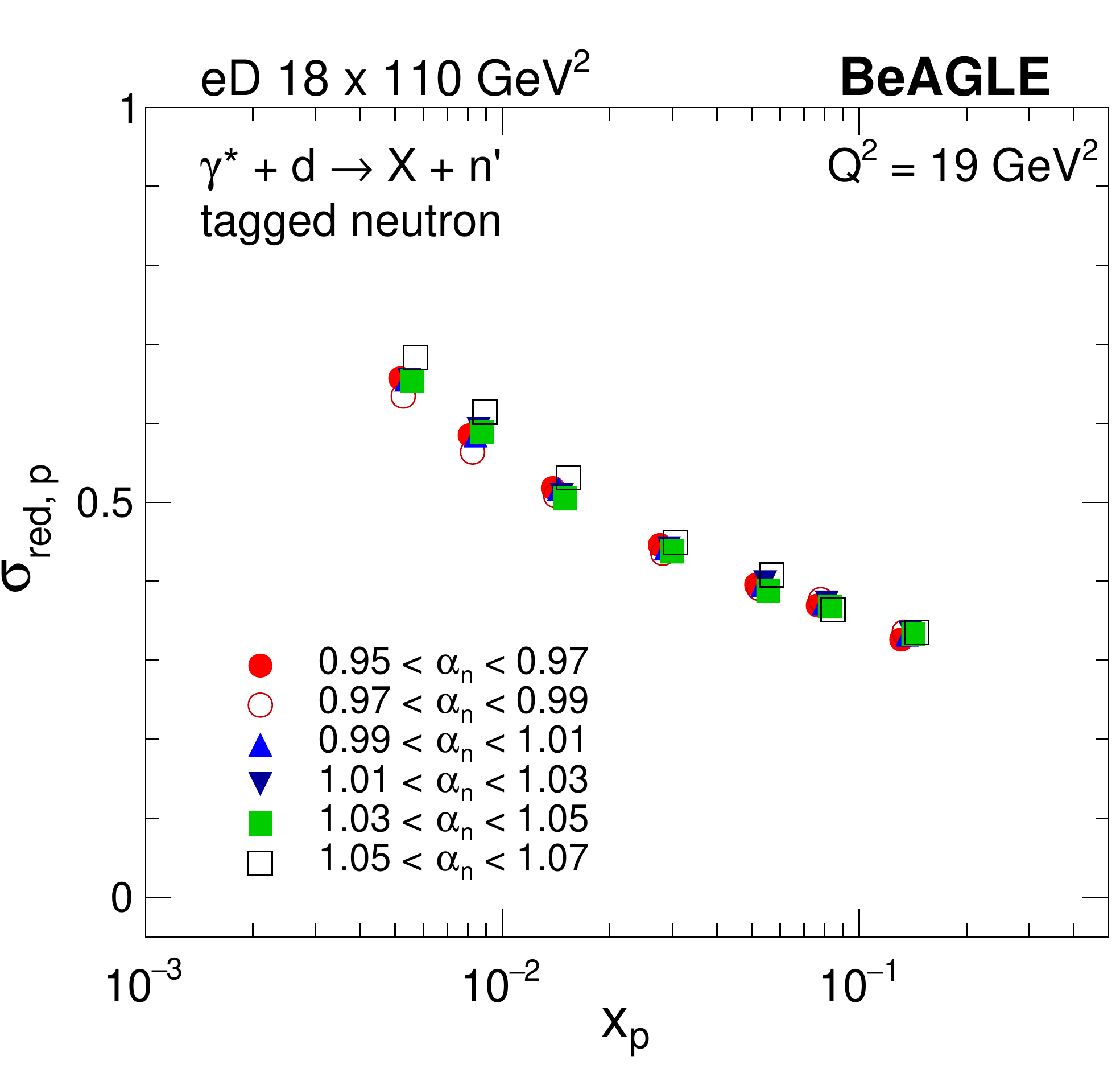}
\caption{The free neutron (top) and proton (bottom) reduced cross sections extracted with pole
extrapolation, as functions of $x_n$ and $x_p$, respectively. The plots show the results of
extractions performed at different $\alpha_p (\alpha_n)$.}
\label{fig:figure_alpha_dependent}
\end{figure}
Figure~\ref{fig:figure_alpha_dependent} shows the free neutron and proton reduced cross sections 
measured via pole extrapolation, Eq.~(\ref{pole_analysis_extrapolation}), at several values of
$\alpha_p$ and $\alpha_n$. The reduced cross sections are presented as 
functions of $x_n$ and $x_p$, Eqs.~(\ref{x_n}) and (\ref{x_p}), the nucleon-level scaling variables 
whose values are fixed by the spectator kinematics. The result shown here have been corrected 
for artifacts resulting from the treatment of the electron-nucleon sub-process kinematics in BeAGLE, 
by applying the factor Eq.~(\ref{beagle_corr}) (see Sec.~\ref{subsec:beagle}; this correction
will not be needed in a real experiment).
An important feature of tagging is that the same value of $x_n (x_p)$ can be realized with different 
combinations of $x$ and $\alpha_p (\alpha_n)$, allowing one to measure the same physical nucleon 
cross section in different settings of the external DIS and spectator kinematics.
Figure~\ref{fig:figure_alpha_dependent} shows that the results obtained at different values of 
$\alpha_p (\alpha_n)$ agree at the level of 5--10\%; the small differences result from the
event-averaged pole-removal procedure and could be reduced by corrections (see Sec.~\ref{subsec:pole}).
This provides a crucial test of the simulations and the robustness of the extraction procedure.
Note that in extractions at $\alpha \neq 1$ the pole extrapolation has to cover a larger distance
than at $\alpha = 1$, because the nucleon pole position Eq.(\ref{a_T_def})
increases quadratically in $(\alpha - 1)$.
\subsection{Validation of nucleon structure extraction}
\label{subsec:validation}
%
%
\begin{figure*}[t]
\includegraphics[width=0.8\textwidth]{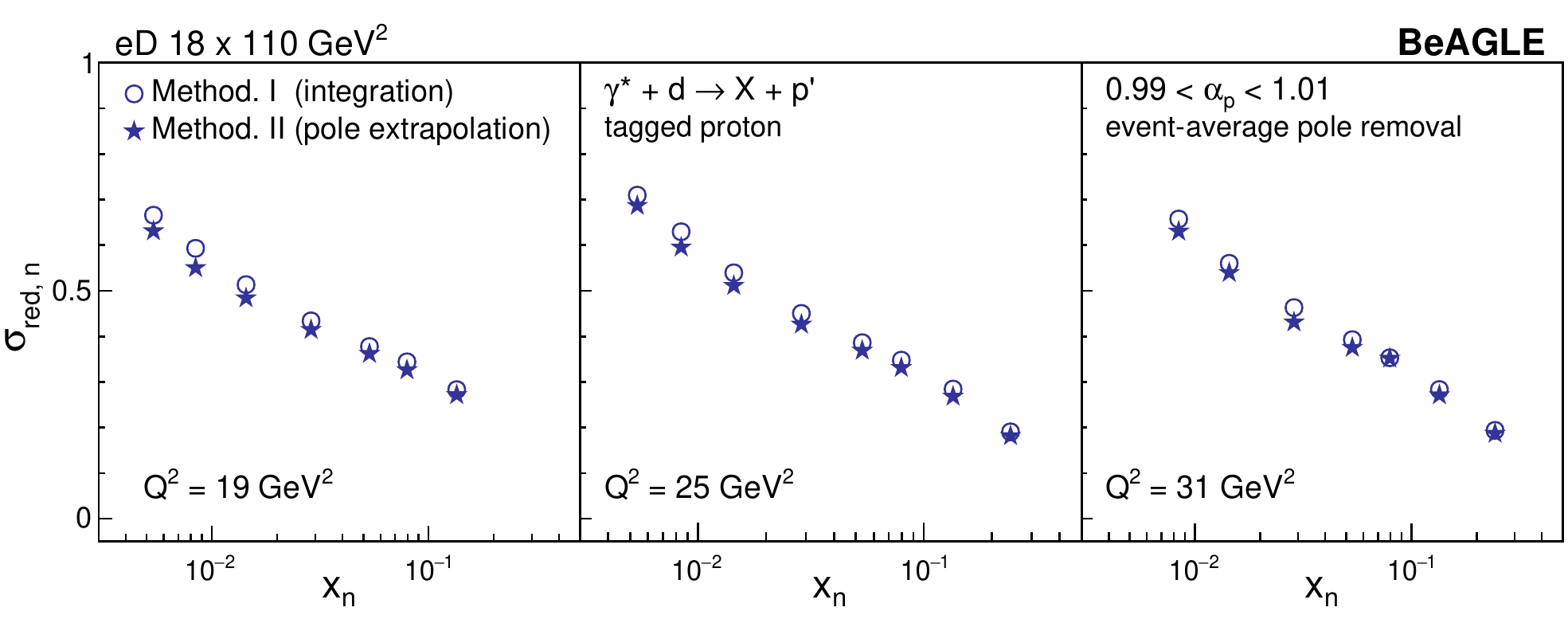}
\includegraphics[width=0.8\textwidth]{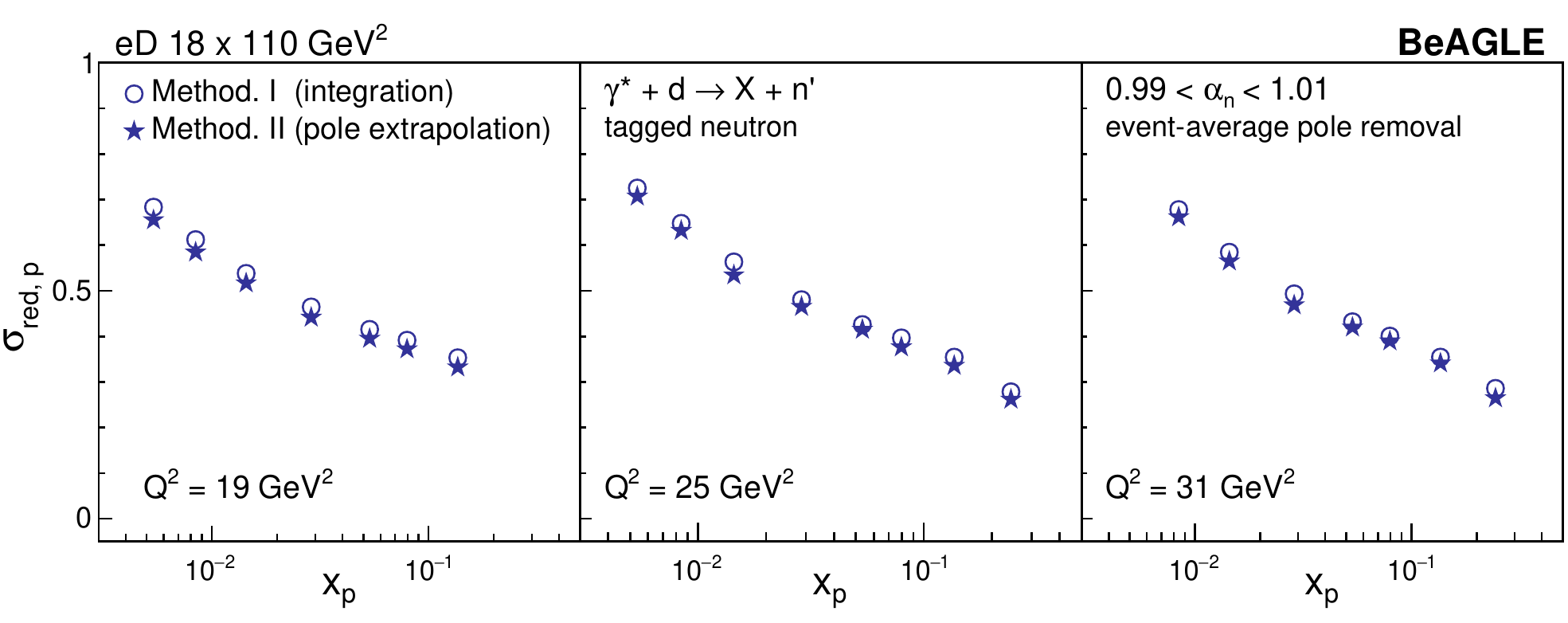}
\caption{Validation of nucleon structure extraction with spectator tagging in BeAGLE.
The plots show the reduced neutron (proton)
cross sections $\sigma_{{\rm red}, n} (\sigma_{{\rm red}, p})$
as functions of $x_n (x_p)$, extracted with two different methods (see Sec.~\ref{subsec:extraction}).
Stars: Integration over spectator momentum (Method I). Circles: Pole extrapolation in spectator momentum (Method II).
Here the event-averaged approach was used in removing the pole factor (see Sec.~\ref{subsec:pole}).}
\label{fig:figure_validation}
\end{figure*}
%
%
\begin{figure*}[t]
\includegraphics[width=0.8\textwidth]{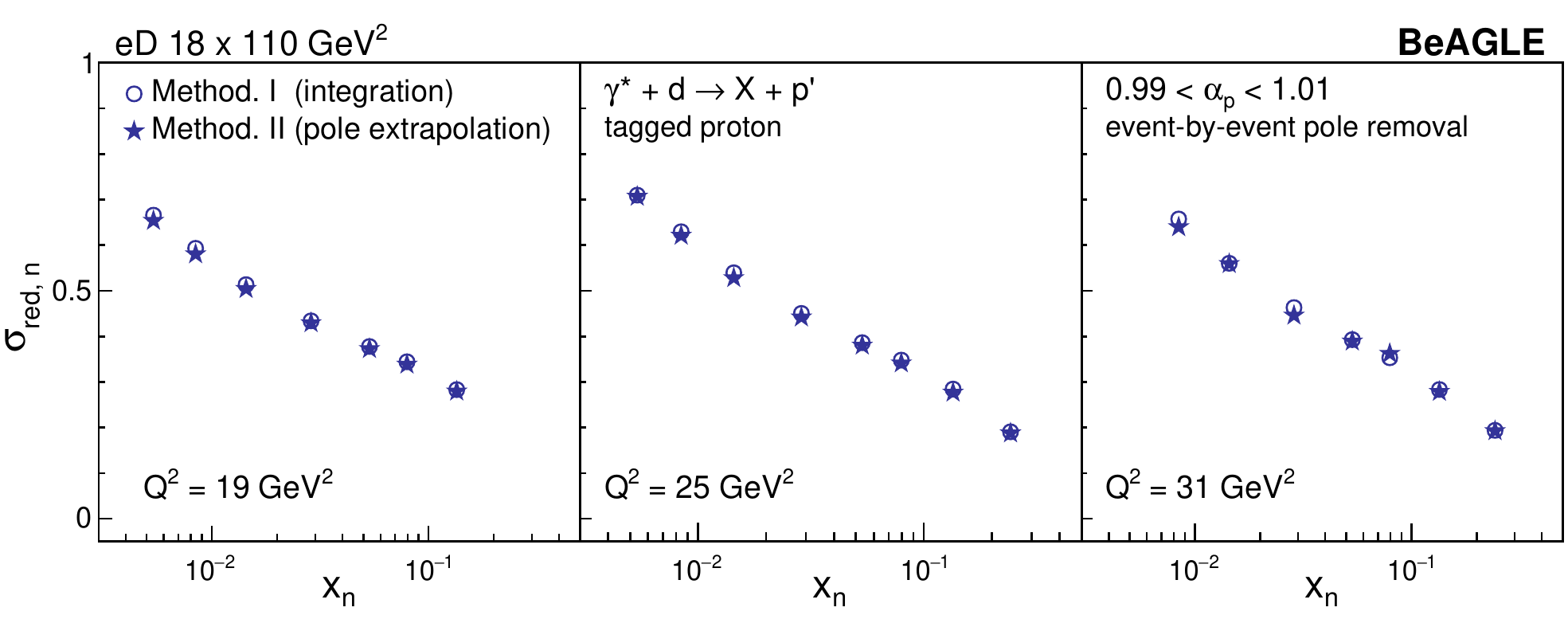}
\includegraphics[width=0.8\textwidth]{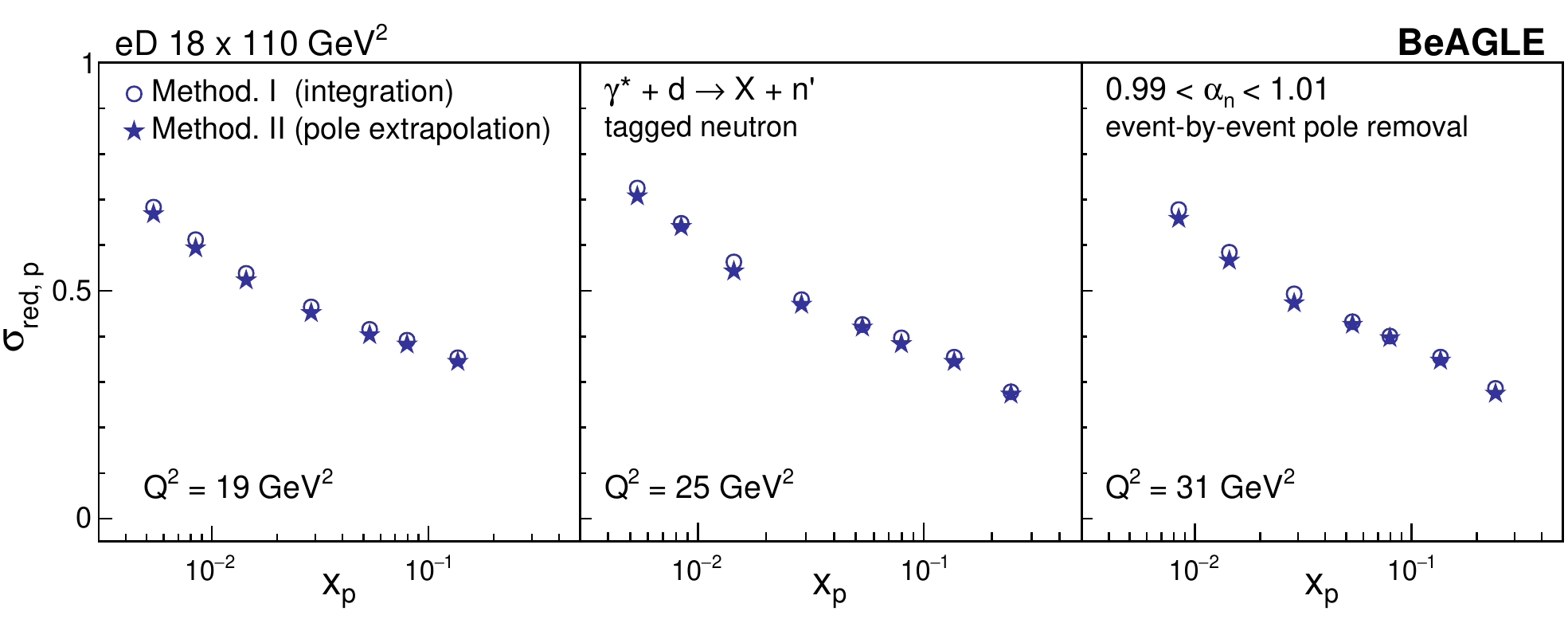}
\caption{Validation of nucleon structure extraction with spectator tagging in BeAGLE (see also
Fig.~\ref{fig:figure_validation}). The plots show the reduced neutron (proton)
cross sections $\sigma_{{\rm red}, n} (\sigma_{{\rm red}, p})$
as functions of $x_n (x_p)$, extracted with two different methods (see Sec.~\ref{subsec:extraction}).
Stars: Integration over spectator momentum (Method I). Circles: Pole extrapolation in spectator momentum 
(Method II). In difference to Fig.~\ref{fig:figure_validation}, here the event-by-event approach was
used in removing the pole factor (see Sec.~\ref{subsec:pole}).}
\label{fig:figure_closuretest}
\end{figure*}
The analysis simulated here extracts the reduced cross section of the free nucleon from the tagged 
DIS data using the method of pole extrapolation. We can validate the results by comparing the extracted 
nucleon cross section with the input at the generator level. In the present study using BeAGLE, 
this validation can be accomplished simply by comparing the result of the pole extrapolation
(``Method II'' of Sec.~\ref{subsec:extraction}) with the nucleon cross section obtained by
integrating over the spectator momentum (``Method I'' of Sec.~\ref{subsec:extraction}).
[As explained in Sec.~\ref{subsec:integration}, this is possible because BeAGLE implements 
the impulse approximation without dynamical initial-state modifications or final-state interactions, so that the 
integration over the spectator momentum recovers the nucleon cross section;
see Eq.~(\ref{integrated_cross_section}).] While specific to this generator, this comparison 
offers a very convenient way to test the result of the pole extrapolation.

Figure~\ref{fig:figure_validation} shows the comparison of the nucleon reduced cross section computed
using integration over the spectator momentum  in BeAGLE (Method I) and the result of the 
pole extrapolation (Method II). The results of the two methods agree within $\sim$2\% 
in the $(x, Q^2)$ range covered in our study. This level of agreement is consistent with
the accuracy of the event-averaged pole removal approach as presently implemented 
(see Sec.~\ref{subsec:analysis_pole_removal}); the accuracy could be improved by applying
bin-centering corrections. For an even more stringent test, Figure~\ref{fig:figure_closuretest} shows the 
the same comparison using the pole extrapolation result obtained with the event-by-event approach, 
which is free of the bin centering corrections (see Sec.~\ref{subsec:analysis_pole_removal}).
Now the agreement between the integration and the pole extrapolation methods is at the level $\lesssim$1\%, 
which is the expected accuracy of the pole extrapolation. This shows that the small discrepancies observed in Figure~\ref{fig:figure_validation} are indeed due to the accuracy of the event-averaged 
pole removal approach. Altogether, these tests show agreement between the integration and 
the pole extrapolation methods at the expected level of accuracy and validate the results
of the pole extrapolation.
\section{Discussion} 
\label{sec:discussion}
\subsection{Experimental uncertainties and effects}
\label{subsec:experimental_uncertainties}
We now discuss the experimental and theoretical uncertainties arising in the proposed tagged DIS 
measurements with EIC and the nucleon structure extraction with pole extrapolation. Because our study refers to simulated measurements with a future facility, the analysis of 
uncertainties is necessarily different from that of actual measurements with an existing facility. Some of the experimental effects included in the simulations of Sec.~\ref{sec:analysis} cannot be fully quantified because they 
depend on the final detector performance, while other effects cannot even be included because
a design of the necessary components is not available for study. In the following discussion we therefore address 
both the status of the modeling of the various effects and their impact on the analysis.
Our goal is to provide an assessment of the uncertainties that is realistic and can be extended and improved with future developments.

\subparagraph{Statistical uncertainties.} 
Tagged DIS has the same rates as inclusive DIS on the deuteron,
only differentiated in the spectator nucleon momentum. In the present study of free nucleon structure
extraction we use spectator momenta $p_{pT} \, (p_{nT}) \lesssim$ 100 MeV/$c$, which correspond to average
nuclear configurations and account for the bulk of the deuteron momentum distribution. The integrated luminosity of 1 fb$^{-1}$ ($\sim 10^8$ events) is more than sufficient for the differential measurements of the
$p_T^2$ distributions in the $(x, Q^2)$ region considered here. The nucleon structure extraction is not limited by statistics and the resulting overall uncertainties are dominated by systematic effects.
The situation will be different in future studies of nuclear modifications, which access
both larger $x \gtrsim 0.3$ and $p_{pT} \, (p_{nT}) \sim$ 300--600 MeV/$c$, where the rates are much lower.

\subparagraph{DIS variable reconstruction.} The DIS variables $x$ and $Q^2$ in tagged DIS are reconstructed in the same way as in standard inclusive DIS. The uncertainties associated with the reconstruction have	been
studied	extensively in inclusive DIS simulations and are described in the Yellow Report~\cite{AbdulKhalek:2021gbh}.
The DIS kinematics covered in the present study is non-exceptional, and the performance of the standard electron method is expected to be at the percent level.

\subparagraph{Spectator momentum reconstruction.} The reconstruction of the far-forward
proton and neutron momenta is affected by various detector and beam effects.
The present simulations include the following effects: 
(i)~Intrinsic detector smearing (both protons and neutrons);
(ii)~Deuteron beam angular divergence; 
(iii)~Deuteron beam momentum spread; 
(iv)~Crab cavity rotations.
These effects have been evaluated with the current EIC accelerator and detector design, and their
aggregate effect on the signal (before correction) is shown in the ``Full Simulation'' results 
in Figs.~\ref{fig:figure_red_deut}, \ref{fig:figure_poleremoval}, and \ref{fig:figure_red_nucleon}. 
The contributions of the individual effects can be seen in the summary plots in Appendix~\ref{app:resolution}.
Note that the impact of the various effects is different for protons and neutrons: 
the dominant effect for protons comes from the angular divergence of the deuteron beam (ii), 
while the neutron suffers mostly from the energy resolution of the Zero-Degree Calorimeter (i).

Several other effects can influence the far-forward nucleon detection but have not yet been included 
in the simulations: (v) Beam pipe design; (vi) Non-linear transport matrix. These effects can be 
included as the technical design or specification of these elements becomes available. 
The beam pipe design (v) will mostly impact the overall detection efficiency; the non-linear 
transport matrix (vi) will affect the assessment of the momentum smearing for protons in the Off-Momentum Detectors.
These effects are not expected to substantially modify the findings of the present simulations.

In the actual experimental analysis the beam and detector effects described here will be corrected 
through an unfolding procedure. The systematic uncertainty in the final physics results is not 
given by the size of the original effects, but by the accuracy 
with which they can be corrected. The unfolding procedure will use apparatus information 
(design, performance) that is not available at present. Progress in detector technology and
correction algorithms in the time until the EIC experiments are performed will significantly
improve the estimated accuracy of the correction procedure. For these reasons we presently cannot
perform a quantitative assessment of the systematic uncertainties after corrections.
The important result of our study is that the aggregate effects before corrections are 
$\sim$10\% for protons and $\sim$30\% for neutrons, see Fig.~\ref{fig:figure_validation}.
A reasonable unfolding procedure is expected to be able to correct these effects with a 
final accuracy at the percent level, which will be sufficient for an impactful physics analysis.

\subparagraph{Backgrounds.} In DIS measurements there are different sources of backgrounds such as beam-gas 
interactions, particles coming from hadronic final states, photoproduction, etc. 
These backgrounds and not studied here, as they are common to all inclusive 
DIS measurements and not specific to spectator tagging \cite{AbdulKhalek:2021gbh}. 
Forward nucleons produced through target fragmentation represent a theoretical background 
and are discussed below.
\subsection{Theoretical and fit uncertainties}
\label{subsec:discussion_theoretical}
\subparagraph{Uncertainties in pole extrapolation.} 
The extraction of the free nucleon structure function through pole extrapolation is subject to
specific uncertainties; see Sec.~\ref{subsec:analysis_pole}. The first uncertainty is related to 
the effects in the event-averaged pole removal approach described in Sec.~\ref{sub:poleremoval}, 
which can be corrected statistically. The error of this correction remains as a source of uncertainty
and is estimated to be $\sim$1--2\%. The second uncertainty is associated with the polynomial fits 
used to perform the extrapolation. The degree of the polynomial and the fitting range are
determined in an optimization process that takes into account the average variation of the function
and the fluctuations of the data; the optimal configuration is approximately stable against variation 
of the degree and range. We have performed fits using polynomials from first- to fifth-order, each
with different fit ranges. Combining all the fit results and comparing to the truth value, the best fit is found to be the first- or second-order polynomial with a fit range between $0 < p^{2}_{T} < 0.01~ \rm{GeV^{2}}$. By comparing to other fit configurations, the final extrapolation uncertainty is 
estimated at $\sim$1--2\% for the fits at $\alpha_{p} \, (\alpha_n) \approx 1$. Note that the distance 
of the pole from the physical region increases for $\alpha_{p} \, (\alpha_n) \neq 1$, 
see Eq.~(\ref{a_T_def}), which causes the extrapolation uncertainty to increase even when the 
fit quality remains the same.
\subparagraph{Deuteron structure in pole extrapolation.} 
The pole extrapolation method of nucleon structure extraction is unique
in that it is theoretically exact (the residue at the pole of the tagged DIS cross section is by
definiton equal to the free nucleon DIS cross section) and permits a model-independent extraction, limited only by the practical ability to perform the extrapolation from the physical region to the pole (see above). The only theoretical input in the final extraction is the residue of the nucleon pole of the deuteron 
wave function, $\Gamma$, Eqs.~(\ref{C_def}) and (\ref{pole_nonrel}), which is determined by low-energy 
nuclear structure calculations and independent low-energy deuteron breakup measurements, 
see Appendix~\ref{app:pole} and Table~\ref{tab:pole} 
and known with an accuracy of $\lesssim 1\%$. In an analysis with actual experimental data, the uncertainty
in $\Gamma$ would cause an overall normalization uncertainty of the extracted cross section proportional 
to $\Gamma^2$. In the present analysis with simulated events, the value of $\Gamma$ in the deuteron
structure model of the BeAGLE generator is known; the same value is used in evaluating the pole factor 
Eq.~(\ref{pole_spectral}) in the analysis; and the simulation and validation procedures do not include
the uncertainty related to this parameter.

\subparagraph{Final-state interactions.}
Effects such as initial-state nuclear modifications and final-state interactions influence only the behavior of the tagged cross section in the physical region $p_{pT}^2 (p_{nT}^2) > 0$, not the result of the extrapolation to the pole. (This is because the nucleon pole singularity is contained exactly in the ``tree graph'' of the impulse approximation, while ``loop graphs'' due to final-state interactions can only produce subleading singularities \cite{Sargsian:2005rm}.) The BeAGLE physics model used in the present study is based on the impulse approximation and does not include dynamical initial-state  modifications or final-state interactions. Theoretical calculations indicate that final-state interactions in tagged DIS at $x \gtrsim 0.1$ change the deuteron spectral function by $\lesssim 10\%$ at $p_{pT} < 100$ MeV/$c$ and have a smooth dependence on $p_{pT}$. The inclusion of these effects
would change the results in Fig.~\ref{fig:figure_red_nucleon} by a relative amount 
$\lesssim 10\%$, while being theoretically constrained to extrapolate to the same free neutron (proton) result.
These effects therefore would not quantitatively affect the quality of the extrapolation or the uncertainties of the extrapolated results. In this sense the use of the impulse approximation in the BeAGLE physics model is justified for the present purpose and does not represent a limitation; simulations of tagged DIS at low transverse momenta $p_{pT} (p_{nT}) < 100$ MeV/$c$ and nucleon structure extraction with pole extrapolation can be safely performed in this approximation. We emphasize that this will be different in studies of nuclear modifications in tagged DIS at higher transverse momenta $p_{pT} (p_{nT}) \sim$ 300--600 MeV/$c$ \cite{Jentsch:inprep}, 
where initial-state modifications and final-state interactions are both of order unity, and these effects need to be included explicitly in the physics model.

\subparagraph{Target fragmentation.} In tagged DIS we require the presence of a forward proton (neutron) with $\alpha_p (\alpha_n) \approx 1$ in the deuteron fragmentation region but have otherwise no information about the hadronic event (see Fig.~\ref{fig:deut_tagged}). 
Such forward protons and neutrons can not only come from the spectator nucleon
in the deuteron breakup but also from the target fragmentation of the active nucleon (baryon production at $x_F \approx -1$). The two mechanisms cannot be distinguished event-by-event and should be treated jointly in the physics analysis. The target fragmentation nucleons have a broad $p_T^2$ distribution with a width $\langle p_T^2 \rangle \approx$ 0.1--0.15 (GeV/$c$)$^2$ \cite{Strikman:2017koc}, much larger than the $p_T^2$ considered in the present study; see Sec.~\ref{subsec:analysis_pole_removal}. The target fragmentation mechanism does not have the 
nucleon pole of the spectator mechanism, so that the pole extrapolation procedure of 
Secs.~\ref{subsec:pole} and \ref{subsec:analysis_pole} eliminates the contribution of 
target fragmentation --- another important advantage of this method.

\subparagraph{Diffractive scattering.}
The present study of nucleon structure extraction from tagged DIS focuses on the region $x \gtrsim 0.1$. In DIS at $x \ll 0.1$, diffractive scattering becomes significant and constitutes 10--15\% of the DIS cross section, as observed in measurements at HERA \cite{Wolf:2009jm,Armesto:2019gxy}. In diffractive events the nucleon remains intact and recoils with a typical momentum transfer $\sim$few 100 MeV/$c$ (in the nucleon rest frame), 
and the other hadrons produced are separated by a rapidity gap. In tagged DIS on the deuteron, diffractive scattering on the nucleons (proton or neutron) creates several effects that are not included in the present simulations\footnote{The present BeAGLE simulations include only the non-diffractive part of the DIS cross section. In the PYTHIA 6 parameters only Process 99 is selected.} and require separate theoretical study: (i) The measurement cannot distinguish between the spectator nucleon and the diffractive nucleon (see above). (ii) The $pn$ state produced in diffractive DIS on the deuteron has small relative momentum $\sim$few 100 MeV/$c$ and the same quantum numbers as the deuteron. There is a large amplitude for this state to remain a bound deuteron, resulting in coherent scattering. If a $pn$ breakup state is measured, its wave function will be strongly distorted by the requirement of
orthogonality to the bound state (large final-state interactions). (iii) Interference between diffractive DIS on the proton and the neutron gives rise to nuclear shadowing \cite{Frankfurt:2003jf,Frankfurt:2006am}. 
These effects are the object of ongoing theoretical studies \cite{Guzey:inprep} and can be included of future simulations of tagged DIS at small $x$. Such measurements present a new opportunity to explore the dynamical origin of leading-twist nuclear shadowing \cite{Frankfurt:2011cs}, which is observed in hard exclusive processes with heavy nuclei \cite{Guzey:2013xba,Guzey:2013qza,Guzey:2020ntc} and governs the small-$x$ behavior of the nuclear PDFs and the approach to gluon saturation. In particular, such measurements could use double tagging -- the detection of both the proton and the neutron resulting from the deuteron breakup -- to completely fix the outgoing nucleonic configuration and enable a differential analysis of the interactions. Deuteron breakup in diffractive $J/\psi$ production
at EIC was studied in Ref.~\cite{Tu:2020ymk}.
\section{Conclusions} 
\label{sec:conclusions}
We have performed a comprehensive study of deuteron DIS with spectator nucleon tagging at EIC,
focusing on the extraction of free neutron and proton structure using the pole extrapolation method.
Our framework combines theoretical methods of light-front nuclear structure, the BeAGLE $eA$ event generator, and a description of the EIC far-forward detector performance based on full GEANT4 simulations. We have simulated the measurement of the reduced cross sections and the extraction of free nucleon structure through pole extrapolation under realistic conditions, including detector acceptance and detector and beam effects on the far-forward momentum resolution. We have quantified the systematic uncertainties to the extent possible at the present stage and discussed possible refinements incorporating future developments. The main conclusions of the study are:

(i)~Detection of far-forward protons and neutrons in the momentum range needed for free nucleon 
structure extraction with spectator tagging, $\alpha_p (\alpha_n) \approx$ 1 
and $p_{pT} (p_{nT}) \lesssim$ 100 MeV/$c$, is possible with nearly full acceptance 
with the baseline EIC far-forward detector design.

(ii)~The steep $p_{pT} (p_{nT})$ dependence of the deuteron spectral function places high demands on the transverse momentum resolution in the tagged cross section measurement. The separation of deuteron and nucleon structure (pole removal) needs to be performed with binned $p_{pT}^2 (p_{nT}^2)$ distributions corrected for detector performance. The overall detector resolution effects on the measured nucleon DIS cross section (before unfolding corrections) are estimated at $\sim$10\% for proton tagging and $\sim$30\% for neutron tagging. With unfolding corrections based on the actual EIC detector implementation, it is expected that a percent-level measurement of the tagged nucleon DIS cross section will be possible.

(iii)~The pole extrapolation of the tagged nucleon DIS cross section can be performed with low-order polynomial fits and gives robust results. In the simulations these results can be validated by comparing with the physics model input. 

(iv)~Systematic uncertainties dominate the tagged deuteron cross section measurement and nucleon structure extraction in the DIS kinematics considered in the present study.
The uncertainties arising from known sources (beam, detector, theory) are estimated at the few-percent level. More detailed estimates will become possible as the EIC detector design advances.

(v)~The pole extrapolation method eliminates both initial-state nuclear modifications and final-state interactions in tagged DIS and minimizes the theoretical uncertainty in nucleon structure extraction. The only theoretical input is the asymptotic normalization constant of the S-state in the deuteron wave function, which is determined with high precision in low-energy nuclear structure calculations and measurements.

Altogether, tagged DIS measurements and free nucleon structure extraction appear feasible with
the EIC accelerator and far-forward detector design, with an experimental accuracy that realizes 
the theoretical potential of the method. 
\section{Extensions}
\label{sec:extensions}
The present study has focused on the application of spectator tagging to inclusive DIS in typical EIC kinematics  ($x \sim 10^{-2}$--$10^{-1}$, $Q^2 \gtrsim$ 10 GeV$^2$) with the goal of extracting the free neutron and proton structure functions. The methods could be extended and applied to other high-energy processes with different physics goals:

\subparagraph{Azimuthal angle dependence.} The $\phi_p (\phi_n)$ dependence
of the tagged DIS cross section provides interesting information on the light-front structure of the deuteron and the dynamics of the breakup process (the angle refers to the photon-deuteron collinear frame, see Sec.~\ref{subsec:kinematic}). T-even (time-reversal-even) azimuthal asymmetries such as $\cos\phi$ and $\cos 2\phi$ are predicted by the impulse approximation, while T-odd asymmetries \cite{Boer:1997nt,Bacchetta:2006tn} are proportional to final-state interactions and can provide sensitive tests of their dynamics (such structures can be formed with polarized electron and unpolarized deuteron beams) \cite{Cosyn:2021inprep}. The $\phi$ dependence of the tagged DIS cross section also needs to be studied as a potential source of uncertainty in the measurement of the $\phi$-integrated cross sections, see Eqs.~(\ref{reduced_cross_section_phi_average}) and (\ref{pole_analysis_reduced_average}), in kinematic regions where the detector acceptance is 
effectively non-uniform in $\phi_p$ \cite{conditions}.

\subparagraph{Flavor tagging with semi-inclusive DIS.} Spectator nucleon tagging in deuteron DIS could be combined with measurements of semi-inclusive hadron production (pions, kaons) in the current fragmentation region of the active nucleon (``flavor tagging''). The possibility of measuring semi-inclusive DIS on the identified neutron as well as the proton would enable new studies of the nucleon flavor decomposition and the meson fragmentation functions, especially regarding unflavored and strange quark fragmentation \cite{Metz:2016swz}. Such measurements could be performed with the tagged neutron and proton in the deuteron as well as with the free proton; the comparison between the three would allow one to separate final-state interactions from 
initial-state structure. Even with the integrated luminosity $\sim$1 fb$^{-1}$ assumed in the present study, simultaneous binning in the spectator nucleon momentum and the semi-inclusive meson momentum should be possible. Such measurements would be attractive even without the full pole extrapolation selecting free nucleon configurations.

\subparagraph{Exclusive processes.} Spectator nucleon tagging in electron scattering on the deuteron could be used to measure hard exclusive processes on the neutron such as meson production and deeply-virtual Compton scattering (DVCS) \cite{Goeke:2001tz}. DVCS measurements on the neutron are important for probing the nucleon GPD $E$ appearing in the angular momentum sum rule, and for the flavor decomposition of the nucleon GPDs in general. In exclusive processes the active nucleon recoils with a momentum transfer $\sim$ few 100 MeV/$c$. When such processes occur in scattering on the deuteron, the recoiling active nucleon and the spectator experience strong final state interactions, which qualitatively change the spectator momentum distribution compared to the impulse approximation (similar to the case of diffractive DIS discussed in Sec.~\ref{subsec:discussion_theoretical}). Including these low-energy final-state interaction effects in the 
physics model and event generator is essential for realistic simulations of tagged exclusive processes. We note that the luminosity required for tagged exclusive processes are much more demanding than for tagged inclusive DIS, because the exclusive processes on the nucleon have low rates in themselves, and tagging with the deuteron further dilutes the statistics.

\subparagraph{Nuclear modifications.} Another class of applications of tagged DIS is the study of nuclear modifications of partonic structure (antishadowing at $x \sim 0.1$; EMC effect at $x \gtrsim$ 0.3). These applications use tagging at higher spectator momenta $p_p (p_n) \sim$ 300--600 MeV/$c$ (in the deuteron rest frame) to select small-size configurations where the nucleons are strongly interacting and the dynamical modifications are expected to be large. At such spectator momenta final-state interactions are generally large and cause qualitiative deviations from the IA. The analysis should focus on strategies to separate the effects of the initial-state modifications from those of final-state interactions, e.g. by using the different kinematic dependence of the effects. These measurements generally require higher luminosity than the extraction of free nucleon structure, because high-momentum tagging uses only a small fraction of the deuteron's momentum distribution, $\lesssim $1\% for $p_p (p_n) >$ 300 MeV/$c$. Simulations of tagged DIS at higher spectator momenta and the exploration of nuclear modifications with EIC will be reported in a forthcoming article \cite{Jentsch:inprep}.

\subparagraph{Polarized deuteron.} Polarized deuteron beams at EIC are regarded as technically possible and considered as a future option \cite{ref:EICCDR}. This would open the possibility of performing measurements of DIS and other high-energy processes on the polarized deuteron with spectator nucleon tagging. The measured spectator momentum controls the D/S wave ratio in the deuteron and thus fixes the spin structure of the $pn$ configuration during the high-energy process. This feature can be used to eliminate D-wave depolarization in the extraction of neutron spin structure, or to maximize the D-wave in the exploration of vector- or tensor-polarized spin
asymmetries \cite{Frankfurt:1983qs,Cosyn:2019hem,Cosyn:2020kwu}. The luminosity and polarization requirements of such measurements are under investigation \cite{AbdulKhalek:2021gbh,Cosyn:2014zfa}. A study of polarized DIS on $^3$He with spectator tagging at EIC has been reported in Ref.~\cite{Friscic:2021oti}.

\begin{acknowledgments}
The authors would like to thank Mark Baker and Elke Aschenauer for
valuable discussions on the BeAGLE event generator and DIS physics;
Mark Strikman, Vadim Guzey, and Wim Cosyn, for collaboration and discussions on
the theoretical treatment of deuteron structure and tagged DIS; and
Rocco Schiavilla, for helpful communication on low-energy deuteron structure. The authors would also like to thank the EIC project interaction region working group for their help in understanding the various impacts of the IR design.

The work of A.~Jentsch is supported by the U.S. Department of Energy under Award
DE-SC0012704. The work of Z.~Tu is supported by LDRD-039, the
U.S. Department of Energy under Award DE-SC0012704, and the Goldhaber
Distinguished Fellowship at Brookhaven National Laboratory. The work
of C.~Weiss is supported by the U.S.~Department of Energy, Office of
Science, Office of Nuclear Physics, under contract DE-AC05-06OR23177.
\end{acknowledgments}
\appendix
\section{Deuteron structure model}
\label{app:deuteron}
\subsection{Light-front spectral function}
\label{app:cm_mometum}
In this appendix we describe the elements of the deuteron structure model used in the event generation and physics analysis in the present study. This includes the construction of the deuteron light-front wave function and spectral function, the non-relativistic approximation, the nucleon pole and its parameters, and a minimal two-pole model of the wave function. These materials can be used in simulations of other high-energy scattering processes with deuteron breakup. Further information can be found in Refs.~\cite{Strikman:2017koc,Cosyn:2020kwu}.

In the light-front description of deuteron structure the $pn$ configurations are characterized by the proton
LF momentum variables $\alpha_p$ and $\bm{p}_{pT}$; see Eq.~(\ref{lf_momentum_def}). An alternative set of variables
is the proton 3-momentum $\bm{k}$ in the center-of-mass (CM) momentum of the $pn$ configuration. The
relation between the variables is
\begin{align}
\alpha_p \; &= \; 1 + \frac{k^z}{E(\bm{k})} ,
\hspace{2em}
\bm{p}_{pT} \; = \; \bm{k}_T ,
\end{align}
or, inversely,
\begin{align}
k^z \; = \; E(\bm{k}) \, (\alpha_p - 1),
\hspace{2em}
\bm{k}_T \; &= \; \bm{p}_{pT} ,
\label{k_from_LF}
\end{align}
where
\begin{align}
E(\bm{k}) \; &\equiv \; \sqrt{|\bm{k}|^2 + m_N^2} \; = \; \left[ \frac{|\bm{p}_{pT}|^2 + m_N^2}{\alpha_p (2 - \alpha_p)}
\right]^{1/2}
\label{E_k}
\end{align}
is the nucleon energy in the CM frame. The integration measures in the variables are related as
\begin{align}
\frac{d\alpha_p \; d^2 p_{pT}}{\alpha_p (2 - \alpha_p)}  
\; = \; \frac{d^3 k}{E(\bm{k})} .
\label{integration_k}
\end{align}
The use of the CM momentum variable provides a rotationally symmetric representation of light-front quantum mechanics
in the 2-body sector. The deuteron light-front wave function is represented in terms of a rotationally symmetric
wave function as
\begin{align}
& \Psi_d (\alpha_p, \bm{p}_{pT} ) \; = \; \widetilde\Psi_d (\bm{k}) ,
\label{wf_k}
\end{align}
with the normalization condition [see Eqs.~(\ref{wf_normalization}) and (\ref{integration_k})]
\begin{align}
& \int \frac{d\alpha_p \; d^2 p_{pT}}{\alpha_p (2 - \alpha_p)} \;
|\Psi_d (\alpha_p, \bm{p}_{pT} )|^2
\nonumber \\[1ex]
\; &= \; \int \frac{d^3 k}{E({\bm k})} \; |\widetilde\Psi_d (\bm{k})|^2
\; = \; 1 .
\label{normalization_k}
\end{align}

The light-front spectral function, which appears in the description of high-energy scattering processes
on the deuteron in the impulse approximation, is defined in terms of the light-front wave function by Eq.~(\ref{spectral_impulse}),
\begin{align}
& \mathcal{S}_d (\alpha_p, \bm{p}_{pT})
\; \equiv \; \frac{|\Psi_d(\alpha_p, \bm{p}_{pT})|^2}{2 - \alpha_p} .
\label{spectral_impulse_alt}
\end{align}
This function is related in a simple way to the density of the rotationally
symmetric wave function. Using Eq.~(\ref{wf_k}) and (\ref{integration_k}) one obtains
\begin{align}
& \mathcal{S}_d (\alpha_p, \bm{p}_{pT}) \; [2 (2\pi)^3] \; d\Gamma_p 
\nonumber \\[1ex]
&= \; |\Psi_d(\alpha_p, \bm{p}_{pT})|^2 \; \frac{d\alpha_p d^2p_{pT}}{\alpha_p (2 - \alpha_p)}
\nonumber \\[1ex]
&= \; |\widetilde\Psi_d(\bm{k})|^2 \; \frac{d^3 k}{E({\bm k})} .
\label{spectral_k}
\end{align}
Thus the product of the light-front spectral function and the light-front phase space element is equal to the
product of the rotationally symmetric momentum distribution and its phase space element.
This illustrates how rotational invariance is recovered in the light-front description and allows
one to make connection with the non-relativistic theory. Using the correspondence Eq.~(\ref{spectral_k}),
the reduced cross section of tagged DIS in the impulse approximation, Eq.~(\ref{sigma_red_ia}), can be expressed
in terms of the CM momentum variable as
\begin{align}
& \bar\sigma_{{\rm red}, d}(x, Q^2; \alpha_p, p_{pT}) \; d\Gamma_p 
\nonumber
\\[1ex]
&= \; \sigma_{{\rm red}, n} (x_n, Q^2) \;  [2(2\pi)^3] \; \mathcal{S}_{d} (\alpha_p, p_{pT}) \; d\Gamma_p 
\nonumber
\\[1ex]
&= \; \sigma_{{\rm red}, n}(x_n, Q^2) \; |\widetilde\Psi_d(\bm{k})|^2 \; \frac{d^3 k}{E({\bm k})} .
\end{align}
\subsection{Proton and neutron momenta}
\label{app:proton_neutron}
In the light-front description the proton and neutron light-front momenta in each $pn$ configuration add up to the total light-front momentum of the deuteron bound state. The neutron light-front momentum in a configuration with
proton light-front momentum $\alpha_p$ and $\bm{p}_{pT}$ is given by
\begin{align}
\alpha_n \; = \; 2 - \alpha_p,  \hspace{2em} \bm{p}_{nT} \; = \; - \bm{p}_{pT}.
\end{align}
The deuteron light-front wave function is symmetric under the exchange of proton and neutron variables
(isospin symmetry)
\begin{align}
\Psi_d (\alpha_n, \bm{p}_{nT}) \; = \;
\Psi_d (2 - \alpha_p, -\bm{p}_{pT} ) \; = \; \Psi_d (\alpha_p, \bm{p}_{pT}).
\label{proton_neutron_wf}
\end{align}
The wave function can therefore equivalently be regarded as a function of the neutron momentum;
because
\begin{align}
\int \frac{d\alpha_p \; d^2 p_{pT}}{\alpha_p (2 - \alpha_p)} \, [...]
\; = \; \int \frac{d\alpha_n \; d^2 p_{pT}}{\alpha_n (2 - \alpha_n)} \, [...] ,
\end{align}
the normalization condition Eq.~(\ref{normalization_k}) takes the same form in the neutron variables.
In terms of the CM momentum variable the neutron has momentum $-\bm{k}$, and the wave function satisfies
\begin{align}
\widetilde\Psi_d (-\bm{k}) \; = \; \widetilde\Psi_d (\bm{k}).
\end{align}
In high-energy scattering on the deuteron with neutron tagging, the impulse approximation involves the spectral function
\begin{align}
& \mathcal{S}_d (\alpha_n, \bm{p}_{nT})
\; \equiv \; \frac{|\Psi_d(\alpha_n, \bm{p}_{nT})|^2}{2 - \alpha_n},
\label{spectral_impulse_alt_neutron}
\end{align}
where $\alpha_n$ and $\bm{p}_{nT}$ are the light-front momentum variables of the detected neutron;
Eq.~(\ref{spectral_impulse_alt_neutron}) is the same mathematical function as the spectral
function for proton tagging, Eq.~(\ref{spectral_impulse_alt}), only evaluated at the neutron
variables $\alpha_n$ and $\bm{p}_{nT}$. Note that the spectral function is not symmetric under
the exchange of proton and neutron momenta,
\begin{align}
\mathcal{S}_d (2 - \alpha_p, -\bm{p}_{pT}) \; \neq \; \mathcal{S}_d (\alpha_p, \bm{p}_{pT}),
\label{proton_neutron_spectral}
\end{align}
because the definition Eq.~(\ref{spectral_impulse_alt}) contains the flux factor $1/(2 - \alpha_p)$.
However, the product of the spectral function and the phase space element is symmetric under
proton-neutron exchange,
\begin{align}
& \mathcal{S}_d (\alpha_p, \bm{p}_{pT}) \; [2 (2\pi)^3] \; d\Gamma_p
\nonumber \\[1ex]
&= \; |\Psi_d(\alpha_p, \bm{p}_{pT})|^2 \; \frac{d\alpha_p d^2p_{pT}}{\alpha_p (2 - \alpha_p)}
\nonumber \\[1ex]
&= \; |\Psi_d(\alpha_n, \bm{p}_{nT})|^2 \; \frac{d\alpha_n d^2p_{nT}}{\alpha_n (2 - \alpha_n)}
\nonumber \\[2ex]
&= \; \mathcal{S}_d (\alpha_n, \bm{p}_{nT}) \; [2 (2\pi)^3] \; d\Gamma_n .
\end{align}
This can also be seen directly from the expression in terms of the $\bm{k}$-vector variable,
Eq.~(\ref{spectral_k}), which is symmetric under $\bm{k} \rightarrow -\bm{k}$.

The treatment of deuteron structure described here assumes isospin symmetry and neglects the
difference between the proton and neutron masses. The nucleon mass in Eq.~(\ref{E_k}) is defined
as the average of the physical proton and neutron masses,
\begin{align}
m_N \; \equiv \; \textstyle{\frac{1}{2}} (m_p + m_n).
\label{nucleon_mass_average}
\end{align}
With this definition the relation between the deuteron mass and the binding energy takes the form
\begin{align}
M_d \; = \; m_p + m_n - \epsilon_d \; = \; 2 m_N - \epsilon_d ,
\end{align}
i.e., the use of the average nucleon mass preserves the	relation between the physical deuteron mass
and the binding energy. This circumstance is important when matching the deuteron structure model
with simulation	codes that use the exact proton and neutron masses.
\subsection{Spin degrees of freedom}
\label{app:spin}
For reference we want to summarize also the treatment of the deuteron and nucleon spin degrees of freedom.
The deuteron light-front wave function depends on the light-front helicity variables of the deuteron and the nucleons,
$\lambda_d = \{1, 0, -1\}$ and $\lambda_{p, n} = \{ \frac{1}{2}, -\frac{1}{2} \}$,
\begin{align}
\Psi_d(\alpha_p, \bm{p}_{pT}; \lambda_p, \lambda_n | \lambda_d).
\end{align}
The representation in terms of the CM momentum variable Eq.~(\ref{wf_k}) exhibits the underlying rotational
invariance of the light-front wave function and allows one to infer its spin structure from that of the rotationally
invariant wve function. It is given by
\begin{align}
& \Psi_d (\alpha_p , \bm{p}_{pT}; \lambda_p, \lambda_n | \lambda_d)
\; = \; \sum_{\lambdapp, \lambdanp}
\widetilde \Psi_d (\bm{k}, \lambdanp, \lambdapp | \lambda_d )  \;
\nonumber \\
& \hspace{3em} \times U^\ast (\bm{k}, \lambdapp, \lambda_p ) \; U^\ast(-\bm{k}, \lambdanp, \lambda_n ) ,
\label{wf_3d}
\end{align}
where ${\lambda'}_{\!\! p, n}$ are the canonical spin variables and $U^\ast$ are the Melosh rotations
connecting the light-front helicity with the canonical spin variables, whose explicit form is given in
Ref.~\cite{Cosyn:2020kwu} and not needed here (the deuteron light-front helicity is the same as
its canonical spin because $\bm{p}_{Td} = 0$). The normalization condition including spins is
\begin{align}
&\sum_{\lambda_p, \lambda_n}
\int \frac{d\alpha_p \; d^2 p_{pT}}{\alpha_p (2 - \alpha_p)} \;
\Psi_d^\ast (\alpha_p, \bm{p}_{pT}; \lambda_p, \lambda_n | \lambdadp)
\nonumber
\\
& \;\; \times \Psi_d (\alpha_p, \bm{p}_{pT}; \lambda_p, \lambda_n | \lambda_d)
\nonumber
\\[1ex]
= \; &\sum_{\lambdapp, \lambdanp}
\int \frac{d^3 k}{E(\bm{k})} \;
\widetilde \Psi_d^\ast (\bm{k}, \lambdanp, \lambdapp | \lambda_d )
\widetilde \Psi_d (\bm{k}, \lambdanp, \lambdapp | \lambda_d )
\nonumber \\[1ex]
= & \;\; \delta (\lambdadp, \lambda_d) .
\end{align}
The unpolarized deuteron spectral function, Eq.~(\ref{spectral_impulse}) and
(\ref{spectral_impulse_alt}) is defined as the sum over the nucleon light-front helicities and the
average over the deuteron light-front helicity,
\begin{align}
& S_d (\alpha_p, \bm{p}_{pT})
\; \equiv \; \frac{1}{3} \sum_{\lambda_d} \sum_{\lambda_p, \lambda_n}
\frac{|\Psi_d (\alpha_p, \bm{p}_{pT}; \lambda_p, \lambda_n | \lambda_d)|^2}{2 - \alpha_p} .
\label{spectral_spin}
\end{align}
The expression in Eq.~(\ref{spectral_k}) becomes
\begin{align}
[...] \; = \; \frac{1}{3} \sum_{\lambda_d} \sum_{\lambdapp, \lambdanp}
|\widetilde \Psi_d (\bm{k}, \lambdanp, \lambdapp | \lambda_d )|^2 \; \frac{d^3 k}{E({\bm k})} .
\label{spectral_k_spin}
\end{align}

The rotationally symmetric wave function can be decomposed into S- and D-wave components with
orbital angular momentum $L = 0$ and $2$ in the $\bm{k}$ vector variable and corresponding spin structure
\cite{Cosyn:2020kwu}
\begin{align}
\widetilde \Psi_d (\bm{k}, \lambdanp, \lambdapp | \lambda_d )
\; \rightarrow \; \widetilde \Psi_d[L = 0] \; + \; \widetilde \Psi_d[L = 2] .
\end{align}
The S- and D-wave of the light-front wave function then follow from Eq.~(\ref{wf_3d}). The D-wave contributes
to the unpolarized spectral function only at large nucleon momenta $|\bm{p}_{p, n}| \gtrsim$ 200 MeV/$c$ (in the deuteron rest frame), which are not used in the present study of free nucleon structure extraction; the only role of the D-wave in this context is its contribution to the normalization of the wave function, which balances that of the S-wave (see below). 

In the text of this article and in the rest of this appendix we suppress the spin degrees of freedom for brevity.
The expressions of the unpolarized spectral function and the impulse approximation cross section are always
understood in the sense of Eq.~(\ref{spectral_spin}).
\subsection{Non-relativistic approximation}
\label{app:nonrelativistic}
The representation in terms of the CM momentum variable can be used to construct an approximation to the
LF wave function in terms of the non-relativistic deuteron wave function. The non-relativistic wave function
$\Phi_d(\bm{p})$ is a function of the ordinary proton 3-momentum $\bm{p} \equiv \bm{p}_p$
in the deuteron rest frame with normalization
\begin{align}
\int d^3 p \; |\Phi_d (\bm{p})|^2 \; = \; 1.
\label{nonrel_normalization}
\end{align}
An approximate relativistic wave function is provided by
\begin{align}
\widetilde\Psi_d(\bm{k}) \; \stackrel{{\rm app}}{=} \; \sqrt{E(\bm{k})} \, \Phi_d(\bm{k}) .
\label{wf_nonrel}
\end{align}
The non-relativistic wave function $\Phi_d$ on the RHS is evaluated at the momentum variable
$\bm{k}$ that is the argument of the relativistic wave function $\widetilde\Psi_{d}$ on the LHS. 
The factor $\sqrt{E({\bm k})}$ ensures that $\widetilde\Psi_{d}$ 
obeys the normalization condition Eq.~(\ref{normalization_k}) if $\Phi_d$
obeys Eq.~(\ref{nonrel_normalization}). 

The non-relativistic approximation is well justified and adequate at nucleon momenta $|\bm{p}_{p, n}|\lesssim$ 200 MeV/$c$ (in the deuteron rest frame), as are considered in the present study of low-momentum spectator tagging. In particular, the approximation Eq.~(\ref{wf_nonrel}) implements the analytic properties of the wave function at small momenta and the nucleon pole, which play an essential role in free nucleon
structure extraction (see Sec.~\ref{subsec:pole}). More generally, the approximation in Eq.~(\ref{wf_nonrel}) allows one to recruit the extensive results on deuteron structure from non-relativistic nuclear theory (with realistic $NN$ interactions) for the description of light-front structure and high-energy scattering processes.
\subsection{Nucleon pole}
\label{app:pole}
The rotationally invariant representation establishes the analytic properties of the deuteron light-front wave function and exhibits the nucleon pole, which dominates the behavior of the wave function at low momenta and plays an essential role in nucleon structure extraction. The pole occurs in the S-wave of the non-relativistic wave function and is of the form
\begin{align}
\Phi_d (\bm{p})[L = 0] \; \sim \; \frac{\Gamma}{|\bm{p}|^2 + a^2} ,
\label{pole_nonrel}
\end{align}
where the pole position is given by
\begin{align}
a^2 \; &\equiv \; \epsilon_d m_N ,
\label{a2_app}
\end{align}
and $\Gamma$ denotes the residue. In the relativistic wave function obtained from
Eq.~(\ref{wf_nonrel}) the pole is of the form
\begin{align}
\widetilde\Psi_{d} (\bm{k})[L = 0] \; \sim \; \frac{\sqrt{m_N} \, \Gamma}{|\bm{k}|^2 + a^2} ,
\label{pole_rel}
\end{align}
where the factor $\sqrt{E({\bm k})}$ at the pole has been approximated by its value at $|\bm{k}|^2 = 0$, 
$\sqrt{m_N}$ (corrections are $\sim\epsilon_d/m_N$ and negligible). 
Substituting $|\bm{k}^2|$ in Eq.~(\ref{pole_rel}) by its expression in terms of the light-front momentum variables, 
Eq.~(\ref{k_from_LF}) and (\ref{E_k}), one obtains the analytic structure of the light-front wave function as
in Eqs.~(\ref{pole_wf}), (\ref{a_T_def}), and (\ref{R_def}):
\begin{align}
\Psi_d (\alpha_p, \bm{p}_{pT})
\; &\sim \; \frac{R(\alpha_p)}{|\bm{p}_{pT}|^2 + a_T^2(\alpha_p)},
\\[1ex]
a_T^2(\alpha_p) \; &= \; (\alpha_p - 1)^2 m_N^2 + \alpha_p (2 - \alpha_p) a^2 ,
\\[2ex]
R(\alpha_p) \; &= \; \alpha_p (2 - \alpha_p) \, \sqrt{m_N} \, \Gamma .
\end{align}
This analytic structure is used for the pole extrapolation in $|\bm{p}_{pT}|^2$ at fixed $\alpha_p$
in Sec.~\ref{subsec:pole}.

The nucleon pole in the non-relativistic momentum-space wave function Eq.(\ref{pole_nonrel})
determines the large-distance behavior of the corresponding coordinate-space wave function.
The residue $\Gamma$ is related to the so-called asymptotic normalization constant of the S-state
wave function, $A_S$. The non-relativistic coordinate-space wave function is defined as
\begin{align}
& \Phi_d (\bm{r}) \; \equiv \; \int \frac{d^3 p}{(2\pi)^3}
\; e^{i\bm{p}\bm{r}} \; \Phi_d (\bm{p}) ,
\\[1ex]
& \int d^3 r \; |\Phi_d (\bm{r})|^2 \; = \; 1.
\end{align}
The asymptotic behavior of the S-state radial wave function is of the form
\begin{align}
& \Phi_d (\bm{r})[L = 0] \; \sim \; \frac{A_S}{\sqrt{4\pi}} \; \frac{e^{-ar}}{r}
\hspace{1em} (r \rightarrow \infty).
\label{A_S_def}
\end{align}
The exponential decay at large distances is determined by the scale
derived from the pole position of the momentum-space wave function, Eqs.(\ref{pole_nonrel})
and (\ref{a2_app}),
\begin{align}
a^{-1} = (\textrm{45 MeV})^{-1} = 4.3 \, \textrm{fm},
\end{align}
known as the Bethe-Peierls radius of the deuteron.
The relation between the constant $A_S$ and the residue $\Gamma$ is
\begin{align}
A_S \; = \; \sqrt{2} \pi \Gamma .
\label{A_S_from_Gamma}
\end{align}
The asymptotic normalization constant $A_S$ is measured in low-energy deuteron breakup 
reactions \cite{Ericson:1988gk}.
It can also be determined with high precision from deuteron bound state calculations using 
empirical $NN$ potentials \cite{Wiringa:1994wb}
or effective field theory- (EFT-) controlled interactions \cite{Piarulli:2014bda}.
This information can be used to determine the value and uncertainty of the residue $\Gamma$ 
needed for the pole extrapolation high-energy scattering. For reference, a compilation of 
the values of $A_S$ and $\Gamma$ from various sources is provided in Table~\ref{tab:pole}.
\begin{table}
\begin{tabular}{l|l|l}
\hline
Source & $A_S$ [fm$^{-1/2}$] & $\Gamma$ [GeV$^{1/2}$]
\\
\hline
Experiment \cite{Ericson:1988gk} & 0.8781 (44) & 0.08772 (44) \\
\hline
AV18 \cite{Wiringa:1994wb} & 0.8850 & 0.0885 \\ 
EFTa \cite{Piarulli:2014bda} & 0.8777 & 0.0877 \\
EFTb \cite{Piarulli:2014bda} & 0.8904 & 0.0890 \\
EFTc \cite{Piarulli:2014bda} & 0.8964 & 0.0896 \\
\hline
CS1995 \cite{CiofidegliAtti:1995qe} & 0.888 & 0.0888 \\
Two-pole Eq.~(\ref{Gamma2_twopole}) & 0.885 & 0.0885 \\
\hline
\end{tabular}
\caption{Values of the asymptotic S-state normalization constant in the deuteron nonrelativistic 
wave function, $A_S$, Eq.~(\ref{A_S_def}), and the corresponding values of the residue of the 
nucleon pole, $\Gamma$, Eqs.~(\ref{pole_nonrel}) and (\ref{A_S_from_Gamma}), obtained from 
experiments or theoretical models of deuteron structure.
Refs.~\cite{Wiringa:1994wb} and \cite{Piarulli:2014bda} are theoretical calculations based
on $NN$ interactions; Refs.~\cite{CiofidegliAtti:1995qe} and the two-pole parametrization
are empirical parametrizations of the deuteron momentum density.}
\label{tab:pole}
\end{table}

The BeAGLE MC generator uses the parametrization of the deuteron momentum density 
of Ref.~\cite{CiofidegliAtti:1995qe} for evaluating the tagged DIS cross section 
via Eqs.~(\ref{spectral_impulse_alt}) and (\ref{spectral_k}) (present code version 1.0, 
see Sec.~\ref{subsec:beagle}). For reference we quote here the nucleon pole parameters 
corresponding to this parametrization. The parametrization is provided by Eq.(74) 
of Ref.~\cite{CiofidegliAtti:1995qe}, for $A = 2$, with the input parameters given in Table A.1
of Ref.~\cite{CiofidegliAtti:1995qe}: $A_1^{(0)} = 157.4 \, \textrm{fm}^3,
B_1^{(0)} = 1.24 \, \textrm{fm}^2, C_1^{(0)} = 18.3 \, \textrm{fm}^2$.
In terms of these parameters the pole parameters in our convention Eq.(\ref{pole_nonrel}) 
are obtained as
\begin{align}
a^2 \; &= \; [C_1^{(0)}]^{-1} \; = \; 0.002121 \, \textrm{GeV}^2,
\label{a2_beagle}
\\[1ex]
\Gamma^2 \; &= \; \frac{\displaystyle A_1^{(0)} \exp [ B_1^{(0)}/C_1^{(0)} ]}
{\displaystyle 4\pi [C_1^{(0)}]^2}
\; = \; 0.007885 \, \textrm{GeV}.
\label{gamma2_beagle}
\end{align}
While the precise numerical values have no physical significance, they can be used for the numerical validation of the pole removal procedure in BeAGLE simulations as performed in the present study. Because the values of $a^2$ and $\Gamma$ used in various deuteron structure models show variations at the percent level (see Table~\ref{tab:pole}), one should use exactly the values Eq.~(\ref{a2_beagle}) and (\ref{gamma2_beagle}) in pole removal studies with BeAGLE, to avoid artifacts from mismatched parameters.
\subsection{Two-pole parametrization}
\label{app:twopole}
A minimal parametrization of the deuteron wave function is obtained by supplementing the nucleon pole in the S-state with a second ``effective'' pole, which effectively accounts for the high-momentum components (two-pole parametrization, or Hulthen wave function). It is of the form 
\begin{align}
\Phi_d (\bm{p})_{\rm two-pole} \; = \; \frac{1}{\sqrt{c}}
\left( \frac{1}{|\bm{p}|^2 + a^2} - \frac{1}{|\bm{p}|^2 + b^2} \right) , \!
\label{twopole}
\end{align}
where $a$ is the nucleon pole position Eq.~(\ref{a2_app}), $b$ is the position of the effective pole,
and $c$ is fixed by the normalization condition Eq.~(\ref{nonrel_normalization}) as
\begin{align}
c \; = \; \frac{\pi^2 (a - b)^2}{ab (a + b)} .
\end{align}
The residue of the nucleon pole in Eq.~(\ref{twopole}) is given by
\begin{align}
\Gamma \; = \; 1/\sqrt{c} .
\label{Gamma2_twopole}
\end{align}
The value of $a$ is calculated from Eq.~(\ref{a2_app}) using $m_N = 0.939$ GeV
and $\epsilon_d = 2.23$ MeV,
\begin{align}
a \; = \; \sqrt{m_N \epsilon_d} \; = \; 0.04576 \, \textrm{GeV} .
\label{a_twopole_value}
\end{align}
The value of $b$ is fixed empirically as \cite{Wong:1994sy}
\begin{align}
b \; = \; 0.2719 \, \textrm{GeV} .
\label{b_twopole_value}
\end{align}

The parametrization Eq.~(\ref{twopole}) embodies the correct analytic properties of the deuteron 
wave function at small momenta. The position and residue of the nucleon pole are given explicitly in terms of the model parameters, see Eq.~(\ref{Gamma2_twopole}). With the parameter values
of Eqs.~(\ref{a_twopole_value}) and (\ref{b_twopole_value}), the value of $\Gamma$
in the two-pole parametrization agrees with the AV18 result within $\ll 1\%$
(see Table~\ref{tab:pole}).
The two-pole parametrization also provides an excellent approximation to the unpolarized
deuteron momentum density obtained with realistic wave functions up to $|\bm{p}| \sim$ 300 MeV. 
Note that the two-pole parametrization uses only the S-wave, while realistic wave functions
have S- and D-waves; the unpolarized momentum densities are nevertheless very close because
the S-wave in the two-pole model is larger than that in realistic models 
at $|\bm{p}| \sim$ few 100 MeV/$c$ (it does not have a node) and makes up for the missing D-wave strength. As such the two-pole parametrization is fully adequate for unpolarized deuteron tagging at low momenta $|\bm{p}| \lesssim$ 100 MeV/$c$ and can be used for analytic and numerical studies of pole extrapolation.
\section{Detector simulations}
\label{app:resolution}
\subsection{Geometric acceptance}
%
%
\begin{figure*}[t]
\includegraphics[width=0.8\textwidth]{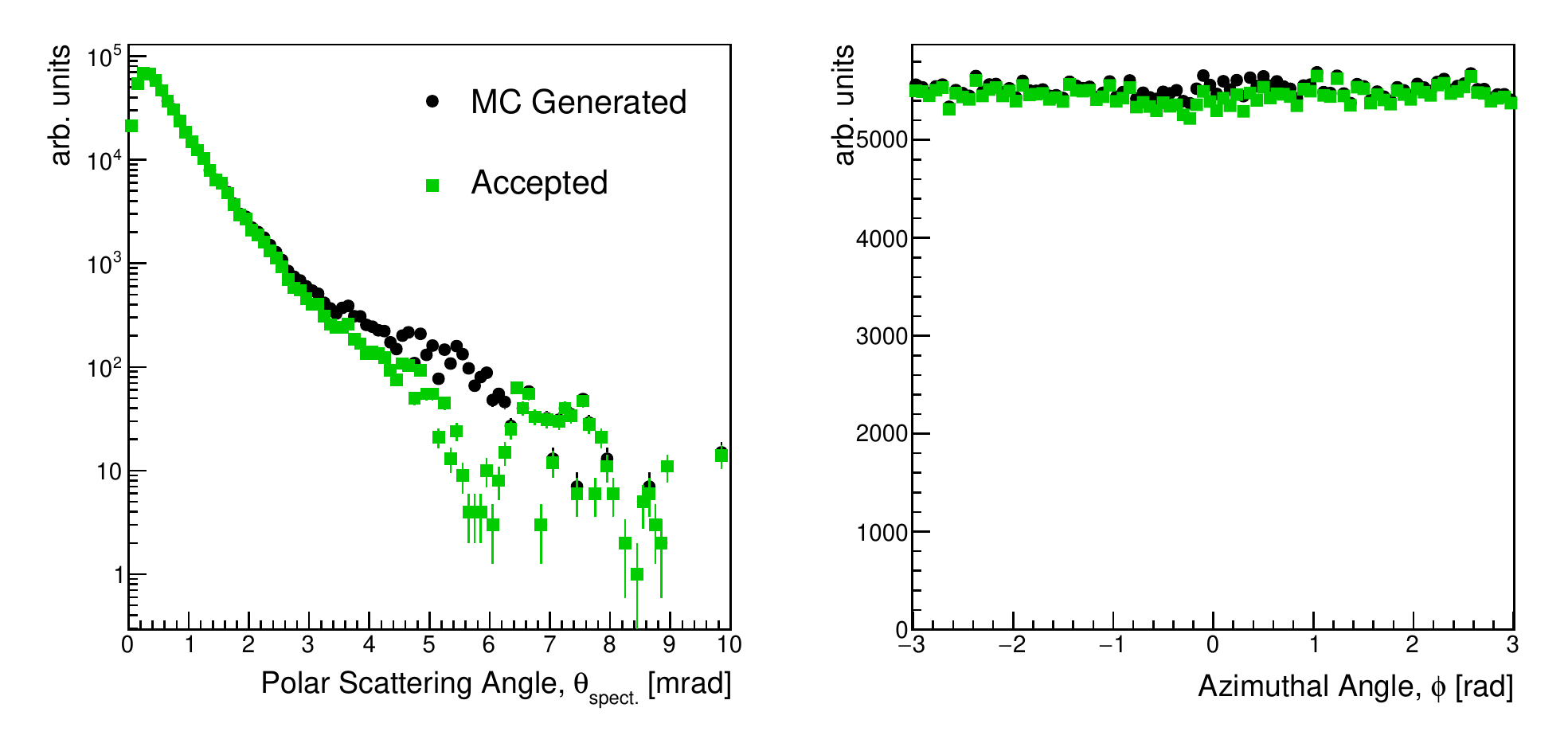}
\caption{Geometric acceptances in polar angle, $\theta_p$ (left), and
azimuthal angle, $\phi_p$ (right) for the spectator protons.}
\label{fig:proton_geo_plot}
\end{figure*}
%
%
\begin{figure*}[htb]
\includegraphics[width=0.8\textwidth]{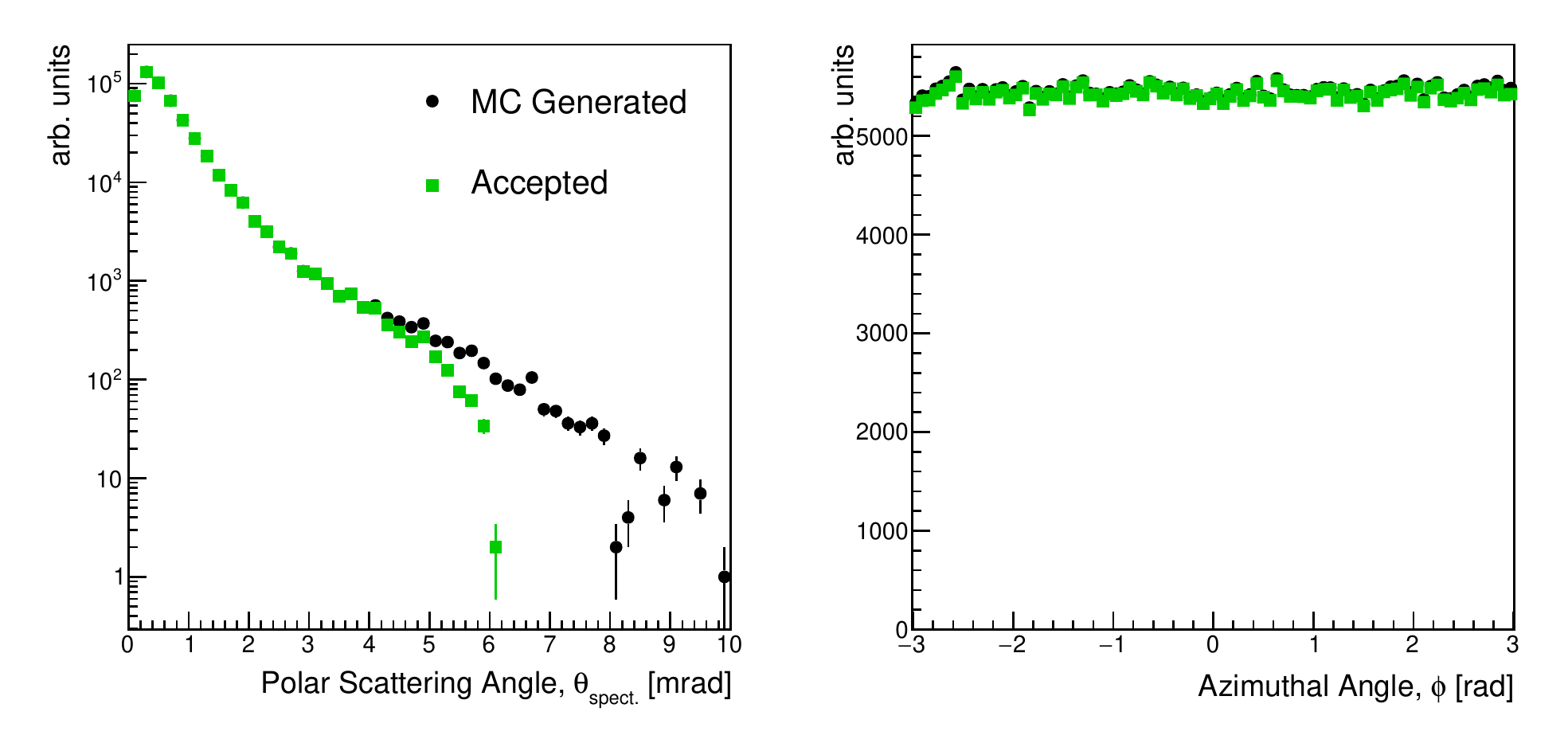}
\caption{Geometric acceptances in polar angle, $\theta_n$ (left), and
azimuthal angle, $\phi_n$ (right) for the spectator neutrons.}
\label{fig:neutron_geo_plot}
\end{figure*}
In this appendix we summarize the results of the full detector simulations that were used to quantify the EIC far-forward acceptance and the detector and beam effects on the momentum resolution. The simulations were performed by processing a subsample of the BeAGLE tagged DIS events with the EicRoot framework \cite{ref:EICROOT}, which implements the far-forward detectors in GEANT4 \cite{GEANT4} in the configuration specified in the EIC Yellow Report \cite{AbdulKhalek:2021gbh}. In the physics study in the main text, the acceptance and resolution were described by parametrizations based on these simulation results. The simulation results summarized here document these parametrizations, provide additional insight into the detector performance (e.g. the relative contribution of various effects on the momentum resolution), and can be used in similar physics studies.

The geometric acceptances were simulated with the full implementation of the EIC interaction region and the far-forward detectors
(see Fig.~\ref{fig:figure_IR}), including the geometry/size of the elements, the dipole and quadrupole fields of the optics,
and the transport of charged particles with magnetic rigidities different from that of the beam.
Figure~\ref{fig:proton_geo_plot} shows the geometric acceptance for spectator protons
in a deuteron beam with 110 GeV/nucleon energy, as used in the present study.
One observes that protons are fully accepted by the detector up to $\sim$2.8 mrad in polar
angle, at which point protons at $\phi = 0$ radians begin to be lost in the quadrupole magnets
due to their lower rigidity compared to the beam. The gap seen between 5 and 6 mrad is the
transition region between the Roman Pots/Off-Momentum Detector acceptance, and the
acceptance of the B0 spectrometer detector.

Figure~\ref{fig:neutron_geo_plot} shows the corresponding acceptance
for spectator neutrons. One sees that the neutron cone is detected with azimuthally symmetric acceptance up to 4.0 mrad, and acceptance up to $\sim$5.5 mrad for $\phi = 0$ radians. For the neutron spectators in the deuteron at 110 GeV/nucleon energy, the overwhelming majority are able to reach the detector, with the acceptance being 100\% for the kinematics relevant for pole extrapolation. The addition of the beam pipe, which is still being designed at the time of this analysis, will reduce the overall efficiency by 5-20\%, depending on the degree of optimization employed in the design (exit window, material, etc.). A study of a simplistic beam pipe design and its associated impact are discussed in Ref.~\cite{Chang:2021jnu}.
\subsection{Momentum resolution}
%
%
\begin{figure*}[htb]
\includegraphics[width=0.8\textwidth]{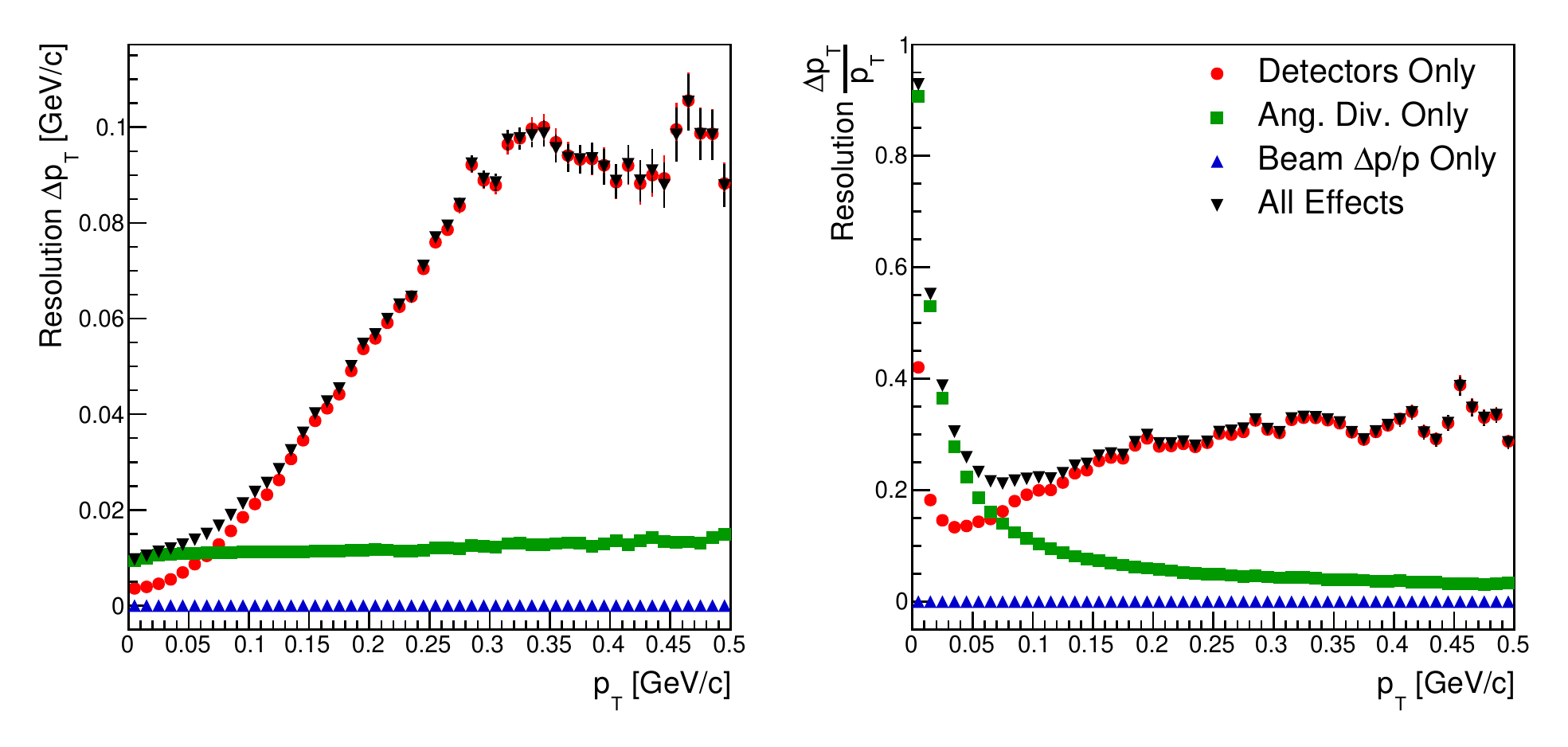}
\caption{Absolute (left) and relative (right) transverse momentum resolution for spectator protons obtained from the simulations. The plots show the contributions of the various effects and the total resolution.}
\label{fig:pt_res_proton}
\end{figure*}
%
%
\begin{figure*}[htb]
\includegraphics[width=0.8\textwidth]{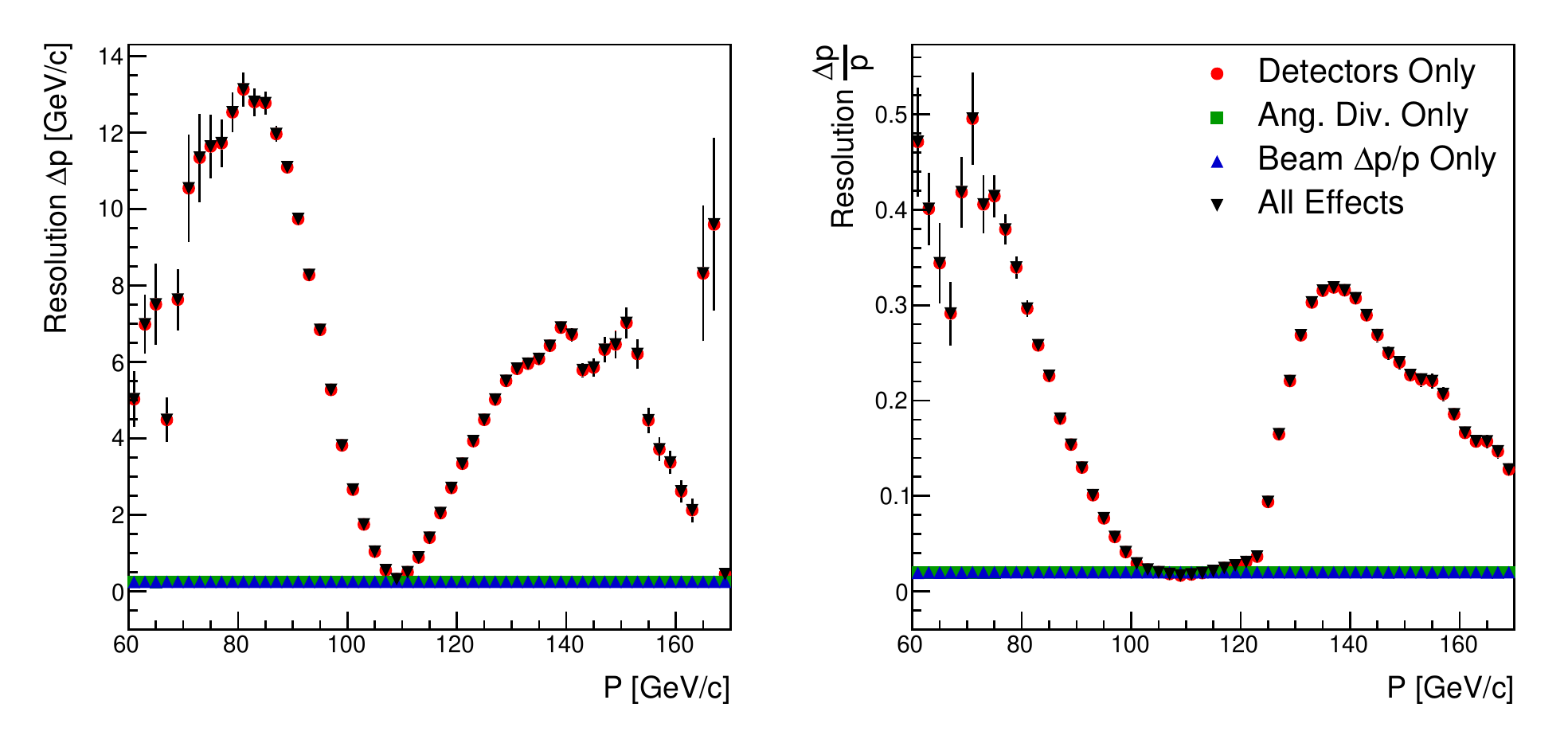}
\caption{Absolute (left) and relative (right) total three-momentum resolution for spectator protons.
The plots show the contributions of the various effects and the
total resolution.}
\label{fig:p_res_proton}
\end{figure*}
%
%
\begin{figure*}[htb]
\includegraphics[width=0.8\textwidth]{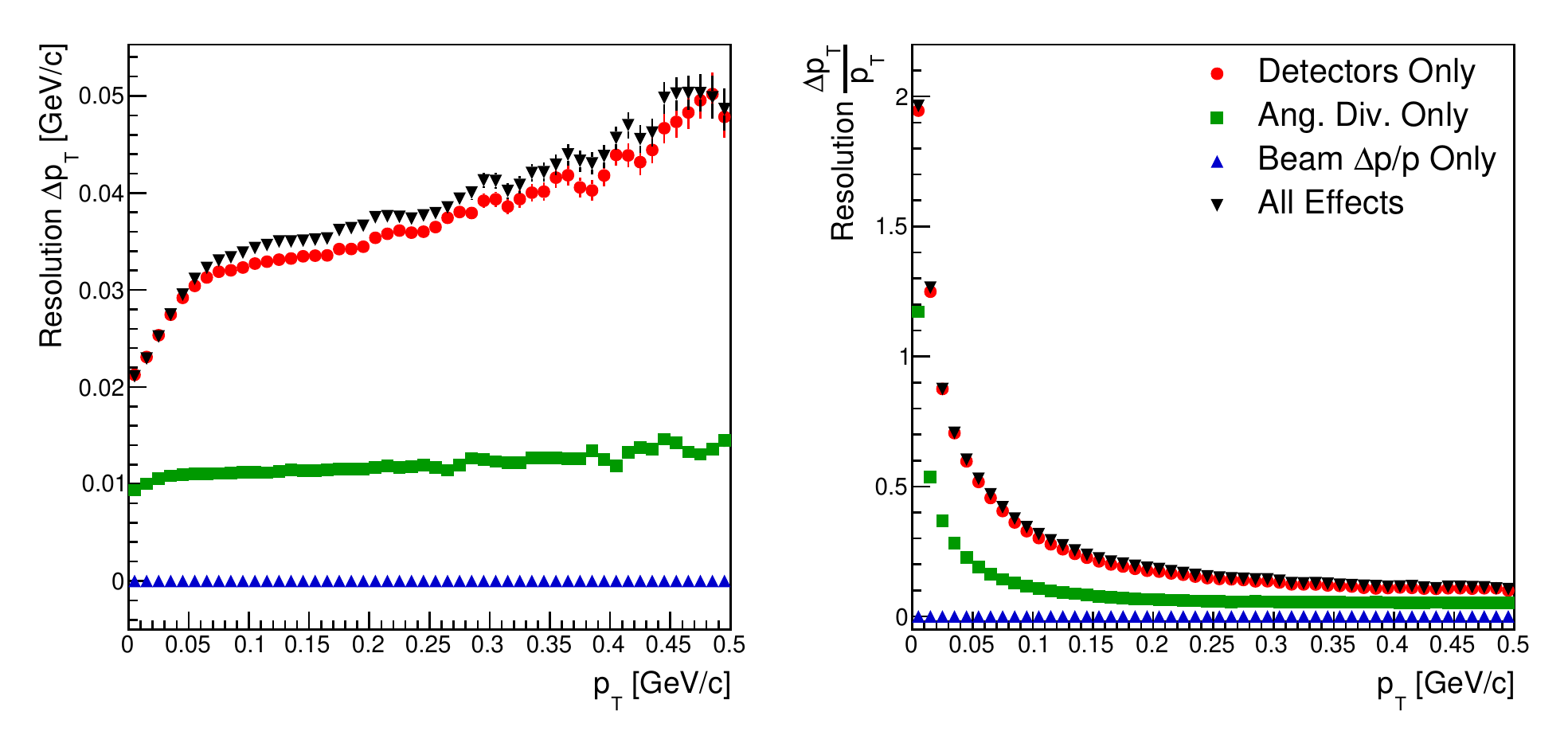}
\caption{Absolute (left) and relative (right) transverse momentum resolution for spectator neutrons.
The plots show the contributions of the various effects and the total resolution.}
\label{fig:pt_res_neutron}
\end{figure*}
%
%
\begin{figure*}[htb]
\includegraphics[width=0.8\textwidth]{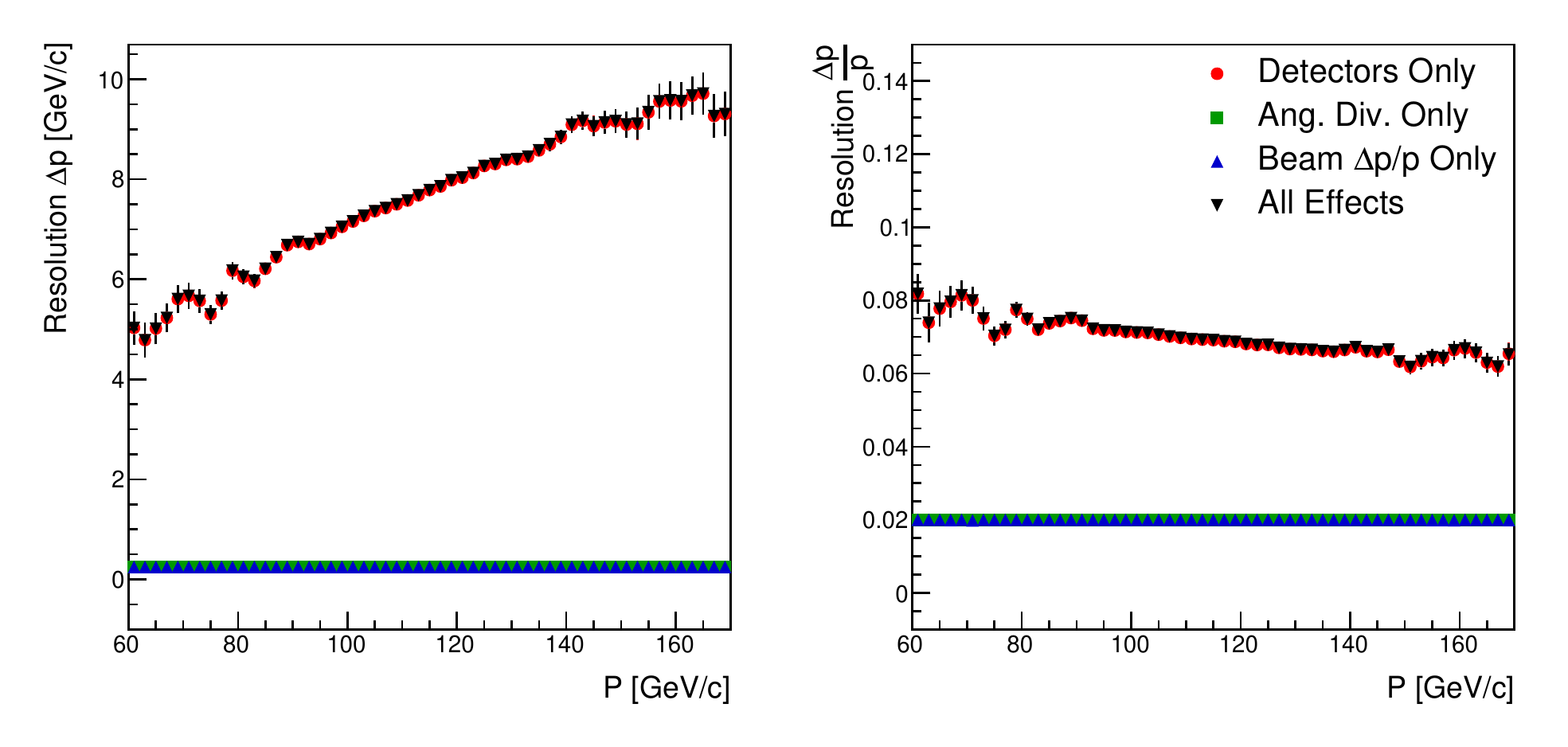} \caption{Absolute
(left) and relative (right) total three-momentum resolution for spectator neutrons.
The plots show the contributions of the various effects and the total resolution.}
\label{fig:p_res_neutron}
\end{figure*}
The resolutions measured using full GEANT4 simulations including effects of detector-level reconstruction (e.g. finite pixel size, energy resolution)
and effects related to the beam, which are modeled by smearing the final-state particle vectors. The angular divergence smearing is included by taking the RMS angular divergence $\Delta\theta_{x,y}$ values from Table 3.4 in the EIC Conceptual Design Report~\cite{ref:EICCDR} as the widths ($\sigma$) for random Gaussian smearing in both $p_{x}$ and $p_{y}$, with a mean of zero (assuming the beams would have no transverse components of momentum without the divergence). Then, the final-state particle vector is boosted to the rest-frame of the unsmeared deuteron (ion rest-frame), which has only a longitudinal momentum component. The particle vector is then boosted back to the lab-frame, but using a boost-vector from the deuteron beam vector now containing the randomly smeared $p_{x}$ and $p_{y}$ components, simulating the effect of having a deuteron beam with initial transverse momentum components being carried to the final-state particle vectors in the lab frame. This approach only applies reconstruction smearing to the transverse momentum.

Figures~\ref{fig:pt_res_proton} and \ref{fig:p_res_proton} summarize the resolution for spectator protons.
Figure~\ref{fig:pt_res_proton} shows the proton transverse momentum resolution.
One observes that at lower values of $p_{T}$ the angular divergence is the
dominant factor in the reconstruction smearing, but quickly becomes sub-dominant to the detector effects at higher $p_{T}$. At higher values of $p_{T}$($>$ 100 MeV/$c$), most of the additional smearing from the detector comes from the assumption of a linear transport matrix with decoupled x and y momentum components, which is an overly simplistic assumption to make in the case where the protons are severely divergent from the beam rigidity. The matrix is tuned for a trajectory from a proton with $\sim$50\% rigidity compared to the beam, allowing the smearing effect to be minimized at that point. This will be corrected in a future analysis as more time allows for a more sophisticated
approach to be developed, but does not affect the kinematic region relevant to the
pole-extrapolation presented in this work.

Figures~\ref{fig:pt_res_neutron} and \ref{fig:p_res_neutron} summarize the corresponding resolutions for neutron spectators. Figure~\ref{fig:pt_res_neutron} shows the neutron transverse momentum resolution.
One observes that the detector resolution accounts for the majority of the overall
reconstruction smearing, with the angular divergence contributing to increased reconstruction
smearing especially at lower values of $p_{T}$. The beam-momentum uncertainty contributes a
negligible amount to the overall reconstruction smearing. Figure~\ref{fig:p_res_neutron}
shows the neutron total (longitudinal) momentum resolution. Since
the angular divergence only acts on the transverse momentum components, the three-momentum is
dominated by the resolution of the detector. The overall momentum smearing in the case of the
neutrons is worse than for the protons due to the energy resolution under consideration for the
Zero-Degree Calorimeter for the EIC, as discussed in Sec.\ref{subsec:detectors}. As seen in the case
of the proton, the beam-momentum uncertainty plays a negligible role in the overall reconstruction
smearing for neutrons.

\begin{figure*}[htb]
\includegraphics[width=0.4\textwidth]{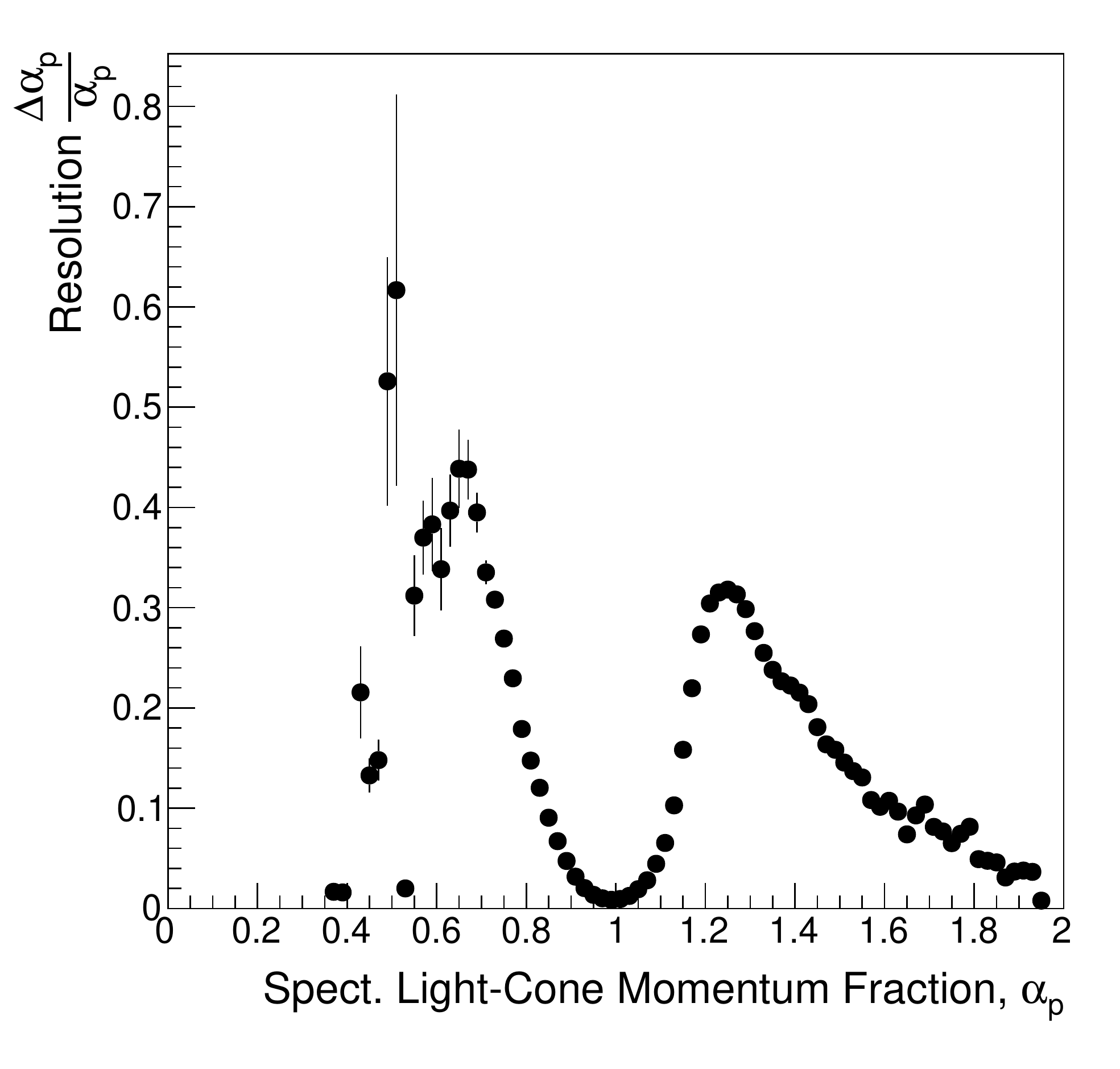}
\includegraphics[width=0.4\textwidth]{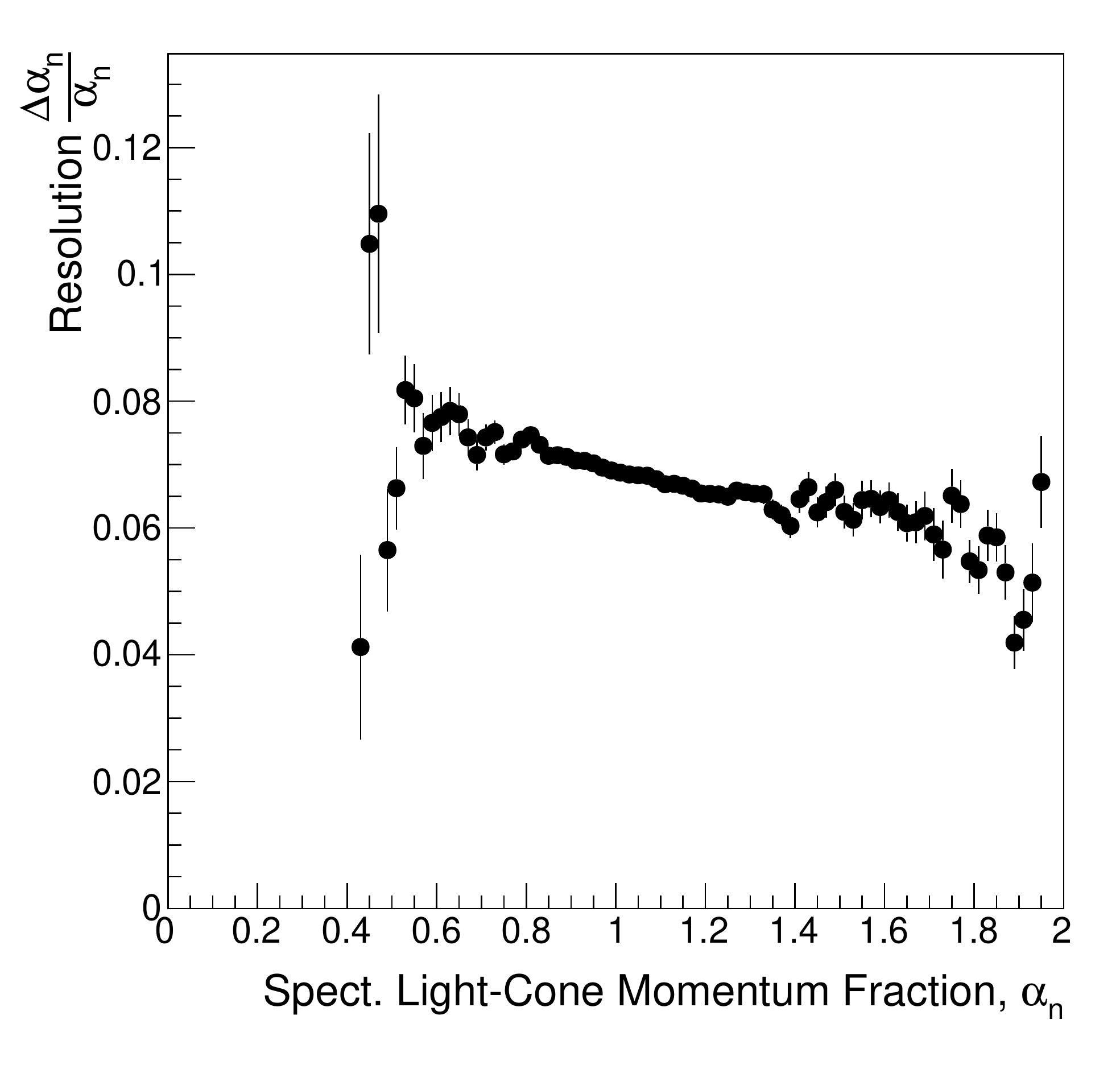}
\caption{Resolution in the light-front momentum fraction, $\alpha$, for protons (left) and neutron
(right).}
\label{fig:alpha_reso_plots}
\end{figure*}

Figure~\ref{fig:alpha_reso_plots} shows the longitudinal momentum resolution for proton and neutron
spectators directly in the physics variables $\alpha_p$ and $\alpha_n$, see Eqs.~(\ref{theta_zeta_from_collinear}).
One observes that, in the area of interest to the present physics study, $\alpha \approx 1$, the resolution
for protons is $<$1\%, while for neutrons it is closer to $\sim$7\%. This difference is essential in
assessing the tagged cross section measurements and the different approaches to pole removal in the
present study (see Sec.~\ref{subsec:pole} and Fig.~\ref{fig:figure_poleremoval}).
We note that in the physics study the simulated resolutions in Fig.~\ref{fig:alpha_reso_plots}
were not applied directly, but were used to construct a smearing function that was applied to the data.
\bibliography{reference}
%
\end{document}